\documentclass[reprint, prx, notitlepage, showkeys,showpacs,superscriptaddress]{revtex4-1}
\usepackage{braket}
\usepackage{amsmath,amssymb,times,color,graphicx,multirow,amsfonts,color}
\usepackage[colorlinks=true,citecolor=blue,linkcolor=blue]{hyperref}
\usepackage{physics,bm,here}

\def\im{\mbox{Im }}
\def\mod{{\rm\ mod\ }}

\def\ker{\mbox{Ker }}

\newcommand{\bk}{\bm{k}}
\newcommand{\br}{\bm{r}}

\newcommand{\XBS}{X^{\text{BdG}}}
\newcommand{\XBSb}{X^{\text{BdG}}}
\newcommand{\AIb}{\{\text{AI}\}^{\text{BdG}}}
\newcommand{\mZ}{\mathbb{Z}}
\newcommand{\calT}{\mathcal{T}}

\begin{document}

\title{Refined symmetry indicators for topological superconductors in all space groups}
\author{Seishiro Ono}
\affiliation{Institute for Solid State Physics, University of Tokyo, Kashiwa 277-8581, Japan}

\author{Hoi Chun Po}
\affiliation{Department of Physics, Massachusetts Institute of Technology, Cambridge, Massachusetts 02139, USA}

\author{Haruki Watanabe}
\affiliation{Department of Applied Physics, University of Tokyo, Tokyo 113-8656, Japan}

\begin{abstract}
Topological superconductors are exotic phases of matter featuring robust surface states that could be leveraged for topological quantum computation. 
A useful guiding principle for the search of topological superconductors is to relate the topological invariants with the behavior of the pairing order parameter on the normal-state Fermi surfaces. The existing formulas, however, become inadequate for the prediction of the recently proposed classes of topological crystalline superconductors.
In this work, we advance the theory of symmetry indicators for topological (crystalline) superconductors to cover all space groups. Our main result is the exhaustive computation of the indicator groups for superconductors under a variety of symmetry settings.
We further illustrate the power of this approach by analyzing four-fold symmetric superconductors with or without inversion symmetry, and show that the indicators can diagnose topological superconductors with surface states of different dimensionalities or dictate gaplessness in the bulk excitation spectrum.
\end{abstract}

\maketitle

\section{Introduction}
\label{sec:intro}
Unconventional pairing symmetry in a superconductor indicates a departure from the  well-established BCS paradigm for superconductivity. Such systems, exemplified by the high-temperature superconductors like the cuprate, typically display a wealth of intricate, oftentimes mysterious, phenomena which are of great theoretical, experimental, and technological interest \cite{Norman196}. 
The physics of unconventional superconductors has gained a new dimension in the past decade thanks to the bloom in the understanding of topological quantum materials \cite{RevModPhys.83.1057, Ando-Fu, Sato_2017}. A hallmark of topological superconductors (TSCs) is the presence of robust surface states which correspond to Majorana fermions --- an exotic emergent excitation that can loosely be described as being half of an ordinary electron. These Majorana excitations might be harvested for topological quantum computation, and much effort has been paid to the experimental realization of such exotic phases of matter \cite{RevModPhys.87.137}.

The intense research effort on topological quantum materials has resulted in an ever increasing arsenal of experimentally verified topological (crystalline) insulators and semimetals, but the discovery of TSCs has proven to be much more challenging.
The theoretical landscape, however, has evolved rapidly in the recent years.
On the one hand, the complex problem of how the diverse set of spatial symmetries in a crystal can both prohibit familiar topological phases and protect new ones has largely been solved, with the theoretical efforts culminating in the production of general classifications for topological crystalline phases in a variety of symmetry settings \cite{Freed2013, PhysRevX.7.011020, PhysRevX.7.041069, PhysRevX.8.011040,PhysRevB.96.205106, PhysRevX.8.031070, Shiozaki2018, Song2018}. On the other hand, general theories for how crystalline symmetries can be used to identify topological materials has been developed \cite{Po2017, TQC}.
In particular, the method of symmetry indicators (SIs) \cite{Po2017} has enabled comprehensive surveys of topological materials among existing crystal structure databases and thousands of materials candidates have been uncovered \cite{Zhang2019, Vergniory2019,Tang2019}.

It is natural to ask if the theory of SIs could be employed to facilitate the discovery of TSCs. There are two main difficulties: first, unconventional superconductivity emerges out of strong electronic correlations, and for such systems  theoretical treatments employing different approximation schemes rarely converge to the same answers. 
Such debates could only be settled by meticulous experimental studies which could take years to be completed.
Second, even within the simplifying assumption that a mean-field Bogoliubov--de-Gennes (BdG) provides a satisfactory treatment for the system, the original theory of SIs falls short in identifying key examples of TSCs like the 1D Kitaev chain \cite{1811.08712,Skurativska2019,1907.13632} and its higher-dimensional analogues like the higher-order TSCs in 2D \cite{PhysRevLett.119.246401, PhysRevB.97.205136}.
We remark that alternative formulas relating the signs of the pairing order parameters on different Fermi surfaces and topological invariants also exist in the literature, but this approach requires more detailed knowledge on the system then just the symmetry  representations \cite{PhysRevB.81.134508}. Furthermore, the extension of these formulas for other crystalline and higher-order TSCs has only been achieved for specific examples \cite{PhysRevB.98.165144,1904.06361,Ahn2019}.

In this work, we address the second part of the problem by extending the theory of SIs to the study of TSCs described by a mean-field BdG Hamiltonian in any space group.
This is achieved by a refinement of the SI for TSCs, which was previously proposed in Refs.\ \onlinecite{Skurativska2019,1907.13632} and analyzed explicitly for inversion-symmetric systems. Technically, our results do not rely on the weak pairing assumption, which states that the superconducting gap scale is much smaller than the normal state bandwidth \cite{PhysRevLett.105.097001,PhysRevB.81.134508, PhysRevB.81.220504, 1811.08712}. In practice, however, the prediction from this method is most reliable when the assumption is valid. For such weakly paired superconductors, only two pieces of data are required to diagnose a TSC: (i) the normal-state symmetry representations of the filled bands at the high-symmetry momenta, and (ii) the pairing symmetry.

Our key result is the exhaustive computation of the refined SI groups for superconductors with or without time-reversal symmetry and spin-orbit coupling, which are tabulated in Appendix~\ref{XBSTable}. In the main text, we will first review the topology of superconductors in Sec.~\ref{sec:top}, followed by Sec.~\ref{sec:SI} in which we give an interpretative elaboration for the SI refinement proposed in Ref.\ \onlinecite{Skurativska2019,1907.13632}. As an example of the results, we will provide an in-depth discussion on the refined SIs for class DIII systems with $C_4$ rotation symmetry in Sec.~\ref{sec:C4}, and a summary of the SIs for other key symmetry groups is provided in Appendix~\ref{KeySGs}. 

Curiously, we discover that the refined $C_4$ SI is, like the Fu-Kane parity formula \cite{PhysRevB.76.045302} and the corresponding version for odd-parity TSC \cite{ PhysRevLett.105.097001,PhysRevB.81.220504},  linked to the $\mathbb Z_2$ quantum spin Hall (QSH) index in the ten-fold way classification of TSC. 
This link is established in Sec.~\ref{sec:WTSC}, and is perhaps surprising given the SI refinement captured TSCs with corner modes in systems with inversion symmetry \cite{Skurativska2019, 1907.13632}. To our knowledge, this also represents the first instance of diagnosing a QSH phase using a proper rotation symmetry. Instead of a reduction of the wave-function based formula for the topological index to the symmetry representations, as was done in the original Fu-Kane approach
\cite{PhysRevB.76.045302}, our argument relies on an introduction of a class of phases which we dub ``Wannierizable TSCs.'' We will conclude and highlight a few future directions in Sec.~\ref{sec:dis}.

\section{Topology of superconductors}
\label{sec:top}
In this section we review the framework of describing TSCs by Bogoliubov--de-Gennes (BdG) Hamiltonians as a preparation for formulating SIs in Sec.~\ref{sec:SI}. Our discussion elucidates the possibility of marginally topological SCs, which may be called fragile TSCs.

\subsection{Symmetry of Bogoliubov--de-Gennes Hamiltonian}
Let us consider the Hamiltonian $H_{\bm{k}}$ of the normal phase, which we assume to be a $D$-dimensional Hermitian matrix. We take a superconducting gap function $\Delta_{\bm{k}}$ that satisfies $\Delta_{\bm{k}}=-\xi\Delta_{-\bm{k}}^T$, which is also a square matrix with the same dimension.  The parameter $\xi$ can be either $+1$ or $-1$ depending on the physical realization. 
We then we form the $2D$-dimensional BdG Hamiltonian 
\begin{align}
\label{BdG}
H_{\bm{k}}^{\text{BdG}} \equiv \begin{pmatrix}
H_{\bm{k}}& \Delta_{\bm{k}} \\
\Delta_{\bm{k}}^{\dagger} & -H_{-\bm{k}}^{*}
\end{pmatrix}.
\end{align}
This Hamiltonian always possesses the particle-hole symmetry 
\begin{align}
&\Xi_DH_{\bm{k}}^{\text{BdG}}{}^*\Xi_D^\dagger=-H_{-\bm{k}}^{\text{BdG}},\\
&\Xi_D\equiv\begin{pmatrix}
&\xi\openone_D\\
+\openone_D&
\end{pmatrix}.
\label{PHS}
\end{align}
Here $\openone_D$ stands for the $D$-dimensional identity matrix. Throughout this work, all blank entries of a matrix should be understood as $0$.  The particle-hole symmetry in Eq.~\eqref{PHS} satisfies $\Xi_D^2=+\xi$. 
To see the consequence of the particle-hole symmetry, suppose that $\psi_{\bm{k}}^{\text{BdG}}$ is an eigenstate of $H_{\bm{k}}^{\text{BdG}}$ with an eigenvalue $E_{\bm{k}}$. Then the particle-hole symmetry implies that $\Xi_D\psi_{\bm{k}}^{\text{BdG}}{}^*$ is an eigenstate of $H_{-\bm{k}}^{\text{BdG}}$ with eigenvalue $-E_{\bm{k}}$.  We call the eigenvalue $E_{\bm{k}}$ the quasi-particle spectrum. The BdG Hamiltonian is gapped when the quasi-particle spectrum has a gap around $E=0$ for all $\bm{k}$.

Suppose that the Hamiltonian of the normal phase has a space group symmetry $G$. Each element $g\in G$ is represented by a unitary matrix $U_{\bm{k}}(g)$ that satisfies
\begin{align}
U_{\bm{k}}(g)H_{\bm{k}}U_{\bm{k}}(g)^\dagger=H_{g\bm{k}}.
\end{align}
If the gap function satisfies 
\begin{align}
U_{\bm{k}}(g)\Delta_{\bm{k}}U_{-\bm{k}}(g)^T= \chi_g \Delta_{g\bm{k}},
\end{align}
the spatial symmetry is encoded in the BdG Hamiltonian as
\begin{align}
&U_{\bm{k}}^{\text{BdG}}(g)H_{\bm{k}}^{\text{BdG}}U_{\bm{k}}^{\text{BdG}}(g){}^\dagger=H_{g\bm{k}}^{\text{BdG}},\\
&U_{\bm{k}}^{\text{BdG}}(g) \equiv \begin{pmatrix}
U_{\bm{k}}(g) &  \\
& \chi_g U_{-\bm{k}}^{*}(g)
\end{pmatrix},\label{UBdG}\\
&\Xi_DU_{\bm{k}}^{\text{BdG}}(g){}^*\Xi_D^\dagger=\chi_g^*U_{-\bm{k}}^{\text{BdG}}(g)
\label{UBdG2}
\end{align}
The one-dimensional representation $\chi_g$ of $G$ defines the symmetry property of the superconducting gap $\Delta_{\bm{k}}$. 

Finally, the BdG Hamiltonian has the time-reversal symmetry if there exists $U_{\mathcal{T}}$ such that
\begin{align}
U_{\mathcal{T}}H_{\bm{k}}^*U_{\mathcal{T}}^{\dagger} = H_{-\bm{k}},\quad U_{\mathcal{T}}\Delta_{\bm{k}}^*U_{\mathcal{T}}^{T} =\Delta_{-\bm{k}}.
\end{align}
The representation $\chi_g$ must be either $\pm1$ for all $g\in G$.  Then the representation of the time-reversal symmetry in the BdG Hamiltonian is
\begin{align}
&U_{\mathcal{T}}^{\text{BdG}}H_{\bm{k}}^{\text{BdG}}{}^*U_{\mathcal{T}}^{\text{BdG}}{}^{\dagger}=H_{-\bm{k}}^{\text{BdG}},\\
&U_{\mathcal{T}}^{\text{BdG}}= \begin{pmatrix} U_{\mathcal{T}} &  \\  & U_{\mathcal{T}}^*\end{pmatrix}.
\end{align}

When $\xi=+1$, which is usually the case for electrons, the BdG Hamiltonians without time-reversal symmetry fall into class D of the ten-fold Altland--Zirnbauer (AZ) symmetry classification. When the time-reversal symmetry is present and satisfies $U_{\mathcal{T}}^{\text{BdG}}U_{\mathcal{T}}^{\text{BdG}}{}^*=+1$, the symmetry class becomes BDI; when it satisfies $U_{\mathcal{T}}^{\text{BdG}}U_{\mathcal{T}}^{\text{BdG}}{}^*=-1$ instead the symmetry class is DIII. If we discard the particle-hole symmetry from class D, BDI, and DIII, they respectively reduce to class A, AI, and AII. In the presence of spin SU(2) symmetry for spinful electrons, $\xi$ effectively becomes $-1$~\cite{PhysRevB.78.195125,SatoFujimoto}. Then the system without TRS is class C and with the time-reversal symmetry $U_{\mathcal{T}}^{\text{BdG}}U_{\mathcal{T}}^{\text{BdG}}{}^*=+1$ is class CI. We can formally consider the case $U_{\mathcal{T}}^{\text{BdG}}U_{\mathcal{T}}^{\text{BdG}}{}^*=-1$, which is classified as class CII, but it may be difficult to be realized in electronic systems. The general discussions of this work applies to all of these symmetry classes with the particle-hole symmetry regardless of $\xi=+1$ or $-1$.

\subsection{Stacking of BdG Hamiltonians}
\label{stacked}
To carefully define the trivial SCs, let us introduce the formal stacking of two SCs by the direct sum of two BdG Hamiltonians
\begin{align}
H_{\bm{k}}^{\text{BdG}}\oplus H_{\bm{k}}^{\text{BdG}}{}',
\end{align}
in which $H_{\bm{k}}$ and $\Delta_{\bm{k}}$ in Eq.~\eqref{BdG} are respectively replaced with
\begin{align}
\begin{pmatrix}
H_{\bm{k}} &  \\
& H_{\bm{k}}'
\end{pmatrix},\quad
\begin{pmatrix}
\Delta_{\bm{k}}&  \\
& \Delta_{\bm{k}}'
\end{pmatrix}.
\end{align}
When the dimension of $H_{\bm{k}}'$ is $D'$, the stacked BdG Hamiltonian is $2(D+D')$-dimensional and has the particle-hole symmetry $\Xi_{D+D'}.$  

We furthermore assume that $H_{\bm{k}}^{\text{BdG}}$ and $H_{\bm{k}}^{\text{BdG}}{}'$ have the same spatial symmetry $G$. Their representations can be different but $\chi_g$ must be common. We define $U_{\bm{k}}^{\text{BdG}}(g)\oplus U_{\bm{k}}^{\text{BdG}}(g)'$ by replacing $U_{\bm{k}}(g)$ in Eq.~\eqref{UBdG} with
\begin{align}
\begin{pmatrix}
U_{\bm{k}}(g) &  \\
& U_{\bm{k}}(g)'
\end{pmatrix}.
\end{align}
The possible time-reversal symmetry of the stacked SC is defined in the same way.

\subsection{Trivial superconductors}
Let us now define the topologically trivial class of SCs. Our discussion is inspired by the recent proposal in Refs.~\onlinecite{Skurativska2019,1907.13632}.

Suppose that the BdG Hamiltonian $H_{\bm{k}}^{\text{BdG}}$ is gapped. We say $H_{\bm{k}}^{\text{BdG}}$ is strictly trivial when it can be smoothly deformed to either
\begin{align}
H^{\text{vac}}\equiv
\begin{pmatrix}
+\openone_D&  \\
& -\openone_D
\end{pmatrix},\label{trivial1v}
\end{align}
which describes the vacuum state where all electronic levels are unoccupied, or 
\begin{align}
H^{\text{full}}\equiv
\begin{pmatrix}
-\openone_D&  \\
& +\openone_D
\end{pmatrix},\label{trivial1f}
\end{align}
which represents the fully occupied state.  They are physically equivalent to the chemical potential $\mu=\pm\infty$ limit of $H_{\bm{k}}^{\text{BdG}}$.  Here the smooth deformation is defined by an interpolating BdG Hamiltonian $H_{\bm{k}}^{\text{BdG}}(t)$ with
\begin{align}
H_{\bm{k}}^{\text{BdG}}(0)=H_{\bm{k}}^{\text{BdG}},\quad H_{\bm{k}}^{\text{BdG}}(1)=H^{\text{vac}}\text{ or }H^{\text{full}}
\end{align}
that maintains both the gap in the quasi-particle spectrum and all the assumed symmetries for all $t\in[0,1]$.  Note that we \emph{do not} modify the representations, such as $\Xi_D$, $U_{\bm{k}}^{\text{BdG}}(g)$, and $U_{\mathcal{T}}^{\text{BdG}}$, of assumed symmetries during the process.  When a smooth deformation to $H^{\text{vac}}$ exists, we write
\begin{align}
H_{\bm{k}}^{\text{BdG}}\sim H^{\text{vac}}.\label{trivial12}
\end{align}
Similarly, 
\begin{align}
H_{\bm{k}}^{\text{BdG}}\sim H^{\text{full}}\label{trivial13}
\end{align}
when there is an adiabatic path to $H^{\text{full}}$. Under a space group symmetry, conditions \eqref{trivial12} and \eqref{trivial13} are generally inequivalent. The SC is strictly trivial when at least one of the two conditions are fulfilled.

\begin{figure}
\begin{center}
\includegraphics[width=0.99\columnwidth]{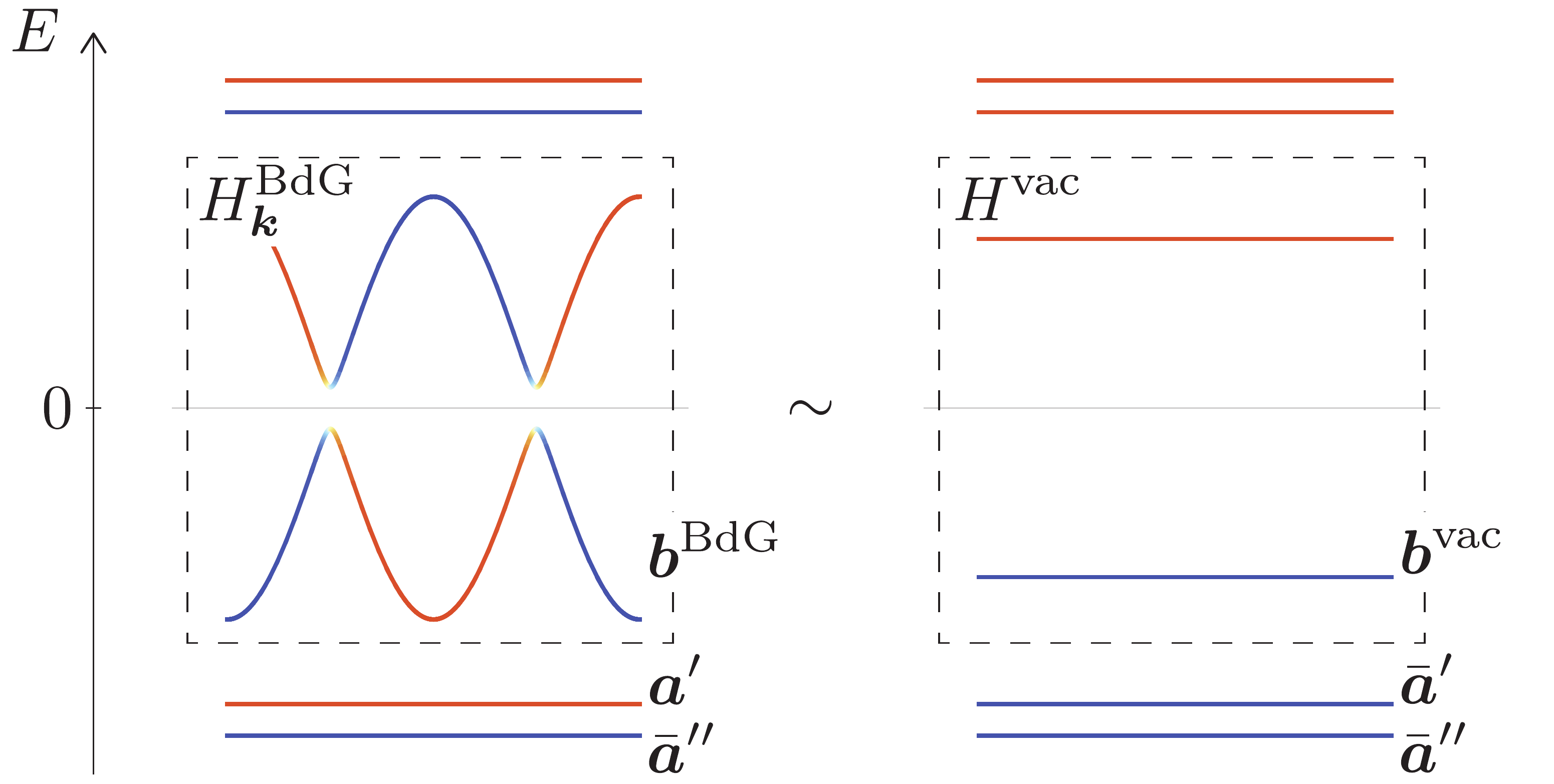}
\caption{\label{figequivalence}
Illustration of the equivalence relation in Eq.~\eqref{trivial2}.  Electron-like states are colored in red and hole-like states are colored in blue. States outside of the dashed box represent trivial DOFs included in the deformation process.   See Sec.~\ref{sec:SI} for the definition of vectors in this figure.}
\end{center}
\end{figure}

The above definition of trivial SCs is, however, sometimes too restrictive, especially under a spatial symmetry. One instead has to allow for adding trivial degrees of freedom (DOFs).  Using the notation summarized in Sec.~\ref{stacked}, we ask if
\begin{align}
&H_{\bm{k}}^{\text{BdG}}
\oplus\begin{pmatrix}
-\openone_{D'}&\\
& +\openone_{D'}
\end{pmatrix}
\oplus\begin{pmatrix}
+\openone_{D''}&\\
& -\openone_{D''}
\end{pmatrix}\notag\\
&\sim H^{\text{vac}}
\oplus\begin{pmatrix}
+\openone_{D'}&  \\
& -\openone_{D'}
\end{pmatrix}
\oplus\begin{pmatrix}
+\openone_{D''}&  \\
& -\openone_{D''}
\end{pmatrix}.\label{trivial2}
\end{align}
See Fig.~\ref{figequivalence} for the illustration.  The right-hand side of this equation is the same as Eq.~\eqref{trivial1v} but the identity matrix is enlarged to $+\openone_{D+D'+D''}$. 
We leverage the freedom in the choice of the matrix size $D'$, $D''$ and the symmetry representation $U_{\bm{k}}^{\text{BdG}}(g)'$, $U_{\bm{k}}^{\text{BdG}}(g)''$ of the trivial DOFs.
If, however, there does not exist any smooth path in Eq.~\eqref{trivial2} for whatever choice of trivial DOFs, then we say  $H_{\bm{k}}^{\text{BdG}}$ is stably topological. It might look unnatural to assign the flipped signs of $\openone_{D'}$ between the left and right hand side of Eq.~\eqref{trivial2}, but this choice is, in fact, necessary in the presence of space group symmetry in general. This point will become clear in Sec.~\ref{obstruction}.  Although Eq.~\eqref{trivial2} describes a smooth deformation to the vacuum state, one can equally consider a deformation to the fully-occupied state, which is mathematically equivalent to Eq.~\eqref{trivial2} as long as trivial DOFs are freely chosen.

These definitions of ``strictly trivial SCs" and ``stably topological SCs" leave a possibility of fragile topological phases~\cite{PhysRevLett.121.126402,PhysRevB.99.045140}, which becomes trivial if and only if appropriate trivial DOFs are added. We will discuss examples of these cases in Sec.~\ref{defexamples}.

\subsection{Examples}
\label{defexamples}
As an example of what we explained so far, let us discuss the odd-parity SC in the Kitaev chain~\cite{Kitaev_2001}.  

\subsubsection{Class D}
The BdG Hamiltonian  of a single Kitaev chain is given by Eq.~\eqref{BdG} with
\begin{align}
H_k=-\cos k,\quad 
\Delta_k=i\sin k.
\label{singleKitaev}
\end{align}
This model falls into the $\mathbb{Z}_2$ nontrivial phase in class D and is stably topological.

Let us take two copies of this model by setting
\begin{align}
H_k=-\cos k\,\openone_2,\quad
\Delta_k=i\sin k\,\openone_2.
\label{doubledKitaev}
\end{align}
For the doubled BdG Hamiltonian, there exists an adiabatic path for both Eq.~\eqref{trivial12} and \eqref{trivial13} given by the interpolating Hamiltonian
\begin{align}
H_k^{\text{BdG}}(t)&=\cos(\pi t/2)H_k^{\text{BdG}}\pm\sin(\pi t/2)
\begin{pmatrix}
+\openone&\\
&-\openone
\end{pmatrix}\notag\\
&\quad+\sin(\pi t)
\begin{pmatrix}
&\tau_2\\
\tau_2&
\end{pmatrix},
\label{kitaevpath1}
\end{align}
which preserves the particle-hole symmetry and the gap in the quasi-particle spectrum. Here $\tau_i$'s are the Pauli matrices.  
Therefore, the two copies of the Kitaev chain is strictly trivial.

\subsubsection{Class D with inversion symmetry}
Let us now take into account the inversion symmetry of the Kitaev chain. For the doubled model, the representation of the inversion symmetry is given by Eq.~\eqref{UBdG} with
\begin{align}
U_k(I)=\openone_2,\quad \chi_I=-1,
\end{align}
where $\chi_I = -1$ indicates odd-parity pairing. Under the inversion symmetry, the adiabatic path in the sense of Eq.~\eqref{trivial12} or Eq.~\eqref{trivial13} no longer exists even for the doubled Kitaev model. This can be easily seen by looking at the inversion parity of the quasi-particle spectrum below $E<0$. On the one hand, in the initial BdG Hamiltonian specified by Eq.~\eqref{doubledKitaev}, the inversion parity of two $E<0$ levels is both $+1$ at $k=0$ and $-1$ at $k=\pi$. (The opposite parity at $k=0$ and $\pi$ is a consequence of $\chi_I=-1$.) On the other hand, in the final trivial Hamiltonian $H^{\text{vac}}$ with $D=2$, the inversion parities of $E=-1$ levels are all $-1$. If we use $H^{\text{full}}$ instead as the trivial Hamiltonian,  the inversion parities are all $+1$.  This mismatch of inversion parities serves as an obstruction for any inversion-symmetric adiabatic deformation. Indeed, the path in Eq.~\eqref{kitaevpath1} breaks the inversion symmetry for $t\in (0,1)$.

To resolve the obstruction we introduce trivial DOFs with an appropriate inversion property. Specifically, we set $D'=2$, $D''=0$ and 
\begin{equation}
U_k(I)'=\begin{pmatrix}
-1&\\
&-e^{ik}
\end{pmatrix}.
\end{equation}
Now, the inversion parities of $E<0$ levels on both sides of Eq.~\eqref{trivial2} agree: two $+1$'s and two $-1$'s at $k=0$ and one $+1$ and three $-1$'s at $k=\pi$. In fact there exists an interpolating Hamiltonian
\begin{align}
&H_k^{\text{BdG}}(t)=\cos(\pi t/2)H_k^{\text{BdG}}
\oplus\begin{pmatrix}
-\openone_{2}&\\
&+\openone_{2}
\end{pmatrix}\notag\\
&\quad+\sin(\pi t/2)\begin{pmatrix}
+\openone_{4}&\\
& -\openone_{4}
\end{pmatrix}+\sin(\pi t)
\begin{pmatrix}
&\tilde{\Delta}_k\\
\tilde{\Delta}_k^\dagger&
\end{pmatrix},
\label{kitaevpath2}\\
&\tilde{\Delta}_k\equiv i
\begin{pmatrix}
0&0&0&1+e^{ik}\\
0&0&1&0\\
0&-1&0&1-e^{ik}\\
-1-e^{-ik}&0&-1+e^{-ik}&0.
\end{pmatrix}.
\end{align}
This confirms that the two copies of Kitaev chains with inversion symmetry becomes trivial if and only if proper trivial DOFs are added.

\section{Refined symmetry-indicators for superconductors}
\label{sec:SI}
In this section we discuss the formalism of SIs for SCs.
Our goal is to systematically diagnose the topological properties of SCs described by BdG Hamiltonians using their space group representation.
We also clarify the difference between the present approach extending the idea of Refs.~\cite{Skurativska2019, 1907.13632} and the previous approach in Refs.~\cite{PhysRevB.98.115150,1811.08712}.

\subsection{Symmetry representations of BdG Hamiltonians}
\label{repsBdG}
Let us consider a BdG Hamiltonian $H_{\bm{k}}^{\text{BdG}}$ in Eq.~\eqref{BdG} with a space group symmetry $G$ represented by $U_{\bm{k}}^{\text{BdG}}(g)$ in Eq.~\eqref{UBdG}.  We assume that the spectrum of $H_{\bm{k}}^{\text{BdG}}$ is gapped at least at all high-symmetry momenta.  

Suppose that $\psi_{\bm{k}}^{\text{BdG}}$ is an eigenstate of $H_{\bm{k}}^{\text{BdG}}$ and belongs to an irreducible representation $u_{\bm{k}}^\alpha$ of the little group $G_{\bm{k}}\leq G$ of $\bm{k}$ ($\alpha$ labels distinct irreducible representations). Then the particle-hole symmetry implies that $\Xi_D\psi_{\bm{k}}^{\text{BdG}}{}^*$ belongs to an irreducible representation $\chi_g(u_{\bm{k}}^\alpha)^*$ of $G_{-\bm{k}}$, which can be seen by Eq.~\eqref{UBdG2}. We write this correspondence among irreducible representations as
\begin{equation}
u_{\bm{k}}^{\bar{\alpha}}\equiv\chi_g(u_{-\bm{k}}^\alpha)^*\label{ufk}.
\end{equation}

The SI is formulated in terms of integers $(n_{\bm{k}}^{\alpha})^{\text{BdG}}$ that count the number of irreducible representations $u_{\bm{k}}^\alpha(g)$ of $G_{\bm{k}}$ appearing in the $E<0$ quasi-particle spectrum.  In other words, the little group representation formed by all the eigenstates with $E<0$ can be decomposed into irreducible representations as
\begin{align}
\oplus_{\alpha}(n_{\bm{k}}^{\alpha})^{\text{BdG}}u_{\bm{k}}^\alpha(g).
\end{align}
We denote by $K$ the collection of inequivalent high-symmetry momenta in the Brillouin zone. We compute $(n_{\bm{k}}^{\alpha})^{\text{BdG}}$ for all $\alpha$ and $\bm{k}\in K$ and form a vector $\bm{b}^{\text{BdG}}$ whose components are given by $(n_{\bm{k}}^{\alpha})^{\text{BdG}}$.  

For a later purpose, let us also define $(\bar{n}_{\bm{k}}^{\alpha})^{\text{BdG}}$ and $\bar{\bm{b}}^{\text{BdG}}$ using the $E>0$ quasi-particle spectrum in the same way.  Due to the particle-hole symmetry, we find
\begin{align}
(\bar{n}_{\bm{k}}^{\alpha})^{\text{BdG}}=(n_{-\bm{k}}^{\bar{\alpha}})^{\text{BdG}}\label{barn}.
\end{align}

Next, let us examine a trivial BdG Hamiltonian $H^{\text{vac}}$ in Eq.~\eqref{trivial1v} for which the space group $G$ is represented by the same matrix $U_{\bm{k}}^{\text{BdG}}(g)$ in Eq.~\eqref{UBdG} as for $H_{\bm{k}}^{\text{BdG}}$.  Observe that $E>0$ levels of $H^{\text{vac}}$ utilize $U_{\bm{k}}(g)$ as the representation of $G_{\bm{k}}$ for every $\bm{k}\in K$. Similarly, $E<0$ levels of $H^{\text{vac}}$ use $\chi_gU_{-\bm{k}}(g)^*$ as the representation of $G_{\bm{k}}$.  We define the integers $(\bar{n}_{\bm{k}}^{\alpha})^{\text{vac}}$ and $(n_{\bm{k}}^{\alpha})^{\text{vac}}$ by the irreducible decomposition
\begin{align}
U_{\bm{k}}(g)&=\oplus_{\alpha}(\bar{n}_{\bm{k}}^{\alpha})^{\text{vac}}u_{\bm{k}}^{\alpha}(g),\label{irrepdec1}\\
\chi_gU_{-\bm{k}}(g)^*&=\oplus_{\alpha}(n_{\bm{k}}^{\alpha})^{\text{vac}}u_{\bm{k}}^{\alpha}(g),\label{irrepdec2}
\end{align}
and construct vectors $\bar{\bm{b}}^{\text{vac}}$ and $\bm{b}^{\text{vac}}$ respectively using integers $(\bar{n}_{\bm{k}}^{\alpha})^{\text{vac}}$ and $(n_{\bm{k}}^{\alpha})^{\text{vac}}$.  
By construction, we have 
\begin{equation}
\bm{b}^{\text{BdG}}+\bar{\bm{b}}^{\text{BdG}}=\bm{b}^{\text{vac}}+\bar{\bm{b}}^{\text{vac}},\label{key1}
\end{equation}
since both sides of this equation denote the total representation counts in $U_{\bm{k}}^{\text{BdG}}(g)$.

Finally, we consider additional trivial DOFs described by
\begin{align}
H^{\text{full}}{}'\equiv\begin{pmatrix}
-\openone_{D'}&  \\
& +\openone_{D'}
\end{pmatrix}.
\end{align}
Suppose that the space group $G$ is represented by $U_{\bm{k}}^{\text{BdG}}(g)'$ in $H^{\text{full}}{}'$.  Then we perform the irreducible decomposition as in Eqs.~\eqref{irrepdec1} and \eqref{irrepdec2} and define $\bm{a}'$ and $\bar{\bm{a}}'$ using the coefficients for $U_{\bm{k}}(g)'$ and $\chi_gU_{-\bm{k}}(g)'{}^*$, respectively.

\subsection{Symmetry obstructions}
\label{obstruction}
Now we are ready to derive several obstructions for the smooth deformation in Eqs.~\eqref{trivial12}, \eqref{trivial13}, and \eqref{trivial2}.  A necessary (but not generally sufficient) condition for the existence of adiabatic paths in Eqs.~\eqref{trivial12} and \eqref{trivial13} is, respectively,
\begin{align}
\bm{b}^{\text{BdG}}&=\bm{b}^{\text{vac}},\label{condition12}\\
\bm{b}^{\text{BdG}}&=\bar{\bm{b}}^{\text{vac}}.\label{condition13}
\end{align}
When both of these conditions are violated, the representation counts in the $E<0$ spectrum of the initial and final BdG Hamiltonian in the deformation process do not agree and a smooth symmetric deformation is prohibited. We have seen this already in the doubled Kitaev model with inversion symmetry in Sec.~\ref{defexamples}.

Similarly, comparing the representation counts in the $E<0$ spectrum of the two end of the adiabatic path of Eq.~\eqref{trivial2}, we find the condition (see Fig.~\ref{figequivalence})
\begin{equation}
\bm{b}^{\text{BdG}}+\bm{a}'+\bar{\bm{a}}''=\bm{b}^{\text{vac}}+\bar{\bm{a}}'+\bar{\bm{a}}''.
\end{equation}
Therefore, a necessary condition for this adiabatic deformation is the existence of $\bm{a}'$ such that
\begin{equation}
\bm{b}^{\text{BdG}}-\bm{b}^{\text{vac}}=\bar{\bm{a}}'-\bm{a}'.\label{condition2}
\end{equation}
In other words, the mismatch in Eq.~\eqref{condition12} of the form $\bar{\bm{a}}'-\bm{a}'$ can be resolved by including trivial DOFs.  This is also what we have done for the doubled Kitaev model in Sec.~\ref{defexamples}.   
Note that $\bar{\bm{a}}''$ is canceled out from Eq.~\eqref{condition2}. Therefore, the trivial DOF in Eq.~\eqref{trivial2} with the same sign of $\openone_{D''}$ on both sides of the equation does not help as far as space group representations are concerned.

\subsection{Completeness of trivial limits}
\label{complete}
The above vector $\bm{a}'$ corresponds to the atomic limit of an insulator in class A, AI, or AII depending on the assumption on the time-reversal symmetry in $H_{\bm{k}}^{\text{BdG}}$.
As discussed in detail in Ref.~\cite{Po2017}, there are generally a variety of distinct atomic insulators in the presence of spatial symmetries. An atomic insulator can be specified by the position of the localized orbitals and the orbital character. These choices specify a representation $U_{\bm{k}}(g)'$ of $G_{\bm{k}}$ for each atomic insulator, and we write its representation count as $\bm{a}_j$ ($j$ labels distinct atomic insulators).  The set
\begin{equation}
\{\text{AI}\}=\Big\{\sum_{j}\ell_{j}\bm{a}_{j}|\ell_j\in\mathbb{Z}\Big\}\label{AI}
\end{equation}
is like a finite-dimensional vector space, except that the scalars are integers. We take a basis $\bm{a}_i$ ($i=1,2,\ldots,d$) of $\{\text{AI}\}$.  

Viewed as the representation counts in the valence bands of an insulator, it was proven in Ref.~\cite{Po2017} that 
 $\bm{b}^{\text{BdG}}$  can always be expanded in terms of $\bm{a}_i$'s using fractional (or integer) coefficients:
\begin{equation}
\bm{b}^{\text{BdG}}=\sum_iq_i\bm{a}_i,\quad q_i\in\mathbb{Q}.\label{bexpand}
\end{equation}
Since the left-hand side is integer valued, only special values of rational numbers are allowed.  The relation \eqref{bexpand} was the fundamental basis of the SIs for topological insulators.  Here we extend the argument for TSCs by proving that $\bm{b}^{\text{BdG}}-\bm{b}^{\text{vac}}$ can always be expanded in the following form:
\begin{align}
\bm{b}^{\text{BdG}}-\bm{b}^{\text{vac}}=\sum_ic_i(\bm{a}_i-\bar{\bm{a}}_i),\quad c_i\in\mathbb{Q}.\label{key2}
\end{align}
Readers not interested in the detail of the proof can skip to Sec.~\ref{quotient}.

To demonstrate Eq.~\eqref{key2}, note first that $\bm{b}^{\text{vac}}$ belongs to $\{\text{AI}\}$ and thus can be expanded as
\begin{equation}
\bm{b}^{\text{vac}}=\sum_ip_i\bm{a}_i,\quad p_i\in\mathbb{Z}.\label{aexpand}
\end{equation}
Also, Eqs.~\eqref{bexpand} and \eqref{aexpand} imply that $\bar{\bm{b}}^{\text{BdG}}=\sum_iq_i\bar{\bm{a}}_i$ and $\bar{\bm{b}}^{\text{vac}}=\sum_ip_i\bar{\bm{a}}_i$, which can be verified using Eq.~\eqref{barn}.  Then it follows that
\begin{align}
&\bm{b}^{\text{BdG}}-\bm{b}^{\text{vac}}=\sum_i(q_i-p_i)\bm{a}_i\notag\\
&=\frac{1}{2}\sum_i[(q_i-p_i)(\bm{a}_i-\bar{\bm{a}}_i)+(q_i-p_i)(\bm{a}_i+\bar{\bm{a}}_i)].
\end{align}
The second term vanishes because $\sum_i(q_i-p_i)(\bm{a}_i+\bar{\bm{a}}_i)=(\bm{b}^{\text{BdG}}+\bar{\bm{b}}^{\text{BdG}})-(\bm{b}^{\text{vac}}+\bar{\bm{b}}^{\text{vac}})$ and Eq.~\eqref{key1}. Therefore, $c_i$ in Eq.~\eqref{key2} is given by $(q_i-p_i)/2$.

\subsection{Quotient group}
\label{quotient}
Given a BdG Hamiltonian $H_{\bm{k}}^{\text{BdG}}$ with a set of assumed symmetries, we can separately compute $\bm{b}^{\text{BdG}}$ and $\bm{b}^{\text{vac}}$ and deduce $\bm{b}^{\text{BdG}}-\bm{b}^{\text{vac}}$.  Distinct BdG Hamiltonians with the same symmetry setting may have different values of $\bm{b}^{\text{BdG}}-\bm{b}^{\text{vac}}$.  Let us introduce the group
\begin{equation}
\{\text{BS}\}^{\text{BdG}}
\end{equation}
as the set of all possible $\bm{b}^{\text{BdG}}-\bm{b}^{\text{vac}}$ realizable using a BdG Hamiltonian in this symmetry class.   

The discussion in Sec.~\ref{obstruction} clarified that, as far as the symmetry obstruction in Eq.~\eqref{condition2} is concerned, the difference in $\{\text{BS}\}^{\text{BdG}}$ by the combination $\bar{\bm{a}}'-\bm{a}'$ is unimportant.  Hence, it makes sense to introduce the following subgroup of $\{\text{BS}\}^{\text{BdG}}$
\begin{equation}
\{\text{AI}\}^{\text{BdG}}=\Big\{\sum_{i}\ell_{i}(\bm{a}_{i}-\bar{\bm{a}}_{i})|\ell_i\in\mathbb{Z}\Big\}.
\end{equation}
When $\bm{b}^{\text{BdG}}-\bm{b}^{\text{vac}}\in\{\text{BS}\}^{\text{BdG}}$ does not belong to $\{\text{AI}\}^{\text{BdG}}$, the condition~\eqref{condition2} is violated and any smooth deformation in Eq.~\eqref{trivial2} is prohibited. Such nontrivial values of $\bm{b}^{\text{BdG}}-\bm{b}^{\text{vac}}$ can be classified by the quotient group
\begin{equation}
X^{\text{BdG}}\equiv\frac{\{\text{BS}\}^{\text{BdG}}}{\{\text{AI}\}^{\text{BdG}}}.\label{XBSBdG}
\end{equation}
This is what we call the refined SI group in this work, which extends the idea in Ref.~\cite{Skurativska2019} to more general symmetry class.

As we proved in the previous section, $\bm{b}^{\text{BdG}}-\bm{b}^{\text{vac}}$ of a given BdG Hamiltonian $H_{\bm{k}}^{\text{BdG}}$ can be expanded as Eq.~\eqref{key2}.  Conversely, a vector $\bm{b}^{\text{BdG}}-\bm{b}^{\text{vac}}$ given in the form of right-hand side of Eq.~\eqref{key2} has a realization using some BdG Hamiltonian $H_{\bm{k}}^{\text{BdG}}$ as far as $\bm{b}^{\text{BdG}}-\bm{b}^{\text{vac}}$ is integer-valued and is consistent with the time-reversal symmetry.  This implies that $X^{\text{BdG}}$ takes the form $\mathbb{Z}_{n_1}\times\mathbb{Z}_{n_2}\times\cdots$ (i.e., it contains only torsion factors) and that the actual calculation of $X^{\text{BdG}}$ can be done by the Smith decomposition of $\{\text{AI}\}^{\text{BdG}}$ without explicitly constructing  $\{\text{BS}\}^{\text{BdG}}$~\cite{Po2017}.

\subsection{Relation to previous approach}
In previous works~\cite{PhysRevB.98.115150,1811.08712}, $\bm{b}^{\text{BdG}}$ was viewed as the representation counts in the valence bands of an insulator and was analyzed in the same way as for class A, AI, or AII. In this approach, $\bm{b}^{\text{BdG}}$ is directly compared against atomic limits $\bm{a}_i$ (discussed in Sec.~\ref{complete}) of the same symmetry setting. When $\bm{b}^{\text{BdG}}$ cannot be written as a superposition of $\bm{a}_i$'s with integer coefficients (i.e., $\bm{b}^{\text{BdG}}\notin\{\text{AI}\}$), then it is said to be nontrivial. Indeed, this is a sufficient condition for violating all of \eqref{condition12}, \eqref{condition13}, and \eqref{condition2}. However, this requirement may be too strong in that, even when $\bm{b}^{\text{BdG}}\in\{\text{AI}\}$, it would still be possible that $\bm{b}^{\text{BdG}}-\bm{b}^{\text{vac}}\notin\{\text{AI}\}^{\text{BdG}}$ and $\bm{b}^{\text{BdG}}-\bm{b}^{\text{vac}}$ belongs to the nontrivial class of $X^{\text{BdG}}$. We will see an example of this in Sec.~\ref{defexamples2}.

\subsection{Weak pairing assumption}
When applying these methods in the actual search for candidate materials of TSCs, it would be more useful if the input data is only the representation count in the band structure of the normal phase described by $H_{\bm{k}}$, not in the quasi-particle spectrum of $H^{\text{BdG}}_{\bm{k}}$.  Such a reduction is achieved in Ref.~\cite{1811.08712} relying on the weak pairing assumption~\cite{PhysRevB.81.134508, PhysRevB.81.220504, PhysRevLett.105.097001}. This assumption states that $(n_{\bm{k}}^{\alpha})^{\text{BdG}}$ in the superconducting phase does not change even if the limit $\Delta_{\bm{k}}\rightarrow 0$ is taken.

To explain how it works, let $\psi_{\bm{k}}$ be an eigenstate of $H_{\bm{k}}$ with the energy $\epsilon_{\bm{k}}$ belonging to the representation $u_{\bm{k}}^\alpha$ of $G_{\bm{k}}$. Then the eigenstate $\psi_{-\bm{k}}^*$ of $-H_{-\bm{k}}^*$ has the energy $-\epsilon_{-\bm{k}}$ and the representation $u_{\bm{k}}^{\bar{\alpha}}$ of $G_{\bm{k}}$, defined in Eq.~\eqref{ufk}.  Thus, representations appearing in the negative-energy quasi-particle spectrum of $H^{\text{BdG}}_{\bm{k}}$ can be decomposed into the occupied bands (occ) of $H_{\bm{k}}$ and unoccupied bands (unocc) of $H_{-\bm{k}}$:
\begin{align}
(n_{\bm{k}}^\alpha)^{\text{BdG}}&=n_{\bm{k}}^\alpha\big|_{\text{occ}}+n_{-\bm{k}}^{\bar{\alpha}}\big|_{\text{unocc}}. \notag\\
&=(n_{\bm{k}}^\alpha-n_{-\bm{k}}^{\bar{\alpha}})\big|_{\text{occ.}}+n_{-\bm{k}}^{\bar{\alpha}}\big|_{\text{occ}+\text{unocc}}.\label{weak-pair}
\end{align}
The last term of this expression is precisely $(n_{\bm{k}}^\alpha)^{\text{vac}}$ defined in Eq.~\eqref{irrepdec2}. This was pointed out recently by Ref.~\cite{Skurativska2019} for the case of inversion symmetry, and we see here that it applies to more general symmetry setting.  After all, components of $\bm{b}^{\text{BdG}}-\bm{b}^{\text{vac}}$ are given by
\begin{align}
(n_{\bm{k}}^\alpha)^{\text{BdG}}-(n_{\bm{k}}^\alpha)^{\text{vac}}=(n_{\bm{k}}^\alpha-n_{-\bm{k}}^{\bar{\alpha}})\big|_{\text{occ.}}.
\end{align}
The last expression is purely the occupied band contribution of the normal phase band structure, which may be calculated, for example, using the density-functional theory~\cite{1811.08712}.

We remark that our sense of ``weak pairing'' is less stringent than that used in Ref.\ \onlinecite{PhysRevB.81.134508}, in that arbitrary inter-Fermi surface pairing is allowed so long as the normal-state energy at the high-symmetry momenta are sufficiently far away from the Fermi surface when compared to the pairing scale.

\subsection{Example}
\label{defexamples2}
As an example of SIs for SCs, let us discuss again the Kitaev chain focusing on its inversion parities.   
Similar exercise has already been performed in Refs.~\onlinecite{Skurativska2019,1907.13632}, but here we repeat it in our notation to clarify the difference in the present and previous approaches.  

For the Kitaev chain with the inversion symmetry, the BdG Hamiltonian is given by $H_k^{\text{BdG}}$ with Eq.~\eqref{singleKitaev} and the symmetry representation is Eq.~\eqref{UBdG} with $U_k(I)=1$ and $\chi_I=-1$.  For this model, we get
\begin{equation}
\bm{b}^{\text{BdG}} =  (n_{0}^{+}, n_{0}^{-}, n_{\pi}^{+}, n_{\pi}^{-}) =(1,0,0,1),
\end{equation}
where $\alpha=\pm$ of $n_{k}^{\alpha}$ corresponds to the inversion parity.
The vacuum limit uses $\chi_IU_k(I)^*=-1$ at both $k=0$ and $\pi$ and thus has $\bm{b}^{\text{vac}}= (0,1,0,1)$. Therefore, we find
\begin{equation}
\bm{b}^{\text{BdG}}-\bm{b}^{\text{vac}}= (1,0,0,1)-(0,1,0,1)=(1,-1,0,0).
\end{equation}

For inversion symmetric one-dimentional models in class A, $\{\text{AI}\}$ is a three dimensional space spanned by
\begin{align}
\bm{a}_1= (1,0,1,0),\,\,\,\bm{a}_2 = (0,1,0,1),\,\,\,\bm{a}_3 = (1,0,0,1).
\end{align}
For these basis vectors, we find
\begin{align}
\bm{a}_1-\bar{\bm{a}}_1&= (1,0,1,0) - (0,1,0,1)= (1,-1,1,-1),\\
\bm{a}_2-\bar{\bm{a}}_2&= (0,1,0,1) - (1,0,1,0)= (-1,1,-1,1),\\
\bm{a}_3-\bar{\bm{a}}_3&= (1,0,0,1) - (0,1,1,0)=  (1,-1,-1,1).
\end{align}
Since $\bm{a}_1-\bar{\bm{a}}_1=-(\bm{a}_2-\bar{\bm{a}}_2)$, $\{\text{AI}\}^{\text{BdG}}$ is a two dimensional space spanned by $\bm{a}_1-\bar{\bm{a}}_1$ and $\bm{a}_3-\bar{\bm{a}}_3$. We find
\begin{align}
\bm{b}^{\text{BdG}}-\bm{b}^{\text{vac}}&= \frac{1}{2}(\bm{a}_1-\bar{\bm{a}}_1)+\frac{1}{2}(\bm{a}_3-\bar{\bm{a}}_3)\notin\{\text{AI}\}^{\text{BdG}}.
\end{align}
The fractional coefficients imply the nontrivial topology of $H_{\bm{k}}^{\text{BdG}}$. In other words, the quotient group in Eq.~\eqref{XBSBdG} is $X^{\text{BdG}}=\mathbb{Z}_2$ and $\bm{b}^{\text{BdG}}-\bm{b}^{\text{vac}}$ of the present model belongs to the nontrivial class of $X^{\text{BdG}}$.

In contrast, we see that
\begin{equation}
\bm{b}^{\text{BdG}} =\bm{a}_3\in\{\text{AI}\}.
\end{equation}
More generally, the quotient group in one dimention is always trivial for class A, AI, or AII~\cite{Po2017}, meaning that all $\bm{b}^{\text{BdG}}$ vectors can be expanded by $\bm{a}_i$'s with integer coefficients.  This implies that one cannot detect the nontrivial topology of $H^{\text{BdG}}_{\bm{k}}$ in one dimension based on representations alone in the previous approach.

\section{Interpretation of computed symmetry-indicators for superconductors}
\label{sec:C4}
Using the refined scheme explained in Sec.~\ref{sec:SI}, we perform a comprehensive computation of $X^{\text{BdG}}$ for all space groups $G$ and one dimensional representations $\chi_g$ of superconducting gap functions.  The full list of results are included in Appendix~\ref{XBSTable} for both spinful and spinless electrons with or without TRS.  The corresponding AZ symmetry classes are listed in Table~\ref{AZlist}.  Most of the nontrivial entries of $X^{\text{BdG}}$ can be understood as supergroup of a countable number of key space groups discussed in Appendix~\ref{KeySGs}.

\begin{table}
\begin{center}
\caption{\label{AZlist} Settings used in the calculation for refined SIs and the corresponding AZ symmetry classes}
\begin{tabular}{c|c|c}
\hline\hline
\quad Spin \quad\quad & \quad TRS \quad\quad  & \quad AZ classes \quad\quad \\\hline
\quad Spinful  \quad\quad  & \quad $-1$ \quad\quad& \quad DIII \quad\quad\\
\quad Spinless  \quad\quad &\quad $+1$ \quad\quad & \quad  BDI, CI \quad\quad\\
\quad Spinful  \quad\quad &\quad $0$ \quad\quad&   \quad  D \quad\quad\\
\quad Spinless  \quad\quad  &\quad $0$ \quad\quad& \quad D, C \quad\quad\\
\hline\hline
\end{tabular}
\end{center}
\end{table}

In this section, we discuss the meaning of $X^{\text{BdG}}$ using two illuminating examples of $G=P4$ and $P4/m$ in class DIII. 
Below we write the component of $\bm{b}^{\text{BdG}}-\bm{b}^{\text{vac}}$ as
\begin{equation}
N_{\bm{k}}^{\alpha}\equiv(n_{\bm{k}}^{\alpha})^{\text{BdG}}-(n_{\bm{k}}^{\alpha})^{\text{vac}}.
\label{defN}
\end{equation}
Also we use the standard labeling of irreducible representations in the literature for $\chi_g$~\cite{Bradley}.

\subsection{$P4$ with $B$ representation}
\label{P4}
The space group $P4$ contains the four-fold rotation symmetry $C_4$ in addition to the lattice translation symmetries. 
The $B$ representation refers to the 1D representation of $C_4$ with $\chi_{C_4}=1$.

In two spatial dimensions, we find $X^{\text{BdG}}=\mathbb{Z}_2$. To see the meaning of this, let us introduce
\begin{align}
\nu_{C_4} \equiv \frac{1}{2\sqrt{2}}\sum_{\bm{k}\in \Gamma, M}\sum_{\alpha=\pm1,\pm3}e^{i\tfrac{\pi\alpha}{4}}N_{\bm{k}}^\alpha\mod 2,
\label{P4indicator}
\end{align}
where $N_{\bm{k}}^{\alpha}$ represents the number of irreducible representations in Eq.~\eqref{defN} with the four-fold rotation eigenvalue $e^{i \frac{\pi\alpha}{4}}$ at $\Gamma=(0,0)$ and $M=(\pi,\pi)$. As we discuss in Sec~\ref{sec:WTSC}, $\nu_{C_4}$ turns out to measure the $\mathbb{Z}_2$ QSH index. This was unexpected, because known diagnosis of the QSH index in class AII required either the inversion symmetry $I$~\cite{PhysRevB.76.045302} or the rotoinversion symmetry $S_4$~\cite{PhysRevX.8.031070,QuantitativeMappings}, and any proper rotation was not sufficient.  Second-order TSCs with a corner Majorna zero mode are also stable under this symmetry setting but are not diagnosed by representations alone, as we discuss in Sec~\ref{sec:WTSC}.

In three spatial dimensions, $X^{\text{BdG}}=\mathbb{Z}_2$ detects the weak topological phase of two-dimensional TSCs stacked along the rotation axis $z$.
The strong $\mathbb{Z}_2$ phase of class DIII is prohibited because the $\mathbb{Z}_2$ index of $k_z=0$ and $k_z=\pi$ planes are forced to be the same by the rotation symmetry $C_4$.

\subsection{$P4/m$ }
\label{P4m}
The space group $P4/m$ contains both the inversion $I$ about the origin and the four-fold rotation $C_4$ around the $z$ axis. The mirror symmetry about $xy$ plane and four-fold rotoinversion symmetry are given as their products.
There are four real one-dimensional representations $A_g$ ($\chi_{C_4} = +1$, $\chi_I=+1$), $A_u$ ($\chi_{C_4} = +1$, $\chi_I=-1$), $B_u$ ($\chi_{C_4} = -1$, $\chi_I=-1$), and $B_g$ ($\chi_{C_4} = -1$, $\chi_I=+1$). For the $A_g$ representation, $X^{\text{BdG}}$ is trivial. We discuss the other three representations one by one.  

\subsubsection{$A_{u}$ representation $(\chi_{C_4} = +1, \chi_I=-1)$}
Let us start with the $A_u$ representation. Although $X^{\text{BdG}}$ in 3D is quite large (see Table~\ref{XBSlist}), many factors can be attributed to lower dimensions. 

In a one-dimensional system along the rotation axis, we find $X^{\text{BdG}} =(\mathbb{Z}_2)^2$, which can be characterized by
\begin{align}
\label{P4/m1D-1}
\nu_{1/2}^{1\text{D}} &= \tfrac{1}{4}\sum_{\bm{k}\in \Gamma, Z}\sum_{\alpha=\pm 1} \left(N_{\bk}^{\alpha,+} - N_{\bk}^{\alpha,-}\right)\mod2,\\
\label{P4/m1D-2}
\nu_{3/2}^{1\text{D}} &= \tfrac{1}{4}\sum_{\bm{k}\in \Gamma, Z}\sum_{\alpha=\pm 3} \left(N_{\bk}^{\alpha,+} - N_{\bk}^{\alpha,-}\right)\mod2,
\end{align}
where $N_{\bm{k}}^{\alpha, \beta}$ represents the number of irreducible representations with the $C_4$ eigenvalue $e^{i \frac{\alpha\pi}{4}}$ and the inversion parity $\beta=\pm1$ at $\Gamma=(0,0,0)$ and $Z=(\pi,\pi,\pi)$.  They measure the number of Majorana edge modes with different rotation eigenvalues.

\begin{figure}
\begin{center}
\includegraphics[width=0.99\columnwidth]{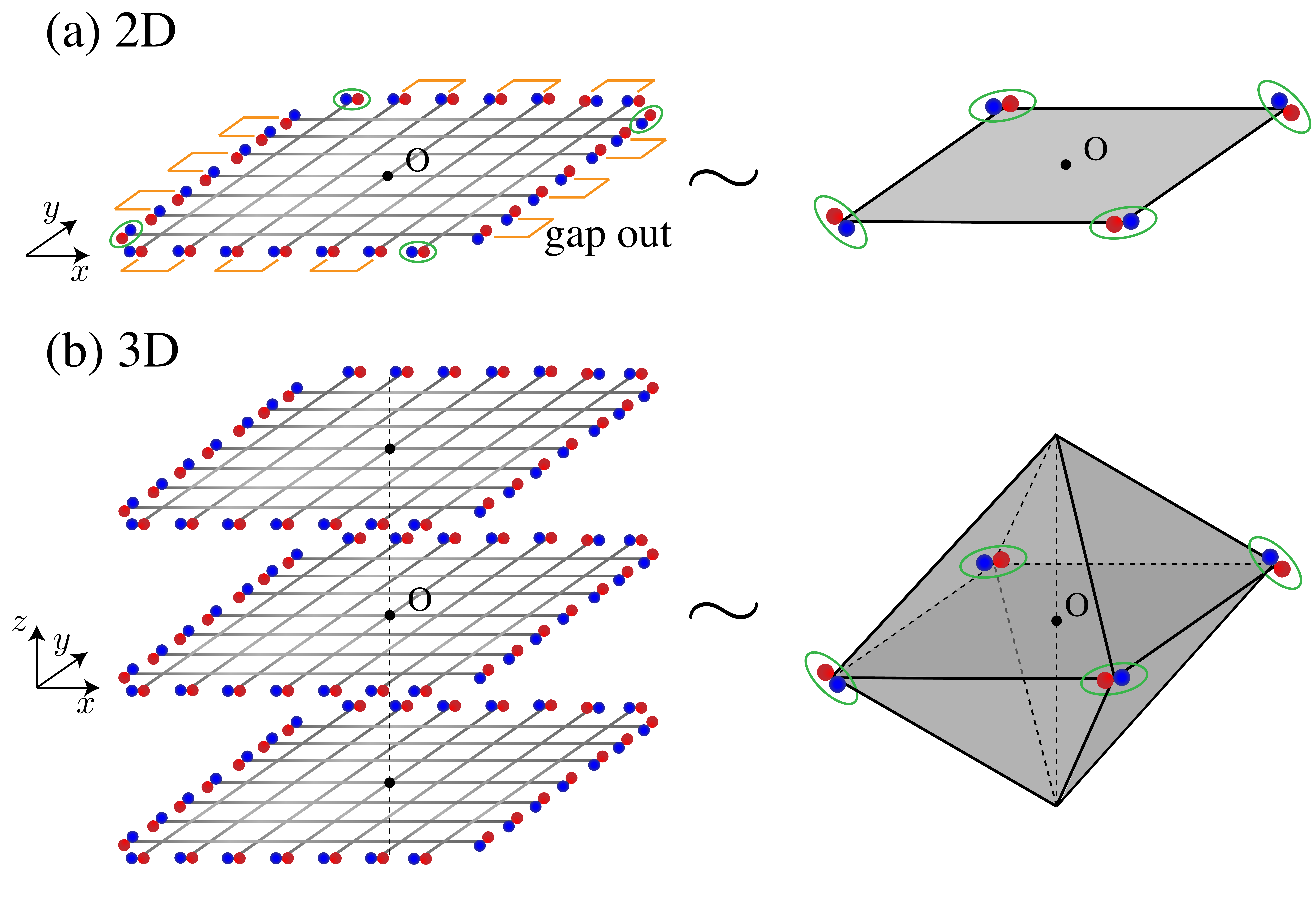}
\caption{\label{WCP4m}
Examples of wire construction for $P4/m$ realizing higher-order TSCs with Majorana corner modes (circled by green ellipses). (a) $(\nu_{x}^{1\text{D}}, z_8) = (1,4)$ phase in 2D. Orange marks represent translation-breaking perturbations on the edge, gapping out pairs of zero modes. (b)  $(z_{8, k_z= 0}, z_{8, k_z= \pi}, z_{16}) = (4, 4, 8)$ phase in 3D. }
\end{center}
\end{figure}

In mirror-invariant 2D planes orthogonal to the $z$ axis, we find $X^{\text{BdG}} = \mathbb{Z}_2 \times \mathbb{Z}_8$.  The $\mathbb{Z}_2$ factor is given by
\begin{align}
\nu^{1\text{D}}_{x} = \frac{1}{4}\sum_{\bk =(0,0), (\pi, 0)}\sum_{\alpha=\pm1,\pm3} \left(N_{\bk}^{\alpha,+} - N_{\bk}^{\alpha,-}\right).
\end{align}
The phase with $\nu^{1\text{D}}_{x}$ corresponds to 1D Kitaev chains stacked along $x$ and $y$ axes.  The $\mathbb{Z}_8$ factor is related to TSCs with mirror Chern number $C_{M}$ and second-order TSCs. It is given by~\cite{QuantitativeMappings}
\begin{align}
z_8=& -\frac{3}{2} N^{ -3, +}_{\Gamma} +\frac{3}{2} N^{ 3, -}_{\Gamma} + \frac{1}{2}N^{ 1, +}_{\Gamma} - \frac{1}{2}N^{ -1, -}_{\Gamma}\notag\\
 &-\frac{3}{2} N^{ -3, +}_{M} +\frac{3}{2} N^{ 3, -}_{M} + \frac{1}{2}N^{ 1, +}_{M}\notag\\
 & - \frac{1}{2}N^{ -1, -}_{M}  - N^{2, +}_{X} + N^{-2, -}_{X} \mod 8.
\end{align}
This expression rewrites the formula for the mirror Chern number in the form of summation. Thus $z_8$ equals to $C_{M}$ mod 4~\cite{PhysRevB.86.115112}.
To see the meaning of $z_8 = 4$ mod $8$, we generate an example of phases with $(\nu_{x}^{1\text{D}}, z_8) = (1, 4)$ by the wire construction as illustrated in Fig.~\ref{WCP4m} (a). Keeping both the inversion and the rotation symmetry but breaking the translation symmetry, one can gap out edges, realizing a second-order TSC that features two zero modes with different chirality at each of four corners.  We can also generate a phase with $(\nu_{x}^{1\text{D}}, z_8) = (0, 4)$ by stacking four copies of mirror Chern insulator with $C_{M}=1$.  From these observations, we conclude
\begin{equation}
C_M = 4\nu_{2} + z_8 +8\mathbb{Z}, 
\end{equation}
where $\nu_2=0$ and $1$ is the index for the second-order TSC.

\begin{table}
\begin{center}
\caption{\label{XBSlist}The list of $X^{\text{BdG}}$ for class DIII systems with $P\bar{1}$~\cite{Skurativska2019}, $P4$, and $P4/m$ symmetry in each spatial dimension.}
\begin{tabular}{c|ccc}
\hline\hline
SG (rep of $\Delta_{\bm{k}}$) & \quad 1D \quad\quad & \quad\quad 2D \quad\quad\quad & 3D \\\hline
$P\bar{1}$  ($A_u$) & $\mathbb{Z}_2$ & $(\mathbb{Z}_2)^2\times\mathbb{Z}_4$ & $(\mathbb{Z}_2)^3\times(\mathbb{Z}_4)^3\times\mathbb{Z}_8$\\
$P4$  ($B$) & $\mathbb{Z}_1$ & $\mathbb{Z}_2$ & $\mathbb{Z}_2$\\
$P4/m$ ($A_u$) &$(\mathbb{Z}_2)^2$ &$\mathbb{Z}_2 \times \mathbb{Z}_8$ &$(\mathbb{Z}_2)^4 \times \mathbb{Z}_4 \times \mathbb{Z}_8 \times \mathbb{Z}_{16} $ \\
$P4/m$ ($B_u$) & $\mathbb{Z}_1$ &$\mathbb{Z}_2 \times \mathbb{Z}_8$ &$\mathbb{Z}_2 \times(\mathbb{Z}_4)^2\times \mathbb{Z}_8 $ \\
$P4/m$ ($B_g$) & $\mathbb{Z}_1$ &$(\mathbb{Z}_2)^2$ &$(\mathbb{Z}_2)^3$ \\
\hline\hline
\end{tabular}
\end{center}
\end{table}

Finally, for three dimensional systems, we introduce a strong $\mathbb{Z}_{16}$ index defined by
\begin{align}
&z_{16}=\kappa_1 -2\kappa_4\mod 16,\\
&\kappa_1=\frac{1}{4}\sum_{\bm{k}\in K_1}\sum_{\alpha=\pm1,\pm3} \left(N_{\bk}^{\alpha,+} - N_{\bk}^{\alpha,-}\right),\label{kappa1}\\
&\kappa_4 = \frac{1}{2\sqrt{2}}\sum_{\bm{k}\in K_4}\sum_{\alpha=\pm1,\pm3}(e^{i\frac{\pi\alpha}{4}} N_{\bk}^{\alpha,+}-e^{i\frac{\pi\alpha}{4}} N_{\bk}^{\alpha,-}),\label{kappa4}
\end{align}
where $K_1$ refers to the set of eight time-reversal invariant momenta.
We find that $z_{16}=0$ mod $16$ holds for all elements in \{AI\}$^{\text{BdG}}$.
This $z_{16}$ invariant, when focused on its ``mod 8" part, has the same implication as for class AII systems~\cite{PhysRevX.8.031070,QuantitativeMappings}. Namely,
$z_{16}$ mod $4$ agrees with the mirror Chern number and $z_{16}=4$ mod $8$ implies a second-order TSC.
To investigate the topology of phases with $z_{16} = 8$ mod 16, we stack 2D second-order TSCs with $z_8=4$ and $C_{M} = 0$ as illustrated in Fig.~\ref{WCP4m} (b).  Since $z_{16}$ is relate to $z_{8, k_z=0}$ and $z_{8, k_z=\pi}$ as
\begin{align}
\label{z16_rel}
z_{16} = -(z_{8,k_z=0} + z_{8,k_z=\pi}),
\end{align}
the value of $z_{16}$ depends on how to stack the 2D layers. If the inversion center is contained in a layer (i.e., there exists a single layer left invariant under the inversion), $z_{16} = 8$ and the system exhibits Majorana corner states.
If, on the other hand, the inversion center is not contained in any layer, $z_{16} = 0$ and the surface can be completely gapped without breaking symmetries or closing the bulk gap.  Phases with $z_{16}=8$ mod 16 can also be mirror Chern TSCs just like in the 2D case.

\subsubsection{$B_{u}$ representation $(\chi_{C_4} = -1, \chi_I=-1)$}
Next, let us consider the $B_{u}$ representation. In one dimensions, we found $X^{\text{BdG}}$ is trivial. This is because the four-fold rotation symmetry together with the choice $\chi_{C_4}=-1$ implies that the $\mathbb{Z}_2$ index of class DIII is trivial. 

In two dimensions, the interpretation of $X^{\text{BdG}}=\mathbb{Z}_2 \times \mathbb{Z}_8$ is the same as the $A_u$ representation, but the formula for $z_8$ index must be replaced by
\begin{align}
z'_{8} &= \sum_{\bk \in \Gamma, M} \left(N^{-3, +}_{\bk} + 3N^{3, -}_{\bk} \right) + 2N^{-2, -}_{X} \mod 8.
\end{align}

In three dimensions, the $\mathbb{Z}_2 \times \mathbb{Z}_4 \times \mathbb{Z}_8$ part of $X^{\text{BdG}}$ is weak indices.  The remaining $\mathbb{Z}_4$ factor can be explained by $\kappa_1$ in Eq.~\eqref{kappa1}.  The QSH indices at $k_z =0$ and $k_z = \pi$ planes must be the same due to the compatibility relation associated with the $C_4$ symmetry. Therefore $\kappa_1$ (defined modulo 8) is restricted to be even and characterizes the $\mathbb{Z}_4$ factor. For the $B_{u}$ representation, $\kappa_4$ always vanishes and $z_{16}=\kappa_1$.  Since Eq.~\eqref{z16_rel} still holds, $\kappa_1=2$ mod 4 implies a nontrivial mirror Chern number. As discussed in Appendix~\ref{WTSC-P4/m}, there are no third-order TSCs in this symmetry setting. With these results, we conclude that $\kappa_1 = 4$ mod 8 also indicates a nonzero mirror Chern number.

\begin{figure*}[t]
\begin{center}
\includegraphics[width=1.69\columnwidth]{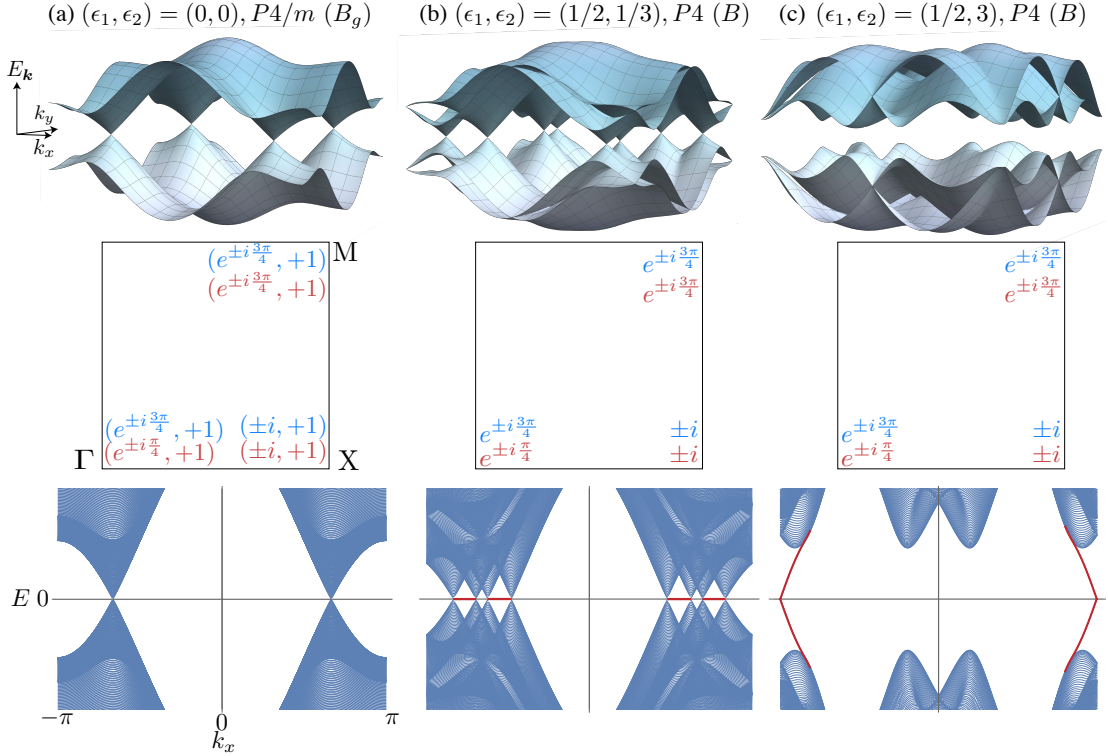}
\caption{\label{fig:example} The quasi-particle spectrum (top), symmetry representations of $E<0$ states (middle), and the surface band structure (bottom) of $H_{\bm{k}}^{\text{BdG}}$ with Eqs.~\eqref{TB-1} and \eqref{TB-2}. In (a), each bands are doubly degenerate due to the inversion and the time-reversal symmetry.  For (b) and (c), inversion-breaking perturbations in Eq.~\eqref{pert} are included.  Symmetry representations colored in red is for $H_{\bm{k}}^{\text{BdG}}$ and those in blue is for $H^{\text{vac}}$. In the surface band structure, states localized near edges are colored in red.
}
\end{center}
\end{figure*}

\subsubsection{$B_{g}$ representation $(\chi_{C_4} = -1, \chi_I=+1)$}
\label{P4m_Bg}
Finally, let us discuss $B_{g}$ representation.  In this case, the mirror symmetry commutes with the particle-hole symmetry ($\chi_{M}=\chi_{C_4}^2\chi_I=+1$) and the mirror Chern number must vanish~\cite{PhysRevB.88.075142}.

In two dimensions, we find $X^{\text{BdG}}= \mathbb{Z}_2 \times \mathbb{Z}_2$. These class are characterized by
\begin{align}
\label{SI_Bg-1}
\nu_{C_4}^{+} &= \frac{1}{2\sqrt{2}}\sum_{\bm{k}\in \Gamma, M}\sum_{\alpha=\pm1,\pm3}e^{i\tfrac{\alpha\pi}{4}}N_{\bm{k}}^{\alpha,+}\mod2,\\
\label{SI_Bg-2}
\nu_{C_4}^{-} &= \frac{1}{2\sqrt{2}}\sum_{\bm{k}\in \Gamma, M}\sum_{\alpha=\pm1,\pm3}e^{i\tfrac{\alpha\pi}{4}}N_{\bm{k}}^{\alpha,-}\mod2,
\end{align}
where $N_{\bm{k}}^{\alpha, \beta}$ represents the number of irreducible representations with the $C_4$ eigenvalue $e^{i \frac{\alpha\pi}{4}}$ and the inversion parity $\beta=\pm1$ at $\Gamma=(0,0)$ and $M=(\pi,\pi)$.   Phases with $(\nu_{C_4}^{+},\nu_{C_4}^{-})=(1,0)$ or $(0,1)$ turn out to be gapless.  An example of the quasi-particle spectrum is shown in  Fig.~\ref{fig:example} (b). To see why, recall that $\nu_{C_4}=\nu_{C_4}^{+}+\nu_{C_4}^{-}=1$ mod 2 implies the nontrivial $\mathbb{Z}_2$ QSH index as discussed in Sec.~\ref{P4}. However, when $\chi_I=+1$, the QSH index cannot be nontrivial~\cite{PhysRevLett.105.097001}.  The only way out is gap closing of the quasi-particle spectrum which invalidates the definition of QSH index.  Phases with $(1,1)\in \mathbb{Z}_2 \times \mathbb{Z}_2$ can be gapped and  they are second-order TSCs.

Finally we discuss  three dimensional systems. For $B_{g}$ representation, $\kappa_1$ should always be 0, while $\kappa_4$ can be nonzero. 
It is restricted to be even, explaining the strong $\mathbb{Z}_2$ factor in $X^{\text{BdG}}$. 
As explained in the Appendix~\ref{KeySGs}, $\kappa_4=2$ mod 4 indicates second-order TSCs.

\subsection{Example}
To demonstrate the prediction of SIs, let us discuss a simple two-dimensional model with $P4/m$ symmetry. The BdG Hamiltonian $H_{\bk}^{\text{BdG}}$ is given by Eq.~\eqref{BdG} with
\begin{align}
\label{TB-1}
H_{\bk} &=-\left(\cos k_x +\cos k_y+1\right)\openone_2,\\
\label{TB-2}
\Delta_{\bk} &= (\cos k_x - \cos k_y) i \sigma_2.
\end{align}
Here $\sigma_i$'s are the Pauli matrices.  
The inversion symmetry and the four-fold rotation symmetry are represented by
\begin{align}
&U_{\bk} (I) = \openone_2,\quad \chi_I = +1,\\
&U_{\bk} (C_4) = e^{-i\tfrac{\pi}{4}\sigma_3},\quad \chi_{C_4} = -1.
\end{align}
Thus the model belongs to the $B_g$ representation of $P4/m$ discussed in Sec.~\ref{P4m_Bg}.  We compute the indices in Eqs.~\eqref{SI_Bg-1} and \eqref{SI_Bg-2} and get $(\nu_{C_4}^{+}, \nu_{C_4}^{-}) = (0,1)$, which suggests nodal points in the quasi-particle spectrum [see Fig.~\ref{fig:example} (a)] as discussed in Sec.~\ref{P4m_Bg}.


If inversion-breaking perturbations are added, the space group symmetry is reduced to $P4$.  Here we consider the following term.
\begin{align}
V_{\bk}^{\text{BdG}} &= \epsilon_1(\sin k_x \sigma_1 \tau_0 + \sin k_y \sigma_2 \tau_3)\nonumber\\
\label{pert}
&\quad \quad \quad \quad \quad +\epsilon_2(\sin k_x \sigma_0 \tau_2 + \sin k_y \sigma_3 \tau_1)
\end{align}
The four-fold rotation symmetry remains intact and $\chi_{C4} = -1$.  
Thus $\nu_{C_4} = \nu_{C_4}^{+}+\nu_{C_4}^{-} = 1$ is still well-defined. For small perturbations, the bulk spectrum remain gapless and flat surface bands (Andreev bound states~\cite{PhysRevB.83.224511}) appear in the surface spectrum [Fig.~\ref{fig:example} (b)].  As the perturbation strength is increased, the system gets fully gapped without closing gap at high-symmetry points and becomes a helical TSC [Fig.~\ref{fig:example} (c)] .  

\section{Indicators for Wannierizable Topological Superconductors}
\label{sec:WTSC}
So far, we have mostly focused on the formalism and physical meaning of the refined SIs for superconductors. 
From the discussions on the Kitaev chain in Refs.~\onlinecite{Skurativska2019,1907.13632} and Sec.~\ref{sec:SI}, one might expect the main power of the SI refinement is to capture TSCs with zero-dimensional surface states. Interestingly, this is untrue: in Sec.~\ref{P4}, we have already asserted that the $C_4$ refined SI, $\nu_{C_4}$, actually detects either a gapless phase or the helical TSC in class DIII in two dimensions. 

We will substantiate our claim in this section. A general approach for physically interpreting the SIs is to first construct a general set of topological phases protected by the symmetries, and then evaluate their SIs to establish the relations between the two \cite{QuantitativeMappings,PhysRevX.8.031070}.
We will follow a similar scheme: first, we introduce the notion of ``Wannierizable TSC'' (WTSC), which, like the Kitaev chain, reduces to an atomic state once the particle-hole symmetries are broken; next, we discuss how particle-hole symmetry restricts the possible associated atomic states that could correspond to a WTSC; lastly, we specialize our discussion to 2D superconductors with $\chi_{C_4}=-1$, and show that the nontrivial refined SI {\it does not} indicate a WTSC. Our claim follows when the arguments above are combined with the established classification of class AII topological (crystalline) insulators \cite{Freed2013, PhysRevX.7.011020,PhysRevX.8.011040,PhysRevB.96.205106, PhysRevX.8.031070, Shiozaki2018, Song2018}.

Before moving on, we remark that, insofar as our claim on the physical meaning of $\nu_{C_4}$ is concerned, there is probably a simpler approach in which one relates the nontrivial SI $\nu_{C_4}= 1$ to the Fermi-surface invariant in Ref.~\onlinecite{PhysRevB.81.134508} under the weak-pairing assumption. Our approach, however, is more general in that the weak-pairing assumption is not required, and that the analysis of the SIs corresponding to WTSCs also helps one understand the physical meaning of the SIs, as can be seen in the $P4/m$ examples.

\subsection{Wannierizable topological superconductors}
Let us begin by introducing the notion of WTSCs. Consider a gapped Hamiltonian. To investigate the possible topological nature of the system, we ask if it is possible to remove all quantum entanglement in the many-body ground state while respecting all symmetries. In the context of non-interacting insulators, this question can be rephrased in the notion of Wannier functions, and we say a phase is topological if there is an obstruction for constructing symmetric, exponentially localized Wannier functions out of the Bloch states below the energy gap \cite{Po2017, TQC, PhysRevLett.121.126402}.

For superconductors, the question of ground-state quantum entanglement is more subtle even within a mean-field BdG treatment. 
As a partial diagnosis, we could still apply the same Wannier criterion to the Bloch states below the gap at $E=0$, and we say a BdG Hamiltonian is ``Wannierizable'' when no Wannier obstruction exists \footnote{There is a technical question of whether the addition of trivial states below the gap is allowed or not, which differentiates ``stable'' topological phases from ``fragile'' ones \cite{PhysRevLett.121.126402}. Since our starting point is a BdG Hamiltonian, the corresponding physical system does not have charge conservation symmetry and it is more natural to focus on stable topological phases. We will take this perspective and always assume appropriate trivial degrees of freedom could be supplied to resolve any possible fragile obstructions in a model.
}. 
A non-Wannierizable BdG Hamiltonian is necessarily topological, and phases like the 2D helical TSC in class DIII can be diagnosed that way. However, as the mentioned Wannier criterion uses only Bloch states with energy $E_{\bm k}<0$, it inherently ignores the presence of particle-hole symmetry when we consider obstructions to forming localized, symmetric Wannier functions, i.e., in the Wannierization we only demand the subgroup of symmetries which commute with the single-particle Hamiltonian.  Because of this limitation, the Wannier criterion does not detect TSCs whose BdG Hamiltonians become trivial when the particle-hole symmetry is ignored, like the Kitaev chain. When a Wannierizable BdG Hamiltonian is topological (in the sense defined in Sec.~\ref{sec:top}), we call it a  ``Wannierizable TSC.''

\subsection{Constraints on the associated atomic insulators}
By definition, given any Wannierizable BdG Hamiltonian in class DIII, we can define an associated atomic insulator in class AII. Based on the recently developed paradigms for the classifications of topological crystalline insulators~\cite{PhysRevX.7.011020, PhysRevX.8.011040, PhysRevB.96.205106, PhysRevB.99.125122}, we can consider the associated atomic insulator $\psi$  as an element of a finitely generated Abelian group $\mathcal C_{\text{AI}}$. More concretely, let $H^{\text{BdG}} $ be Wannierizable, and let $\psi^{\text{BdG}} \in \mathcal C_{\text{AI}}$ be the associated atomic insulator. Similar to the formalism for the refined SI, we also consider the limit when the chemical potential approaches $ - \infty$. 
The vacuum is clearly Wannierizable, and so we can also define $\psi^{\text{vac}} \in \mathcal C_{\text{AI}}$. In the following, we will again be focusing on the difference $\delta \psi \equiv \psi^{\text{BdG}} - \psi^{\text{vac}} \in \mathcal C_{\text{AI}}$.

Although we have ignored the particle-hole symmetry $\Xi$ in discussing the Wannierizability of a BdG Hamiltonian, it casts important constraints on the possible states $\delta \psi \in \mathcal C_{\text{AI}}$. Physically, $\psi^{\text{vac}}$ can be identified with the states forming the hole bands ($E<0$) of an empty lattice, and it is determined by the sites, orbitals, as well as the choice on the superconducting pairing symmetry denoted by $\chi$. We can also consider the states forming the electron bands ($E>0$) of the empty lattice, which are related to $\psi$ by the particle-hole symmetry. More generally, we can define a linear map $\Xi_{\chi}: \mathcal C_{\text{AI}} \rightarrow  \mathcal C_{\text{AI}}$ which relates an atomic insulator with its particle-hole conjugate. Noticing that $\psi^{\text{vac}} + \Xi_{\chi} [\psi^{\text{vac}} ]$ simply describes the full Hilbert space in our BdG description, we must have 
\begin{equation}\begin{split}\label{eq:psi_sum}
\psi^{\text{BdG}}  + \Xi_{\chi} [\psi^{\text{BdG}}]=\psi^{\text{vac}}  + \Xi_{\chi} [\psi^{\text{vac}} ].
\end{split}\end{equation}
We remark that Eq.~\eqref{eq:psi_sum} parallels Eq.~\eqref{key1} in defining the refined SI.
Importantly, Eq.~\eqref{eq:psi_sum} can be rearranged into a condition which $\delta \psi$ has to satisfy:
\begin{equation}\begin{split}\label{eq:psi_cond}
(\openone + \Xi_{\chi}) \delta \psi \equiv \delta \psi + \Xi_{\chi} [\delta  \psi] = {\bm{0}},
\end{split}\end{equation}
where we denote the identity map by $\openone$ and the trivial element of $\mathcal C_{\text{AI}}$ by ${\bm{0}}$. Note that $\Xi_\chi^2 = \openone$. 

An obvious class of solutions to Eq.~\eqref{eq:psi_cond} is to take $\delta \psi = \psi - \Xi_\chi[\psi]$ for any $\psi \in \mathcal C_{\text{AI}}$.
Such solutions arise when we take $\psi^{\text{BdG}} = \Xi_{\chi}[\psi^{\text{vac}} ] $, the fully filled state of the system,  in the definition of $\delta \psi$. Mathematically, we can view them as elements in the image of the map $\openone - \Xi_\chi$, and it is natural for us to quotient out these trivial solutions:
\begin{equation}\begin{split}\label{eq:}
\mathcal{X}^{\text{WTSC}} \equiv \frac{\ker (\openone + \Xi_\chi )}{{\rm Im} (\openone - \Xi_\chi )}.
\end{split}\end{equation}
If $\delta \psi$ belongs to a nontrivial class in $\mathcal{X}^{\text{WTSC}}$, the gap must close when we change the chemical potential to either one of the limits $\mu \rightarrow \pm\infty$, and so the BdG Hamiltonian cannot be trivial.
Physically, we interpret $\mathcal{X}^{\text{WTSC}}$ as an indicator for WTSC. Note that, generally, $\mathcal{X}^{\text{WTSC}}$ is only an indicator, not a classification, of WTSCs. This is because $\psi_1 = \psi_2$ is only a necessary, but not generally sufficient, condition for the existence of a symmetric, adiabatic deformation between two Wannierizable BdG Hamiltonians.

We can now relate $\mathcal{X}^{\text{WTSC}}$ to the refined SI by evaluating the momentum-space symmetry representations of $\delta \psi$. 
If $\delta \psi$ belongs to the trivial class of $\mathcal{X}^{\text{WTSC}}$, 
we can write  $\delta \psi = \psi -\Xi_\chi[\psi]$ for some $\psi \in \mathcal C_{\text{AI}}$. Correspondingly, its representation vector takes the form ${\bm a} -\bar{\bm a} $ for some ${\bm a} \in \{ {\text{AI}}\}$, and so its SI will also be trivial. This implies if two atomic mismatches $\delta \psi_1 $ and $\delta \psi_2 $ belong to the same class in the quotient group $\mathcal{X}^{\text{WTSC}}$, they will have the same refined SI. In other words, the evaluation of the refined SI gives a well-defined map  ${\rm SI}: \mathcal{X}^{\text{WTSC}} \rightarrow X^{\text{BdG}}$.
Note that the symmetry representations may not detect all topological distinctions between atomic states, and so ${\rm SI} [\mathcal{X}^{\text{WTSC}}]$ generally contains less information than $\mathcal{X}^{\text{WTSC}}$.

Observe that  ${\rm SI} [\mathcal{X}^{\text{WTSC}}]$ is a subgroup of $X^{\text{BdG}}$. 
If $H^{\text{BdG}}$ is Wannierizable, its representation vector ${\bm{b}}^{\text{BdG}} - {\bm{b}}^{\text{vac}}$ must have an SI in the subgroup ${\rm SI} [\mathcal{X}^{\text{WTSC}}]$. Conversely, any SI which does not belong to this subgroup is inconsistent with any WTSC.

 \subsection{Interpretation of $\nu_{C_4}$}
We can now apply the formalism to show that a 2D BdG Hamiltonian in class DIII with $\nu_{C_4}=1$ cannot be Wannierizable, and hence it must be either gapless or has a nontrivial $\mathbb Z_2$ QSH index \cite{PhysRevX.8.031070, Shiozaki2018, Song2018}. 
Following the general plan described above, we will first compute the group $\mathcal C_{\text{AI}}$ classifying the associated atomic insulators, construct the map $\Xi_{\chi}$ corresponding to $\chi_{C_4}=-1$, and finally show that a phase with  $\nu_{C_4}=1$ cannot be Wannierizable as ${\rm SI} [\mathcal{X}^{\text{WTSC}}] = \mathbb Z_1$, the trivial group. 

To classify the associated atomic insulators, we first consider the set of possible lattices and orbitals. 
In 2D with $C_4$ rotation symmetry, there are four Wyckoff positions: $\mathcal W_a = \{(0,0) \}$, $\mathcal W_b= \{(1/2,1/2) \}$, $\mathcal W_c = \{ (1/2,0), (0,1/2)\}$, and $\mathcal W_d = \{ (x,y), (-y,x), (-x,-y), (y,-x)\}$ being the general position. 
A site in $\mathcal W_a$ or $\mathcal W_b$ is symmetric under $C_4$ rotation, and for spinful fermions with time-reversal symmetry we can label the orbitals by $\alpha=\pm1$ or $\pm3$ characterizing the $C_4$ eigenvalue $e^{i\frac{\pi\alpha}{4}}$, where the $\pm$ states form a Kramers pair.  
When the site filling is two, we simply fill one of the two types of orbitals, and we denote the corresponding atomic insulators by $\psi_{a,b}^{\pm 1}$ and $\psi_{a,b}^{\pm 3}$. 
Generally, the site-filling may be larger than two, and we denote a state with $2 n$ fermions filling orbitals with $\alpha= \pm 1$  and $2m$ with $\alpha = \pm 3$, both in $\mathcal W_a$, by the expression $n \psi_a^{\pm 1} + m \psi_a^{\pm 3}$.
We can perform the same analysis for $\mathcal W_c$ and $\mathcal W_d$. A site in $\mathcal W_c$ only has $C_2$ rotation symmetry, and the two possible rotation eigenvalues form a Kramers pair. 
Since there is only one orbital type, we will simply denote the corresponding atomic insulator by $\psi_c$.  Similarly, we will let $\psi_d$ denote the atomic insulator living on the general position.

While we have listed a total of six possible atomic insulators with the minimal filling of two fermions per site, these states are not completely independent. To see why, consider setting the free parameters  in the general position $\mathcal W_d$ to $x=y=0$, which corresponds to moving all four sites in the unit cell to the point-group origin. 
As the deformation of sites can be done in a smooth manner, the atomic insulator $\psi_d$  must be equivalent to an appropriate stack of atomic insulators defined on $\mathcal W_a$. Such equivalence can be deduced by studying the point-group symmetry representation furnished by the collapsing sites~\cite{TQC,PhysRevB.99.125122}. 
We can perform a similar analysis by collapsing the sites in $\mathcal W_d$ to the other two Wyckoff positions, and altogether we find the equivalence relations: 
\begin{equation}\begin{split}\label{eq:LH}
\psi_d \sim & 2 \psi_a^{\pm 1}+ 2 \psi_a^{\pm 3}
\sim  2 \psi_b^{\pm 1}+ 2 \psi_b^{\pm 3}
\sim  2 \psi_c.
\end{split}\end{equation}
As such, any atomic insulator $\psi$ in our setting can be formally expanded as
\begin{equation}\begin{split}\label{eq:}
\psi = &n_a \psi_a^{\pm 1}  + n_b \psi_b^{\pm 1} + n_c  \psi_c \\
& + \xi_a \left( \psi_a^{\pm 1}+\psi_a^{\pm 3}  - \psi_c  \right)
+ \xi_b \left( \psi_b^{\pm 1}+\psi_b^{\pm 3}  - \psi_c  \right),
\end{split}\end{equation}
where $n_{a,b,c}\in \mathbb Z$ and $\xi_{a,b} \in \mathbb Z_2$. In other words, the atomic insulators are classified by the group $\mathcal C_{\text{AI}} = \mathbb Z^3 \times (\mathbb Z_2)^2$. 
In this language, we represent any (class of) atomic insulator by the collection of integers $(n_a,n_b,n_c,\xi_a,\xi_b)$. For instance, 
\begin{equation}\begin{split}\label{eq:}
\psi_a^{\pm 1} \mapsto (1,0,0,0,0);~~~
\psi_a^{\pm 3} \mapsto (-1,0,1,1,0).
\end{split}\end{equation}

We are now ready to construct the map $\Xi_\chi$. With the choice of $\chi_{C_4}=-1$, the $C_4$ rotation eigenvalues of local orbitals related by $\Xi$ differ by $-1$. As such, the particle-hole acts on the atomic insulators as follows:
\begin{equation}\begin{split}\label{eq:}
\Xi_\chi [\psi_{a,b}^{\pm 1}] =& \psi_{a,b}^{\pm 3};~~~
\Xi_\chi [\psi_c] = \psi_c,
\end{split}\end{equation}
and recall that $\Xi_{\chi}^2 = \openone$, the identity.  We can equally represent the action of $\Xi_\chi$ by a matrix
\begin{equation}\begin{split}\label{eq:}
\Xi_\chi \left[ 
\left(
\begin{array}{c}
n_a\\ n_b\\ n_c\\ \xi_a \\ \xi_b
\end{array}
\right)
\right]
= 
\left(
\begin{array}{ccccc}
-1 & 0 & 0 & 0 & 0\\
0 & -1 & 0 & 0 & 0\\
1 & 1 & 1 & 0 & 0\\
1 & 0 & 0 & 1 & 0\\
0 & 1 & 0 & 0 & 1
\end{array}
\right)
\left(
\begin{array}{c}
n_a\\ n_b\\ n_c\\ \xi_a \\ \xi_b
\end{array}
\right).
\end{split}\end{equation}

We can now compute $\mathcal{X}^{\text{WTSC}}$. On the one hand, we can parameterize elements in $\ker(\openone+ \Xi_\chi )$ by 
\begin{equation}\begin{split}\label{eq:}
\delta \psi = (2 m_a, 2 m_b, -m_a+m_b, \xi_a, \xi_b),
\end{split}\end{equation}
where each of $m_{a,b},\xi_{a,b}$ corresponds to a generator, i.e.,
\begin{equation}\begin{split}\label{eq:ker2DP4}
\ker(\openone+ \Xi_\chi ) = {\rm span}&\left\{ (2,0,-1,0,0), (0,2,-1,0,0),\right.\\
&~~~~~~~~~~~\left.
 (0,0,0,1,0), (0,0,0,0,1)
\right\}.
\end{split}\end{equation}
This shows that $\ker(\openone+ \Xi_\chi ) \simeq \mathbb Z^2 \times (\mathbb Z_2)^2$. On the other hand, an element $\delta \psi' \in  {\rm Im}(\openone-\Xi_\chi)$ takes the form
\begin{equation}\begin{split}\label{eq:im2DP4}
\delta \psi' =
(2 n_a, 2 n_b, -n_a-n_b, n_a \mod 2,  n_b \mod 2),
\end{split}\end{equation}
and so we can write
\begin{equation}\begin{split}\label{eq:}
\im (\openone-\Xi_\chi ) =  {\rm span}&\left\{ (2,0,-1,1,0), (0,2,-1,0,1)
\right\},
\end{split}\end{equation}
which is abstractly the group $\mathbb Z^2$.
Comparing Eq.~\eqref{eq:im2DP4} against  Eq.~\eqref{eq:ker2DP4}, we find the quotient group
\begin{equation}\begin{split}\label{eq:}
\mathcal{X}^{\text{WTSC}} = (\mathbb Z_2)^2,
\end{split}\end{equation}
and we may take $(0,0,0,1,0)$ and $(0,0,0,0,1)$ as representatives of the generating elements.

Finally, we evaluate ${\rm SI} [\mathcal{X}^{\text{WTSC}} ]$. 
The corresponding representation vectors of the atomic states satisfy the relations
\begin{equation}\begin{split}\label{eq:}
{\bm a}_c 
=  {\bm a}_a^{\pm 1} +{\bm a}_a^{\pm 3}
=  {\bm a}_b^{\pm 1} +{\bm a}_b^{\pm 3}.
\end{split}\end{equation}
From this, we conclude ${\rm SI}[\mathcal{X}^{\text{WTSC}} ] = \mathbb Z_1$, and so $\nu_{C_4}=1$ implies the BdG Hamiltonian cannot be Wannierizable.

\begin{table}
\begin{center}
\caption{The list of $\mathcal{X}^{\text{WTSC}}$ for class DIII systems with $P\bar{1}$, $P4$, and $P4/m$ symmetry in each spatial dimension..\label{tabWTSC}}
\begin{tabular}{c|ccc}
\hline\hline
SG (rep of $\Delta_{\bm{k}}$) & \quad \quad 1D \quad\quad & \quad\quad 2D\quad\quad  & \quad\quad 3D\quad\quad  \\
\hline
$P\bar{1}$  ($A_u$) &  \quad\quad $\mathbb{Z}_2$  \quad\quad &  \quad\quad $(\mathbb{Z}_2)^3$ \quad\quad  & \quad\quad  $(\mathbb{Z}_2)^7$ \quad\quad \\
 $P4$ ($B$) & \quad\quad $\mathbb{Z}_1$ \quad\quad & \quad\quad $\mathbb{Z}_2$\quad\quad  & \quad\quad $\mathbb{Z}_2$\quad\quad  \\
 $P4/m$ ($A_u$) & \quad\quad $(\mathbb{Z}_2)^2$ \quad\quad & \quad\quad $(\mathbb{Z}_2)^2$\quad\quad  & \quad\quad $(\mathbb{Z}_2)^7$\quad\quad  \\
 $P4/m$ ($B_u$) & \quad\quad $\mathbb{Z}_1$\quad\quad  & \quad\quad $(\mathbb{Z}_2)^2$\quad\quad  & \quad\quad $(\mathbb{Z}_2)^3$\quad\quad  \\
 $P4/m$ ($B_g$) & \quad\quad $\mathbb{Z}_1$\quad\quad  & \quad\quad $\mathbb{Z}_2$\quad\quad  & \quad\quad $\mathbb{Z}_2$\quad\quad  \\
\hline\hline
\end{tabular}
\end{center}
\end{table}

While the discussion above focuses on a two dimensional system with $C_4$ rotation symmetry, one can perform the same analysis for any other symmetry setting. 
In particular, we tabulate the results for space group $P\bar{1}$ and $P4/m$ under different SC representations in Table~\ref{tabWTSC}. 
For $P\bar{1}$ and $P4/m$, we found ${\rm SI} [\mathcal{X}^{\text{WTSC}} ]=\mathcal{X}^{\text{WTSC}}$, and nontrivial entries correspond to WTSCs like stacked Kitaev chains and higher-order TSCs. For $P\bar{1}$ and $P4/m$ with the $A_u$ representation, $\mathcal{X}^{\text{WTSC}}$ coincides with the maximal $(\mathbb{Z}_2)^m$ subgroup of $X^{\text{BdG}}$. (For $P4/m$, $m=2$, $2$, $7$ in one, two, and three dimensions.) For $P4/m$ with $B_u$ and $B_g$ representations, $\mathcal{X}^{\text{WTSC}}$ is only a subgroup of the maximal $(\mathbb{Z}_2)^m$ subgroups of $X^{\text{BdG}}$ and we explain the correspondence in Appendix~\ref{WTSC-P4/m}.

\section{Discussion}
\label{sec:dis}
We advanced the theory of SIs for topological superconductors 
and computed the indicator groups explicitly for all space groups and pairing symmetries. 
We showed that the refinement proposed in Refs.\ \onlinecite{Skurativska2019,1907.13632} enables the detection of a variety of phases, including both ``first-order'' (i.e., conventional) and higher-order TSCs .
This is perhaps surprising, as the refinement only captures phases with zero-dimensional Majorana modes in the case of inversion symmetry studied in Refs.\ \onlinecite{Skurativska2019,1907.13632}. Furthermore, we found that the same indicator could correspond to a possibly gapped or a necessarily gapless phase depending on the additional spatial symmetries that are present. Such observations should be contrasted with the familiar case of the Fu-Kane parity criterion for topological insulators \cite{PhysRevB.76.045302}, which is valid independent of the other spatial symmetries in the system. This suggests that caution must be used in diagnosing a TSC using only part of the spatial symmetries, and it is desirable to perform a more comprehensive analysis taking into account the entire space group preserved by the superconductor, as is done in the present work.

As a concrete example, our analysis for systems with $C_4$ rotation symmetry revealed a new $\mathbb Z_2$-valued index, which we denote by $\nu_{C_4}$. We argued that $\nu_{C_4} = 1$ implies the system is a helical TSC when the system is gapped, or indicates a gapless phase when inversion symmetry is present and the superconducting pairing has even parity. Within the weak pairing assumption, this nontrivial index can be realized in systems with $d$-wave pairing and an odd number of filled Kramers pairs in the normal state (Appendix~\ref{P4_SI}). When inversion symmetry is broken such that mixed-parity pairing becomes possible, one could gap out the nodes of the superconducting gap by increasing the $p$-wave component, and the end result will be a helical TSC.  In fact, a similar picture was proposed in Ref.\ \onlinecite{Tada_2009}, although the role of the SI was not recognized there. Such mechanism may be possible for the (proximitized) superconductivity on the surfaces of 3D materials,  
where the surface termination breaks inversion symmetry and can give rise to Rashba spin-orbit coupling. 
If the system has $C_4$ rotational symmetry and a SC pairing with $\chi_{C_4}=-1$ (e.g, $d$-wave) is realized in the bulk, the induced surface superconductor on a $C_4$ preserving surface will be topological when the number of filled surface-Kramers pair at the momenta $\Gamma$ and ${\rm M}$ is odd in the normal state. 
The surface SC, if viewed as a standalone system, will be either a nodal or helical TSC.
Alternatively, one could also replace the innate surface state in the proposal above by an independent 2D system in which superconductivity is induced by proximity coupling to a $d$-wave superconductor.  

More generally, it is interesting to ask how our theory could be applied to surface superconductivity, especially for the anomalous surface states arising from a topological bulk \cite{PhysRevLett.100.096407}.
Conceptually, one can also compute the refined SI of a non-superconducting insulator by assuming an arbitrarily weak pairing amplitude with a chosen pairing symmetry. If the insulator is atomic to begin with (i.e., its ground state is smoothly deformable to a product state of localized electrons), the refined SI is trivial by definition. However, if the insulator is topological, its refined SI may be nontrivial. 
As the pairing can be arbitrarily weak in the bulk, this nontrivial refined SI is a statement on the nature of the TSC realized at the surface. As a concrete example, consider an inversion-symmetric strong TI. If we assume an odd-parity pairing is added to the system, one sees that the refined SI will be nontrivial. This setup is formally realized for an S-TI-S junction with a $\pi$ phase shift, and the helical Majorana mode that appears \cite{PhysRevLett.100.096407} is consistent with the refined SI discussed above. This correspondence between a strong TI and a (higher-order) TSC is quite general, and has been noted earlier in Ref.\ \onlinecite{PhysRevB.99.125149} assuming $C_4$ symmetry. Given the vast majority of TI candidates discovered from materials database searches \cite{Zhang2019, Vergniory2019,Tang2019} are in fact (semi-)metallic, they may have superconducting instability and could realize a TSC based on the analysis above.

On a more technical note, we remark that our theory does not incorporate the Pfaffian invariant discussed in Ref.\ \onlinecite{1907.13632}, although this invariant can be readily related to the number of filled states in the normal-state band structure within the weak-pairing assumption. While it will be interesting to incorporate it into our formalism, the Pfaffian invariant is different from the usual representation counts as it is $\mathbb Z_2$-valued. This will bring about some technical differences in the computation of the SI group, although a systematic computation is still possible \cite{1907.13632}.

Lastly, we note that in our analysis for the physical meaning of $\nu_{C_4}$ we introduced the notion of Wannierizable TSCs, examples of which include the 1D Kitaev chain and 2D higher-order TSCs, as well as weak phases constructed by stacks of them. 
As a more nontrivial example, we note that the set of WTSCs also includes ``first order'' examples like the even entries for the $\mathbb Z$-valued classification of class DIII superconductors in 3D.
While we have developed a formalism for the partial diagnosis of such TSCs, our analysis does not result in a full classification for WTSCs. It will be interesting to explore how the full classification can be obtained, as well as the unique physical properties, if any, that are tied to the notion of WTSCs.

\begin{acknowledgments}
We would like to thank  E.\ Khalaf, T.\ Morimoto, K.\ Shiozaki, A.\ Vishwanath, Y.\ Yanase, and M.\ Zaletel for valuable discussions and earlier collaborations on related topics.
The work of SO is supported by Materials Education program for the future leaders in Research, Industry, and Technology (MERIT).  
The work of HCP is supported by a Pappalardo Fellowship at MIT and a Croucher Foundation Fellowship.
The work of HW is supported by JSPS KAKENHI Grant No.~JP17K17678 and by JST PRESTO Grant No.~JPMJPR18LA.
\end{acknowledgments}
\bibliography{ref}

\begin{thebibliography}{46}%
\makeatletter
\providecommand \@ifxundefined [1]{%
 \@ifx{#1\undefined}
}%
\providecommand \@ifnum [1]{%
 \ifnum #1\expandafter \@firstoftwo
 \else \expandafter \@secondoftwo
 \fi
}%
\providecommand \@ifx [1]{%
 \ifx #1\expandafter \@firstoftwo
 \else \expandafter \@secondoftwo
 \fi
}%
\providecommand \natexlab [1]{#1}%
\providecommand \enquote  [1]{``#1''}%
\providecommand \bibnamefont  [1]{#1}%
\providecommand \bibfnamefont [1]{#1}%
\providecommand \citenamefont [1]{#1}%
\providecommand \href@noop [0]{\@secondoftwo}%
\providecommand \href [0]{\begingroup \@sanitize@url \@href}%
\providecommand \@href[1]{\@@startlink{#1}\@@href}%
\providecommand \@@href[1]{\endgroup#1\@@endlink}%
\providecommand \@sanitize@url [0]{\catcode `\\12\catcode `\$12\catcode
  `\&12\catcode `\#12\catcode `\^12\catcode `\_12\catcode `\%12\relax}%
\providecommand \@@startlink[1]{}%
\providecommand \@@endlink[0]{}%
\providecommand \url  [0]{\begingroup\@sanitize@url \@url }%
\providecommand \@url [1]{\endgroup\@href {#1}{\urlprefix }}%
\providecommand \urlprefix  [0]{URL }%
\providecommand \Eprint [0]{\href }%
\providecommand \doibase [0]{http://dx.doi.org/}%
\providecommand \selectlanguage [0]{\@gobble}%
\providecommand \bibinfo  [0]{\@secondoftwo}%
\providecommand \bibfield  [0]{\@secondoftwo}%
\providecommand \translation [1]{[#1]}%
\providecommand \BibitemOpen [0]{}%
\providecommand \bibitemStop [0]{}%
\providecommand \bibitemNoStop [0]{.\EOS\space}%
\providecommand \EOS [0]{\spacefactor3000\relax}%
\providecommand \BibitemShut  [1]{\csname bibitem#1\endcsname}%
\let\auto@bib@innerbib\@empty
\bibitem [{\citenamefont {Norman}(2011)}]{Norman196}%
  \BibitemOpen
  \bibfield  {author} {\bibinfo {author} {\bibfnamefont {M.~R.}\ \bibnamefont
  {Norman}},\ }\href {\doibase 10.1126/science.1200181} {\bibfield  {journal}
  {\bibinfo  {journal} {Science}\ }\textbf {\bibinfo {volume} {332}},\ \bibinfo
  {pages} {196} (\bibinfo {year} {2011})}\BibitemShut {NoStop}%
\bibitem [{\citenamefont {Qi}\ and\ \citenamefont
  {Zhang}(2011)}]{RevModPhys.83.1057}%
  \BibitemOpen
  \bibfield  {author} {\bibinfo {author} {\bibfnamefont {X.-L.}\ \bibnamefont
  {Qi}}\ and\ \bibinfo {author} {\bibfnamefont {S.-C.}\ \bibnamefont {Zhang}},\
  }\href {\doibase 10.1103/RevModPhys.83.1057} {\bibfield  {journal} {\bibinfo
  {journal} {Rev. Mod. Phys.}\ }\textbf {\bibinfo {volume} {83}},\ \bibinfo
  {pages} {1057} (\bibinfo {year} {2011})}\BibitemShut {NoStop}%
\bibitem [{\citenamefont {Ando}\ and\ \citenamefont {Fu}(2015)}]{Ando-Fu}%
  \BibitemOpen
  \bibfield  {author} {\bibinfo {author} {\bibfnamefont {Y.}~\bibnamefont
  {Ando}}\ and\ \bibinfo {author} {\bibfnamefont {L.}~\bibnamefont {Fu}},\
  }\href {\doibase 10.1146/annurev-conmatphys-031214-014501} {\bibfield
  {journal} {\bibinfo  {journal} {Annual Review of Condensed Matter Physics}\
  }\textbf {\bibinfo {volume} {6}},\ \bibinfo {pages} {361} (\bibinfo {year}
  {2015})}\BibitemShut {NoStop}%
\bibitem [{\citenamefont {Sato}\ and\ \citenamefont {Ando}(2017)}]{Sato_2017}%
  \BibitemOpen
  \bibfield  {author} {\bibinfo {author} {\bibfnamefont {M.}~\bibnamefont
  {Sato}}\ and\ \bibinfo {author} {\bibfnamefont {Y.}~\bibnamefont {Ando}},\
  }\href {\doibase 10.1088/1361-6633/aa6ac7} {\bibfield  {journal} {\bibinfo
  {journal} {Reports on Progress in Physics}\ }\textbf {\bibinfo {volume}
  {80}},\ \bibinfo {pages} {076501} (\bibinfo {year} {2017})}\BibitemShut
  {NoStop}%
\bibitem [{\citenamefont {Elliott}\ and\ \citenamefont
  {Franz}(2015)}]{RevModPhys.87.137}%
  \BibitemOpen
  \bibfield  {author} {\bibinfo {author} {\bibfnamefont {S.~R.}\ \bibnamefont
  {Elliott}}\ and\ \bibinfo {author} {\bibfnamefont {M.}~\bibnamefont
  {Franz}},\ }\href {\doibase 10.1103/RevModPhys.87.137} {\bibfield  {journal}
  {\bibinfo  {journal} {Rev. Mod. Phys.}\ }\textbf {\bibinfo {volume} {87}},\
  \bibinfo {pages} {137} (\bibinfo {year} {2015})}\BibitemShut {NoStop}%
\bibitem [{\citenamefont {Freed}\ and\ \citenamefont
  {Moore}(2013)}]{Freed2013}%
  \BibitemOpen
  \bibfield  {author} {\bibinfo {author} {\bibfnamefont {D.~S.}\ \bibnamefont
  {Freed}}\ and\ \bibinfo {author} {\bibfnamefont {G.~W.}\ \bibnamefont
  {Moore}},\ }\href {\doibase 10.1007/s00023-013-0236-x} {\bibfield  {journal}
  {\bibinfo  {journal} {Annales Henri Poincar{\'e}}\ }\textbf {\bibinfo
  {volume} {14}},\ \bibinfo {pages} {1927} (\bibinfo {year}
  {2013})}\BibitemShut {NoStop}%
\bibitem [{\citenamefont {Song}\ \emph {et~al.}(2017)\citenamefont {Song},
  \citenamefont {Huang}, \citenamefont {Fu},\ and\ \citenamefont
  {Hermele}}]{PhysRevX.7.011020}%
  \BibitemOpen
  \bibfield  {author} {\bibinfo {author} {\bibfnamefont {H.}~\bibnamefont
  {Song}}, \bibinfo {author} {\bibfnamefont {S.-J.}\ \bibnamefont {Huang}},
  \bibinfo {author} {\bibfnamefont {L.}~\bibnamefont {Fu}}, \ and\ \bibinfo
  {author} {\bibfnamefont {M.}~\bibnamefont {Hermele}},\ }\href {\doibase
  10.1103/PhysRevX.7.011020} {\bibfield  {journal} {\bibinfo  {journal} {Phys.
  Rev. X}\ }\textbf {\bibinfo {volume} {7}},\ \bibinfo {pages} {011020}
  (\bibinfo {year} {2017})}\BibitemShut {NoStop}%
\bibitem [{\citenamefont {Kruthoff}\ \emph {et~al.}(2017)\citenamefont
  {Kruthoff}, \citenamefont {de~Boer}, \citenamefont {van Wezel}, \citenamefont
  {Kane},\ and\ \citenamefont {Slager}}]{PhysRevX.7.041069}%
  \BibitemOpen
  \bibfield  {author} {\bibinfo {author} {\bibfnamefont {J.}~\bibnamefont
  {Kruthoff}}, \bibinfo {author} {\bibfnamefont {J.}~\bibnamefont {de~Boer}},
  \bibinfo {author} {\bibfnamefont {J.}~\bibnamefont {van Wezel}}, \bibinfo
  {author} {\bibfnamefont {C.~L.}\ \bibnamefont {Kane}}, \ and\ \bibinfo
  {author} {\bibfnamefont {R.-J.}\ \bibnamefont {Slager}},\ }\href {\doibase
  10.1103/PhysRevX.7.041069} {\bibfield  {journal} {\bibinfo  {journal} {Phys.
  Rev. X}\ }\textbf {\bibinfo {volume} {7}},\ \bibinfo {pages} {041069}
  (\bibinfo {year} {2017})}\BibitemShut {NoStop}%
\bibitem [{\citenamefont {Thorngren}\ and\ \citenamefont
  {Else}(2018)}]{PhysRevX.8.011040}%
  \BibitemOpen
  \bibfield  {author} {\bibinfo {author} {\bibfnamefont {R.}~\bibnamefont
  {Thorngren}}\ and\ \bibinfo {author} {\bibfnamefont {D.~V.}\ \bibnamefont
  {Else}},\ }\href {\doibase 10.1103/PhysRevX.8.011040} {\bibfield  {journal}
  {\bibinfo  {journal} {Phys. Rev. X}\ }\textbf {\bibinfo {volume} {8}},\
  \bibinfo {pages} {011040} (\bibinfo {year} {2018})}\BibitemShut {NoStop}%
\bibitem [{\citenamefont {Huang}\ \emph {et~al.}(2017)\citenamefont {Huang},
  \citenamefont {Song}, \citenamefont {Huang},\ and\ \citenamefont
  {Hermele}}]{PhysRevB.96.205106}%
  \BibitemOpen
  \bibfield  {author} {\bibinfo {author} {\bibfnamefont {S.-J.}\ \bibnamefont
  {Huang}}, \bibinfo {author} {\bibfnamefont {H.}~\bibnamefont {Song}},
  \bibinfo {author} {\bibfnamefont {Y.-P.}\ \bibnamefont {Huang}}, \ and\
  \bibinfo {author} {\bibfnamefont {M.}~\bibnamefont {Hermele}},\ }\href
  {\doibase 10.1103/PhysRevB.96.205106} {\bibfield  {journal} {\bibinfo
  {journal} {Phys. Rev. B}\ }\textbf {\bibinfo {volume} {96}},\ \bibinfo
  {pages} {205106} (\bibinfo {year} {2017})}\BibitemShut {NoStop}%
\bibitem [{\citenamefont {Khalaf}\ \emph {et~al.}(2018)\citenamefont {Khalaf},
  \citenamefont {Po}, \citenamefont {Vishwanath},\ and\ \citenamefont
  {Watanabe}}]{PhysRevX.8.031070}%
  \BibitemOpen
  \bibfield  {author} {\bibinfo {author} {\bibfnamefont {E.}~\bibnamefont
  {Khalaf}}, \bibinfo {author} {\bibfnamefont {H.~C.}\ \bibnamefont {Po}},
  \bibinfo {author} {\bibfnamefont {A.}~\bibnamefont {Vishwanath}}, \ and\
  \bibinfo {author} {\bibfnamefont {H.}~\bibnamefont {Watanabe}},\ }\href
  {\doibase 10.1103/PhysRevX.8.031070} {\bibfield  {journal} {\bibinfo
  {journal} {Phys. Rev. X}\ }\textbf {\bibinfo {volume} {8}},\ \bibinfo {pages}
  {031070} (\bibinfo {year} {2018})}\BibitemShut {NoStop}%
\bibitem [{\citenamefont {{Shiozaki}}\ \emph {et~al.}()\citenamefont
  {{Shiozaki}}, \citenamefont {{Sato}},\ and\ \citenamefont
  {{Gomi}}}]{Shiozaki2018}%
  \BibitemOpen
  \bibfield  {author} {\bibinfo {author} {\bibfnamefont {K.}~\bibnamefont
  {{Shiozaki}}}, \bibinfo {author} {\bibfnamefont {M.}~\bibnamefont {{Sato}}},
  \ and\ \bibinfo {author} {\bibfnamefont {K.}~\bibnamefont {{Gomi}}},\
  }\href@noop {} {\ }\Eprint {http://arxiv.org/abs/1802.06694}
  {arXiv:1802.06694} \BibitemShut {NoStop}%
\bibitem [{\citenamefont {{Song}}\ \emph {et~al.}()\citenamefont {{Song}},
  \citenamefont {{Huang}}, \citenamefont {{Qi}}, \citenamefont {{Fang}},\ and\
  \citenamefont {{Hermele}}}]{Song2018}%
  \BibitemOpen
  \bibfield  {author} {\bibinfo {author} {\bibfnamefont {Z.}~\bibnamefont
  {{Song}}}, \bibinfo {author} {\bibfnamefont {S.-J.}\ \bibnamefont {{Huang}}},
  \bibinfo {author} {\bibfnamefont {Y.}~\bibnamefont {{Qi}}}, \bibinfo {author}
  {\bibfnamefont {C.}~\bibnamefont {{Fang}}}, \ and\ \bibinfo {author}
  {\bibfnamefont {M.}~\bibnamefont {{Hermele}}},\ }\href@noop {} {\ }\Eprint
  {http://arxiv.org/abs/1810.02330} {arXiv:1810.02330} \BibitemShut {NoStop}%
\bibitem [{\citenamefont {Po}\ \emph {et~al.}(2017)\citenamefont {Po},
  \citenamefont {Vishwanath},\ and\ \citenamefont {Watanabe}}]{Po2017}%
  \BibitemOpen
  \bibfield  {author} {\bibinfo {author} {\bibfnamefont {H.~C.}\ \bibnamefont
  {Po}}, \bibinfo {author} {\bibfnamefont {A.}~\bibnamefont {Vishwanath}}, \
  and\ \bibinfo {author} {\bibfnamefont {H.}~\bibnamefont {Watanabe}},\ }\href
  {\doibase 10.1038/s41467-017-00133-2} {\bibfield  {journal} {\bibinfo
  {journal} {Nat. Commun.}\ }\textbf {\bibinfo {volume} {8}},\ \bibinfo {pages}
  {50} (\bibinfo {year} {2017})}\BibitemShut {NoStop}%
\bibitem [{\citenamefont {{Bradlyn}}\ \emph {et~al.}(2017)\citenamefont
  {{Bradlyn}}, \citenamefont {{Elcoro}}, \citenamefont {{Cano}}, \citenamefont
  {{Vergniory}}, \citenamefont {{Wang}}, \citenamefont {{Felser}},
  \citenamefont {{Aroyo}},\ and\ \citenamefont {{Bernevig}}}]{TQC}%
  \BibitemOpen
  \bibfield  {author} {\bibinfo {author} {\bibfnamefont {B.}~\bibnamefont
  {{Bradlyn}}}, \bibinfo {author} {\bibfnamefont {L.}~\bibnamefont {{Elcoro}}},
  \bibinfo {author} {\bibfnamefont {J.}~\bibnamefont {{Cano}}}, \bibinfo
  {author} {\bibfnamefont {M.~G.}\ \bibnamefont {{Vergniory}}}, \bibinfo
  {author} {\bibfnamefont {Z.}~\bibnamefont {{Wang}}}, \bibinfo {author}
  {\bibfnamefont {C.}~\bibnamefont {{Felser}}}, \bibinfo {author}
  {\bibfnamefont {M.~I.}\ \bibnamefont {{Aroyo}}}, \ and\ \bibinfo {author}
  {\bibfnamefont {B.~A.}\ \bibnamefont {{Bernevig}}},\ }\href {\doibase
  10.1038/nature23268} {\bibfield  {journal} {\bibinfo  {journal} {Nature}\
  }\textbf {\bibinfo {volume} {547}},\ \bibinfo {pages} {298} (\bibinfo {year}
  {2017})}\BibitemShut {NoStop}%
\bibitem [{\citenamefont {Zhang}\ \emph {et~al.}(2019)\citenamefont {Zhang},
  \citenamefont {Jiang}, \citenamefont {Song}, \citenamefont {Huang},
  \citenamefont {He}, \citenamefont {Fang}, \citenamefont {Weng},\ and\
  \citenamefont {Fang}}]{Zhang2019}%
  \BibitemOpen
  \bibfield  {author} {\bibinfo {author} {\bibfnamefont {T.}~\bibnamefont
  {Zhang}}, \bibinfo {author} {\bibfnamefont {Y.}~\bibnamefont {Jiang}},
  \bibinfo {author} {\bibfnamefont {Z.}~\bibnamefont {Song}}, \bibinfo {author}
  {\bibfnamefont {H.}~\bibnamefont {Huang}}, \bibinfo {author} {\bibfnamefont
  {Y.}~\bibnamefont {He}}, \bibinfo {author} {\bibfnamefont {Z.}~\bibnamefont
  {Fang}}, \bibinfo {author} {\bibfnamefont {H.}~\bibnamefont {Weng}}, \ and\
  \bibinfo {author} {\bibfnamefont {C.}~\bibnamefont {Fang}},\ }\href {\doibase
  10.1038/s41586-019-0944-6} {\bibfield  {journal} {\bibinfo  {journal}
  {Nature}\ }\textbf {\bibinfo {volume} {566}},\ \bibinfo {pages} {475}
  (\bibinfo {year} {2019})}\BibitemShut {NoStop}%
\bibitem [{\citenamefont {Vergniory}\ \emph {et~al.}(2019)\citenamefont
  {Vergniory}, \citenamefont {Elcoro}, \citenamefont {Felser}, \citenamefont
  {Regnault}, \citenamefont {Bernevig},\ and\ \citenamefont
  {Wang}}]{Vergniory2019}%
  \BibitemOpen
  \bibfield  {author} {\bibinfo {author} {\bibfnamefont {M.~G.}\ \bibnamefont
  {Vergniory}}, \bibinfo {author} {\bibfnamefont {L.}~\bibnamefont {Elcoro}},
  \bibinfo {author} {\bibfnamefont {C.}~\bibnamefont {Felser}}, \bibinfo
  {author} {\bibfnamefont {N.}~\bibnamefont {Regnault}}, \bibinfo {author}
  {\bibfnamefont {B.~A.}\ \bibnamefont {Bernevig}}, \ and\ \bibinfo {author}
  {\bibfnamefont {Z.}~\bibnamefont {Wang}},\ }\href {\doibase
  10.1038/s41586-019-0954-4} {\bibfield  {journal} {\bibinfo  {journal}
  {Nature}\ }\textbf {\bibinfo {volume} {566}},\ \bibinfo {pages} {480}
  (\bibinfo {year} {2019})}\BibitemShut {NoStop}%
\bibitem [{\citenamefont {Tang}\ \emph {et~al.}(2019)\citenamefont {Tang},
  \citenamefont {Po}, \citenamefont {Vishwanath},\ and\ \citenamefont
  {Wan}}]{Tang2019}%
  \BibitemOpen
  \bibfield  {author} {\bibinfo {author} {\bibfnamefont {F.}~\bibnamefont
  {Tang}}, \bibinfo {author} {\bibfnamefont {H.~C.}\ \bibnamefont {Po}},
  \bibinfo {author} {\bibfnamefont {A.}~\bibnamefont {Vishwanath}}, \ and\
  \bibinfo {author} {\bibfnamefont {X.}~\bibnamefont {Wan}},\ }\href {\doibase
  10.1038/s41586-019-0937-5} {\bibfield  {journal} {\bibinfo  {journal}
  {Nature}\ }\textbf {\bibinfo {volume} {566}},\ \bibinfo {pages} {486}
  (\bibinfo {year} {2019})}\BibitemShut {NoStop}%
\bibitem [{\citenamefont {Ono}\ \emph {et~al.}(2019)\citenamefont {Ono},
  \citenamefont {Yanase},\ and\ \citenamefont {Watanabe}}]{1811.08712}%
  \BibitemOpen
  \bibfield  {author} {\bibinfo {author} {\bibfnamefont {S.}~\bibnamefont
  {Ono}}, \bibinfo {author} {\bibfnamefont {Y.}~\bibnamefont {Yanase}}, \ and\
  \bibinfo {author} {\bibfnamefont {H.}~\bibnamefont {Watanabe}},\ }\href
  {\doibase 10.1103/PhysRevResearch.1.013012} {\bibfield  {journal} {\bibinfo
  {journal} {Phys. Rev. Res.}\ }\textbf {\bibinfo {volume} {1}},\ \bibinfo
  {pages} {013012} (\bibinfo {year} {2019})}\BibitemShut {NoStop}%
\bibitem [{\citenamefont {Skurativska}\ \emph {et~al.}()\citenamefont
  {Skurativska}, \citenamefont {Neupert},\ and\ \citenamefont
  {Fischer}}]{Skurativska2019}%
  \BibitemOpen
  \bibfield  {author} {\bibinfo {author} {\bibfnamefont {A.}~\bibnamefont
  {Skurativska}}, \bibinfo {author} {\bibfnamefont {T.}~\bibnamefont
  {Neupert}}, \ and\ \bibinfo {author} {\bibfnamefont {M.~H.}\ \bibnamefont
  {Fischer}},\ }\href {http://arxiv.org/abs/1906.11267} {\ }\Eprint
  {http://arxiv.org/abs/1906.11267} {arXiv:1906.11267} \BibitemShut {NoStop}%
\bibitem [{\citenamefont {Shiozaki}()}]{1907.13632}%
  \BibitemOpen
  \bibfield  {author} {\bibinfo {author} {\bibfnamefont {K.}~\bibnamefont
  {Shiozaki}},\ }\href@noop {} {\ }\Eprint
  {http://arxiv.org/abs/arXiv:1907.13632} {arXiv:1907.13632} \BibitemShut
  {NoStop}%
\bibitem [{\citenamefont {Langbehn}\ \emph {et~al.}(2017)\citenamefont
  {Langbehn}, \citenamefont {Peng}, \citenamefont {Trifunovic}, \citenamefont
  {von Oppen},\ and\ \citenamefont {Brouwer}}]{PhysRevLett.119.246401}%
  \BibitemOpen
  \bibfield  {author} {\bibinfo {author} {\bibfnamefont {J.}~\bibnamefont
  {Langbehn}}, \bibinfo {author} {\bibfnamefont {Y.}~\bibnamefont {Peng}},
  \bibinfo {author} {\bibfnamefont {L.}~\bibnamefont {Trifunovic}}, \bibinfo
  {author} {\bibfnamefont {F.}~\bibnamefont {von Oppen}}, \ and\ \bibinfo
  {author} {\bibfnamefont {P.~W.}\ \bibnamefont {Brouwer}},\ }\href {\doibase
  10.1103/PhysRevLett.119.246401} {\bibfield  {journal} {\bibinfo  {journal}
  {Phys. Rev. Lett.}\ }\textbf {\bibinfo {volume} {119}},\ \bibinfo {pages}
  {246401} (\bibinfo {year} {2017})}\BibitemShut {NoStop}%
\bibitem [{\citenamefont {Khalaf}(2018)}]{PhysRevB.97.205136}%
  \BibitemOpen
  \bibfield  {author} {\bibinfo {author} {\bibfnamefont {E.}~\bibnamefont
  {Khalaf}},\ }\href {\doibase 10.1103/PhysRevB.97.205136} {\bibfield
  {journal} {\bibinfo  {journal} {Phys. Rev. B}\ }\textbf {\bibinfo {volume}
  {97}},\ \bibinfo {pages} {205136} (\bibinfo {year} {2018})}\BibitemShut
  {NoStop}%
\bibitem [{\citenamefont {Qi}\ \emph {et~al.}(2010)\citenamefont {Qi},
  \citenamefont {Hughes},\ and\ \citenamefont {Zhang}}]{PhysRevB.81.134508}%
  \BibitemOpen
  \bibfield  {author} {\bibinfo {author} {\bibfnamefont {X.-L.}\ \bibnamefont
  {Qi}}, \bibinfo {author} {\bibfnamefont {T.~L.}\ \bibnamefont {Hughes}}, \
  and\ \bibinfo {author} {\bibfnamefont {S.-C.}\ \bibnamefont {Zhang}},\ }\href
  {\doibase 10.1103/PhysRevB.81.134508} {\bibfield  {journal} {\bibinfo
  {journal} {Phys. Rev. B}\ }\textbf {\bibinfo {volume} {81}},\ \bibinfo
  {pages} {134508} (\bibinfo {year} {2010})}\BibitemShut {NoStop}%
\bibitem [{\citenamefont {Wang}\ \emph {et~al.}(2018)\citenamefont {Wang},
  \citenamefont {Lin},\ and\ \citenamefont {Hughes}}]{PhysRevB.98.165144}%
  \BibitemOpen
  \bibfield  {author} {\bibinfo {author} {\bibfnamefont {Y.}~\bibnamefont
  {Wang}}, \bibinfo {author} {\bibfnamefont {M.}~\bibnamefont {Lin}}, \ and\
  \bibinfo {author} {\bibfnamefont {T.~L.}\ \bibnamefont {Hughes}},\ }\href
  {\doibase 10.1103/PhysRevB.98.165144} {\bibfield  {journal} {\bibinfo
  {journal} {Phys. Rev. B}\ }\textbf {\bibinfo {volume} {98}},\ \bibinfo
  {pages} {165144} (\bibinfo {year} {2018})}\BibitemShut {NoStop}%
\bibitem [{\citenamefont {Hsu}\ \emph {et~al.}(2019)\citenamefont {Hsu},
  \citenamefont {Cole}, \citenamefont {Zhang},\ and\ \citenamefont
  {Sau}}]{1904.06361}%
  \BibitemOpen
  \bibfield  {author} {\bibinfo {author} {\bibfnamefont {Y.-T.}\ \bibnamefont
  {Hsu}}, \bibinfo {author} {\bibfnamefont {W.~S.}\ \bibnamefont {Cole}},
  \bibinfo {author} {\bibfnamefont {R.-X.}\ \bibnamefont {Zhang}}, \ and\
  \bibinfo {author} {\bibfnamefont {J.~D.}\ \bibnamefont {Sau}},\ }\href@noop
  {} {} (\bibinfo {year} {2019}),\ \Eprint
  {http://arxiv.org/abs/arXiv:1904.06361} {arXiv:1904.06361} \BibitemShut
  {NoStop}%
\bibitem [{\citenamefont {{Ahn}}\ and\ \citenamefont {{Yang}}()}]{Ahn2019}%
  \BibitemOpen
  \bibfield  {author} {\bibinfo {author} {\bibfnamefont {J.}~\bibnamefont
  {{Ahn}}}\ and\ \bibinfo {author} {\bibfnamefont {B.-J.}\ \bibnamefont
  {{Yang}}},\ }\href@noop {} {\ }\Eprint {http://arxiv.org/abs/1906.02709}
  {arXiv:1906.02709} \BibitemShut {NoStop}%
\bibitem [{\citenamefont {Fu}\ and\ \citenamefont
  {Berg}(2010)}]{PhysRevLett.105.097001}%
  \BibitemOpen
  \bibfield  {author} {\bibinfo {author} {\bibfnamefont {L.}~\bibnamefont
  {Fu}}\ and\ \bibinfo {author} {\bibfnamefont {E.}~\bibnamefont {Berg}},\
  }\href {\doibase 10.1103/PhysRevLett.105.097001} {\bibfield  {journal}
  {\bibinfo  {journal} {Phys. Rev. Lett.}\ }\textbf {\bibinfo {volume} {105}},\
  \bibinfo {pages} {097001} (\bibinfo {year} {2010})}\BibitemShut {NoStop}%
\bibitem [{\citenamefont {Sato}(2010)}]{PhysRevB.81.220504}%
  \BibitemOpen
  \bibfield  {author} {\bibinfo {author} {\bibfnamefont {M.}~\bibnamefont
  {Sato}},\ }\href {\doibase 10.1103/PhysRevB.81.220504} {\bibfield  {journal}
  {\bibinfo  {journal} {Phys. Rev. B}\ }\textbf {\bibinfo {volume} {81}},\
  \bibinfo {pages} {220504} (\bibinfo {year} {2010})}\BibitemShut {NoStop}%
\bibitem [{\citenamefont {Fu}\ and\ \citenamefont
  {Kane}(2007)}]{PhysRevB.76.045302}%
  \BibitemOpen
  \bibfield  {author} {\bibinfo {author} {\bibfnamefont {L.}~\bibnamefont
  {Fu}}\ and\ \bibinfo {author} {\bibfnamefont {C.~L.}\ \bibnamefont {Kane}},\
  }\href {\doibase 10.1103/PhysRevB.76.045302} {\bibfield  {journal} {\bibinfo
  {journal} {Phys. Rev. B}\ }\textbf {\bibinfo {volume} {76}},\ \bibinfo
  {pages} {045302} (\bibinfo {year} {2007})}\BibitemShut {NoStop}%
\bibitem [{\citenamefont {Schnyder}\ \emph {et~al.}(2008)\citenamefont
  {Schnyder}, \citenamefont {Ryu}, \citenamefont {Furusaki},\ and\
  \citenamefont {Ludwig}}]{PhysRevB.78.195125}%
  \BibitemOpen
  \bibfield  {author} {\bibinfo {author} {\bibfnamefont {A.~P.}\ \bibnamefont
  {Schnyder}}, \bibinfo {author} {\bibfnamefont {S.}~\bibnamefont {Ryu}},
  \bibinfo {author} {\bibfnamefont {A.}~\bibnamefont {Furusaki}}, \ and\
  \bibinfo {author} {\bibfnamefont {A.~W.~W.}\ \bibnamefont {Ludwig}},\ }\href
  {\doibase 10.1103/PhysRevB.78.195125} {\bibfield  {journal} {\bibinfo
  {journal} {Phys. Rev. B}\ }\textbf {\bibinfo {volume} {78}},\ \bibinfo
  {pages} {195125} (\bibinfo {year} {2008})}\BibitemShut {NoStop}%
\bibitem [{\citenamefont {Sato}\ and\ \citenamefont
  {Fujimoto}(2016)}]{SatoFujimoto}%
  \BibitemOpen
  \bibfield  {author} {\bibinfo {author} {\bibfnamefont {M.}~\bibnamefont
  {Sato}}\ and\ \bibinfo {author} {\bibfnamefont {S.}~\bibnamefont
  {Fujimoto}},\ }\href {\doibase 10.7566/JPSJ.85.072001} {\bibfield  {journal}
  {\bibinfo  {journal} {J. Phys. Soc. Jpn.}\ }\textbf {\bibinfo {volume}
  {85}},\ \bibinfo {pages} {072001} (\bibinfo {year} {2016})}\BibitemShut
  {NoStop}%
\bibitem [{\citenamefont {Po}\ \emph {et~al.}(2018)\citenamefont {Po},
  \citenamefont {Watanabe},\ and\ \citenamefont
  {Vishwanath}}]{PhysRevLett.121.126402}%
  \BibitemOpen
  \bibfield  {author} {\bibinfo {author} {\bibfnamefont {H.~C.}\ \bibnamefont
  {Po}}, \bibinfo {author} {\bibfnamefont {H.}~\bibnamefont {Watanabe}}, \ and\
  \bibinfo {author} {\bibfnamefont {A.}~\bibnamefont {Vishwanath}},\ }\href
  {\doibase 10.1103/PhysRevLett.121.126402} {\bibfield  {journal} {\bibinfo
  {journal} {Phys. Rev. Lett.}\ }\textbf {\bibinfo {volume} {121}},\ \bibinfo
  {pages} {126402} (\bibinfo {year} {2018})}\BibitemShut {NoStop}%
\bibitem [{\citenamefont {Bradlyn}\ \emph {et~al.}(2019)\citenamefont
  {Bradlyn}, \citenamefont {Wang}, \citenamefont {Cano},\ and\ \citenamefont
  {Bernevig}}]{PhysRevB.99.045140}%
  \BibitemOpen
  \bibfield  {author} {\bibinfo {author} {\bibfnamefont {B.}~\bibnamefont
  {Bradlyn}}, \bibinfo {author} {\bibfnamefont {Z.}~\bibnamefont {Wang}},
  \bibinfo {author} {\bibfnamefont {J.}~\bibnamefont {Cano}}, \ and\ \bibinfo
  {author} {\bibfnamefont {B.~A.}\ \bibnamefont {Bernevig}},\ }\href {\doibase
  10.1103/PhysRevB.99.045140} {\bibfield  {journal} {\bibinfo  {journal} {Phys.
  Rev. B}\ }\textbf {\bibinfo {volume} {99}},\ \bibinfo {pages} {045140}
  (\bibinfo {year} {2019})}\BibitemShut {NoStop}%
\bibitem [{\citenamefont {Kitaev}(2001)}]{Kitaev_2001}%
  \BibitemOpen
  \bibfield  {author} {\bibinfo {author} {\bibfnamefont {A.~Y.}\ \bibnamefont
  {Kitaev}},\ }\href {\doibase 10.1070/1063-7869/44/10s/s29} {\bibfield
  {journal} {\bibinfo  {journal} {Physics-Uspekhi}\ }\textbf {\bibinfo {volume}
  {44}},\ \bibinfo {pages} {131} (\bibinfo {year} {2001})}\BibitemShut
  {NoStop}%
\bibitem [{\citenamefont {Ono}\ and\ \citenamefont
  {Watanabe}(2018)}]{PhysRevB.98.115150}%
  \BibitemOpen
  \bibfield  {author} {\bibinfo {author} {\bibfnamefont {S.}~\bibnamefont
  {Ono}}\ and\ \bibinfo {author} {\bibfnamefont {H.}~\bibnamefont {Watanabe}},\
  }\href {\doibase 10.1103/PhysRevB.98.115150} {\bibfield  {journal} {\bibinfo
  {journal} {Phys. Rev. B}\ }\textbf {\bibinfo {volume} {98}},\ \bibinfo
  {pages} {115150} (\bibinfo {year} {2018})}\BibitemShut {NoStop}%
\bibitem [{\citenamefont {Bradley}\ and\ \citenamefont
  {Cracknell}(1972)}]{Bradley}%
  \BibitemOpen
  \bibfield  {author} {\bibinfo {author} {\bibfnamefont {C.~J.}\ \bibnamefont
  {Bradley}}\ and\ \bibinfo {author} {\bibfnamefont {A.~P.}\ \bibnamefont
  {Cracknell}},\ }\href
  {https://global.oup.com/academic/product/the-mathematical-theory-of-symmetry-in-solids-9780199582587?cc=jp&lang=en&#}
  {\emph {\bibinfo {title} {The Mathematical Theory of Symmetry in Solids}}}\
  (\bibinfo  {publisher} {Oxford University Press},\ \bibinfo {year}
  {1972})\BibitemShut {NoStop}%
\bibitem [{\citenamefont {Song}\ \emph {et~al.}(2018)\citenamefont {Song},
  \citenamefont {Zhang}, \citenamefont {Fang},\ and\ \citenamefont
  {Fang}}]{QuantitativeMappings}%
  \BibitemOpen
  \bibfield  {author} {\bibinfo {author} {\bibfnamefont {Z.}~\bibnamefont
  {Song}}, \bibinfo {author} {\bibfnamefont {T.}~\bibnamefont {Zhang}},
  \bibinfo {author} {\bibfnamefont {Z.}~\bibnamefont {Fang}}, \ and\ \bibinfo
  {author} {\bibfnamefont {C.}~\bibnamefont {Fang}},\ }\href {\doibase
  10.1038/s41467-018-06010-w} {\bibfield  {journal} {\bibinfo  {journal} {Nat.
  Commun.}\ }\textbf {\bibinfo {volume} {9}},\ \bibinfo {pages} {3530}
  (\bibinfo {year} {2018})}\BibitemShut {NoStop}%
\bibitem [{\citenamefont {Fang}\ \emph {et~al.}(2012)\citenamefont {Fang},
  \citenamefont {Gilbert},\ and\ \citenamefont
  {Bernevig}}]{PhysRevB.86.115112}%
  \BibitemOpen
  \bibfield  {author} {\bibinfo {author} {\bibfnamefont {C.}~\bibnamefont
  {Fang}}, \bibinfo {author} {\bibfnamefont {M.~J.}\ \bibnamefont {Gilbert}}, \
  and\ \bibinfo {author} {\bibfnamefont {B.~A.}\ \bibnamefont {Bernevig}},\
  }\href {\doibase 10.1103/PhysRevB.86.115112} {\bibfield  {journal} {\bibinfo
  {journal} {Phys. Rev. B}\ }\textbf {\bibinfo {volume} {86}},\ \bibinfo
  {pages} {115112} (\bibinfo {year} {2012})}\BibitemShut {NoStop}%
\bibitem [{\citenamefont {Chiu}\ \emph {et~al.}(2013)\citenamefont {Chiu},
  \citenamefont {Yao},\ and\ \citenamefont {Ryu}}]{PhysRevB.88.075142}%
  \BibitemOpen
  \bibfield  {author} {\bibinfo {author} {\bibfnamefont {C.-K.}\ \bibnamefont
  {Chiu}}, \bibinfo {author} {\bibfnamefont {H.}~\bibnamefont {Yao}}, \ and\
  \bibinfo {author} {\bibfnamefont {S.}~\bibnamefont {Ryu}},\ }\href {\doibase
  10.1103/PhysRevB.88.075142} {\bibfield  {journal} {\bibinfo  {journal} {Phys.
  Rev. B}\ }\textbf {\bibinfo {volume} {88}},\ \bibinfo {pages} {075142}
  (\bibinfo {year} {2013})}\BibitemShut {NoStop}%
\bibitem [{\citenamefont {Sato}\ \emph {et~al.}(2011)\citenamefont {Sato},
  \citenamefont {Tanaka}, \citenamefont {Yada},\ and\ \citenamefont
  {Yokoyama}}]{PhysRevB.83.224511}%
  \BibitemOpen
  \bibfield  {author} {\bibinfo {author} {\bibfnamefont {M.}~\bibnamefont
  {Sato}}, \bibinfo {author} {\bibfnamefont {Y.}~\bibnamefont {Tanaka}},
  \bibinfo {author} {\bibfnamefont {K.}~\bibnamefont {Yada}}, \ and\ \bibinfo
  {author} {\bibfnamefont {T.}~\bibnamefont {Yokoyama}},\ }\href {\doibase
  10.1103/PhysRevB.83.224511} {\bibfield  {journal} {\bibinfo  {journal} {Phys.
  Rev. B}\ }\textbf {\bibinfo {volume} {83}},\ \bibinfo {pages} {224511}
  (\bibinfo {year} {2011})}\BibitemShut {NoStop}%
\bibitem [{Note1()}]{Note1}%
  \BibitemOpen
  \bibinfo {note} {There is a technical question of whether the addition of
  trivial states below the gap is allowed or not, which differentiates
  ``stable'' topological phases from ``fragile'' ones \cite
  {PhysRevLett.121.126402}. Since our starting point is a BdG Hamiltonian, the
  corresponding physical system does not have charge conservation symmetry and
  it is more natural to focus on stable topological phases. We will take this
  perspective and always assume appropriate trivial degrees of freedom could be
  supplied to resolve any possible fragile obstructions in a
  model.}\BibitemShut {Stop}%
\bibitem [{\citenamefont {Else}\ \emph {et~al.}(2019)\citenamefont {Else},
  \citenamefont {Po},\ and\ \citenamefont {Watanabe}}]{PhysRevB.99.125122}%
  \BibitemOpen
  \bibfield  {author} {\bibinfo {author} {\bibfnamefont {D.~V.}\ \bibnamefont
  {Else}}, \bibinfo {author} {\bibfnamefont {H.~C.}\ \bibnamefont {Po}}, \ and\
  \bibinfo {author} {\bibfnamefont {H.}~\bibnamefont {Watanabe}},\ }\href
  {\doibase 10.1103/PhysRevB.99.125122} {\bibfield  {journal} {\bibinfo
  {journal} {Phys. Rev. B}\ }\textbf {\bibinfo {volume} {99}},\ \bibinfo
  {pages} {125122} (\bibinfo {year} {2019})}\BibitemShut {NoStop}%
\bibitem [{\citenamefont {Tada}\ \emph {et~al.}(2009)\citenamefont {Tada},
  \citenamefont {Kawakami},\ and\ \citenamefont {Fujimoto}}]{Tada_2009}%
  \BibitemOpen
  \bibfield  {author} {\bibinfo {author} {\bibfnamefont {Y.}~\bibnamefont
  {Tada}}, \bibinfo {author} {\bibfnamefont {N.}~\bibnamefont {Kawakami}}, \
  and\ \bibinfo {author} {\bibfnamefont {S.}~\bibnamefont {Fujimoto}},\ }\href
  {\doibase 10.1088/1367-2630/11/5/055070} {\bibfield  {journal} {\bibinfo
  {journal} {New Journal of Physics}\ }\textbf {\bibinfo {volume} {11}},\
  \bibinfo {pages} {055070} (\bibinfo {year} {2009})}\BibitemShut {NoStop}%
\bibitem [{\citenamefont {Fu}\ and\ \citenamefont
  {Kane}(2008)}]{PhysRevLett.100.096407}%
  \BibitemOpen
  \bibfield  {author} {\bibinfo {author} {\bibfnamefont {L.}~\bibnamefont
  {Fu}}\ and\ \bibinfo {author} {\bibfnamefont {C.~L.}\ \bibnamefont {Kane}},\
  }\href {\doibase 10.1103/PhysRevLett.100.096407} {\bibfield  {journal}
  {\bibinfo  {journal} {Phys. Rev. Lett.}\ }\textbf {\bibinfo {volume} {100}},\
  \bibinfo {pages} {096407} (\bibinfo {year} {2008})}\BibitemShut {NoStop}%
\bibitem [{\citenamefont {Bultinck}\ \emph {et~al.}(2019)\citenamefont
  {Bultinck}, \citenamefont {Bernevig},\ and\ \citenamefont
  {Zaletel}}]{PhysRevB.99.125149}%
  \BibitemOpen
  \bibfield  {author} {\bibinfo {author} {\bibfnamefont {N.}~\bibnamefont
  {Bultinck}}, \bibinfo {author} {\bibfnamefont {B.~A.}\ \bibnamefont
  {Bernevig}}, \ and\ \bibinfo {author} {\bibfnamefont {M.~P.}\ \bibnamefont
  {Zaletel}},\ }\href {\doibase 10.1103/PhysRevB.99.125149} {\bibfield
  {journal} {\bibinfo  {journal} {Phys. Rev. B}\ }\textbf {\bibinfo {volume}
  {99}},\ \bibinfo {pages} {125149} (\bibinfo {year} {2019})}\BibitemShut
  {NoStop}%
\end{thebibliography}%

\clearpage

\appendix

\onecolumngrid

\section{Full classification of pairing symmetries and $X^{\text{BdG}}$}
\label{XBSTable}
Here we list all results of $X^{\text{BdG}}$. For brevity, we denote ${X^{\text{BdG}}} = \mZ_p \times \mZ_q \times\mZ_r\times\ldots$ by $\{p, q,r,\ldots\}$.

\begin{table}[H]
\caption{The list of $X^{\text{BdG}}$ for spinful electrons with time-reversal symmetry.}
\label{tab:XBS_DIII}
\begin{align*}
\begin{array}{lll}
\mbox{SG}&\mbox{Irrep}&X^{\text{BdG}}\\
\hline 
1 & A & \{\} \\
2 & A_u & \{2,2,2,4,4,4,8\} \\
2 & A_g & \{\} \\
3 & B & \{\} \\
3 & A & \{\} \\
4 & B & \{\} \\
4 & A & \{\} \\
5 & B & \{\} \\
5 & A & \{\} \\
6 & A'' & \{\} \\
6 & A'& \{\} \\
7 & A'' & \{\} \\
7 & A' & \{\} \\
8 & A'' & \{\} \\
8 & A' & \{\} \\
9 & A'' & \{\} \\
9 & A' & \{\} \\
10 & B_u & \{2,2,2,4,4,4,8\} \\
10 & B_g & \{\} \\
10 & A_u & \{2,2,2,4,4,4,8\} \\
10 & A_g & \{\} \\
11 & B_u & \{2,4,4,8\} \\
11 & B_g & \{\} \\
11 & A_u & \{2,4,4,8\} \\
11 & A_g & \{\} \\
12 & B_u & \{2,2,4,4,8\} \\
12 & B_g & \{\} \\
12 & A_u & \{2,2,4,4,8\} \\
12 & A_g & \{\} \\
13 & B_u & \{2,4,4,8\} \\
13 & B_g & \{\} \\
13 & A_u & \{2,4,4,8\} \\
13 & A_g & \{\} \\
14 & B_u & \{2,4,8\} \\
14 & B_g & \{\} \\
14 & A_u & \{2,4,8\} \\
14 & A_g & \{\} \\
15 & B_u & \{2,2,4,8\} \\
15 & B_g & \{\} \\
15 & A_u & \{2,2,4,8\} \\
15 & A_g & \{\} \\
16 & B_3 & \{\} \\
16 & B_2 & \{\} \\
16 & B_1 & \{\} \\
16 & A & \{\} \\
17 & B_3 & \{\} \\
17 & B_2 & \{\} \\
17 & B_1 & \{\} \\
17 & A & \{\} \\
\end{array}&&
\begin{array}{lll}
\mbox{SG}&\mbox{Irrep}&X^{\text{BdG}}\\
\hline 
18 & B_3 & \{\} \\
18 & B_2 & \{\} \\
18 & B_1 & \{\} \\
18 & A & \{\} \\
19 & B_3 & \{\} \\
19 & B_2 & \{\} \\
19 & B_1 & \{\} \\
19 & A & \{\} \\
20 & B_3 & \{\} \\
20 & B_2 & \{\} \\
20 & B_1 & \{\} \\
20 & A & \{\} \\
21 & B_3 & \{\} \\
21 & B_2 & \{\} \\
21 & B_1 & \{\} \\
21 & A & \{\} \\
22 & B_3 & \{\} \\
22 & B_2 & \{\} \\
22 & B_1 & \{\} \\
22 & A & \{\} \\
23 & B_3 & \{\} \\
23 & B_2 & \{\} \\
23 & B_1 & \{\} \\
23 & A & \{\} \\
24 & B_3 & \{\} \\
24 & B_2 & \{\} \\
24 & B_1 & \{\} \\
24 & A & \{\} \\
25 & B_1 & \{\} \\
25 & B_2 & \{\} \\
25 & A_2 & \{\} \\
25 & A_1 & \{\} \\
26 & B_1 & \{\} \\
26 & B_2 & \{\} \\
26 & A_2 & \{\} \\
26 & A_1 & \{\} \\
27 & B_1 & \{\} \\
27 & B_2 & \{\} \\
27 & A_2 & \{\} \\
27 & A_1 & \{\} \\
28 & B_1 & \{\} \\
28 & B_2 & \{\} \\
28 & A_2 & \{\} \\
28 & A_1 & \{\} \\
29 & B_1 & \{\} \\
29 & B_2 & \{\} \\
29 & A_2 & \{\} \\
29 & A_1 & \{\} \\
30 & B_1 & \{\} \\
\end{array}&&
\begin{array}{lll}
\mbox{SG}&\mbox{Irrep}&X^{\text{BdG}}\\
\hline 
30 & B_2 & \{\} \\
30 & A_2 & \{\} \\
30 & A_1 & \{\} \\
31 & B_1 & \{\} \\
31 & B_2 & \{\} \\
31 & A_2 & \{\} \\
31 & A_1 & \{\} \\
32 & B_1 & \{\} \\
32 & B_2 & \{\} \\
32 & A_2 & \{\} \\
32 & A_1 & \{\} \\
33 & B_1 & \{\} \\
33 & B_2 & \{\} \\
33 & A_2 & \{\} \\
33 & A_1 & \{\} \\
34 & B_1 & \{\} \\
34 & B_2 & \{\} \\
34 & A_2 & \{\} \\
34 & A_1 & \{\} \\
35 & B_1 & \{\} \\
35 & B_2 & \{\} \\
35 & A_2 & \{\} \\
35 & A_1 & \{\} \\
36 & B_1 & \{\} \\
36 & B_2 & \{\} \\
36 & A_2 & \{\} \\
36 & A_1 & \{\} \\
37 & B_1 & \{\} \\
37 & B_2 & \{\} \\
37 & A_2 & \{\} \\
37 & A_1 & \{\} \\
38 & B_1 & \{\} \\
38 & B_2 & \{\} \\
38 & A_2 & \{\} \\
38 & A_1 & \{\} \\
39 & B_1 & \{\} \\
39 & B_2 & \{\} \\
39 & A_2 & \{\} \\
39 & A_1 & \{\} \\
40 & B_1 & \{\} \\
40 & B_2 & \{\} \\
40 & A_2 & \{\} \\
40 & A_1 & \{\} \\
41 & B_1 & \{\} \\
41 & B_2 & \{\} \\
41 & A_2 & \{\} \\
41 & A_1 & \{\} \\
42 & B_1 & \{\} \\
42 & B_2 & \{\} \\
\end{array}&&
\begin{array}{lll}
\mbox{SG}&\mbox{Irrep}&X^{\text{BdG}}\\
\hline
42 & A_2 & \{\} \\
42 & A_1 & \{\} \\
43 & B_1 & \{\} \\
43 & B_2 & \{\} \\
43 & A_2 & \{\} \\
43 & A_1 & \{\} \\
44 & B_1 & \{\} \\
44 & B_2 & \{\} \\
44 & A_2 & \{\} \\
44 & A_1 & \{\} \\
45 & B_1 & \{\} \\
45 & B_2 & \{\} \\
45 & A_2 & \{\} \\
45 & A_1 & \{\} \\
46 & B_1 & \{\} \\
46 & B_2 & \{\} \\
46 & A_2 & \{\} \\
46 & A_1 & \{\} \\
47 & B_{3u} & \{2,2,2,4,4,4,8\} \\
47 & B_{3g} & \{\} \\
47 & B_{2u} & \{2,2,2,4,4,4,8\} \\
47 & B_{2g} & \{\} \\
47 & B_{1u} & \{2,2,2,4,4,4,8\} \\
47 & B_{1g} & \{\} \\
47 & A_u & \{2,2,2,4,4,4,8\} \\
47 & A_g & \{\} \\
48 & B_{3u} & \{4,8\} \\
48 & B_{3g} & \{\} \\
48 & B_{2u} & \{4,8\} \\
48 & B_{2g} & \{\} \\
48 & B_{1u} & \{4,8\} \\
48 & B_{1g} & \{\} \\
48 & A_u & \{4,8\} \\
48 & A_g & \{\} \\
49 & B_{3u} & \{2,4,4,8\} \\
49 & B_{3g} & \{\} \\
49 & B_{2u} & \{2,4,4,8\} \\
49 & B_{2g} & \{\} \\
49 & B_{1u} & \{2,4,4,8\} \\
49 & B_{1g} & \{\} \\
49 & A_u & \{2,4,4,8\} \\
49 & A_g & \{\} \\
50 & B_{3u} & \{4,8\} \\
50 & B_{3g} & \{\} \\
50 & B_{2u} & \{4,8\} \\
50 & B_{2g} & \{\} \\
50 & B_{1u} & \{4,8\} \\
50 & B_{1g} & \{\} \\
50 & A_u & \{4,8\} \\
\end{array}
\end{align*}
\end{table}

\begin{table}[H]
\begin{align*}
\begin{array}{lll}
\mbox{SG}&\mbox{Irrep}&X^{\text{BdG}}\\
\hline 
50 & A_g & \{\} \\
51 & B_{3u} & \{2,4,4,8\} \\
51 & B_{3g} & \{\} \\
51 & B_{2u} & \{2,4,4,8\} \\
51 & B_{2g} & \{\} \\
51 & B_{1u} & \{2,4,4,8\} \\
51 & B_{1g} & \{\} \\
51 & A_u & \{2,4,4,8\} \\
51 & A_g & \{\} \\
52 & B_{3u} & \{2,8\} \\
52 & B_{3g} & \{\} \\
52 & B_{2u} & \{2,8\} \\
52 & B_{2g} & \{\} \\
52 & B_{1u} & \{2,8\} \\
52 & B_{1g} & \{\} \\
52 & A_u & \{2,8\} \\
52 & A_g & \{\} \\
53 & B_{3u} & \{2,4,8\} \\
53 & B_{3g} & \{\} \\
53 & B_{2u} & \{2,4,8\} \\
53 & B_{2g} & \{\} \\
53 & B_{1u} & \{2,4,8\} \\
53 & B_{1g} & \{\} \\
53 & A_u & \{2,4,8\} \\
53 & A_g & \{\} \\
54 & B_{3u} & \{4,8\} \\
54 & B_{3g} & \{\} \\
54 & B_{2u} & \{4,8\} \\
54 & B_{2g} & \{\} \\
54 & B_{1u} & \{4,8\} \\
54 & B_{1g} & \{\} \\
54 & A_u & \{4,8\} \\
54 & A_g & \{\} \\
55 & B_{3u} & \{2,4,8\} \\
55 & B_{3g} & \{\} \\
55 & B_{2u} & \{2,4,8\} \\
55 & B_{2g} & \{\} \\
55 & B_{1u} & \{2,4,8\} \\
55 & B_{1g} & \{\} \\
55 & A_u & \{2,4,8\} \\
55 & A_g & \{\} \\
56 & B_{3u} & \{2,8\} \\
56 & B_{3g} & \{\} \\
56 & B_{2u} & \{2,8\} \\
56 & B_{2g} & \{\} \\
56 & B_{1u} & \{2,8\} \\
56 & B_{1g} & \{\} \\
56 & A_u & \{2,8\} \\
56 & A_g & \{\} \\
\end{array}&&
\begin{array}{lll}
\mbox{SG}&\mbox{Irrep}&X^{\text{BdG}}\\
\hline 
57 & B_{3u} & \{4,8\} \\
57 & B_{3g} & \{\} \\
57 & B_{2u} & \{4,8\} \\
57 & B_{2g} & \{\} \\
57 & B_{1u} & \{4,8\} \\
57 & B_{1g} & \{\} \\
57 & A_u & \{4,8\} \\
57 & A_g & \{\} \\
58 & B_{3u} & \{2,2,8\} \\
58 & B_{3g} & \{\} \\
58 & B_{2u} & \{2,2,8\} \\
58 & B_{2g} & \{\} \\
58 & B_{1u} & \{2,2,8\} \\
58 & B_{1g} & \{\} \\
58 & A_u & \{2,2,8\} \\
58 & A_g & \{\} \\
59 & B_{3u} & \{4,8\} \\
59 & B_{3g} & \{\} \\
59 & B_{2u} & \{4,8\} \\
59 & B_{2g} & \{\} \\
59 & B_{1u} & \{4,8\} \\
59 & B_{1g} & \{\} \\
59 & A_u & \{4,8\} \\
59 & A_g & \{\} \\
60 & B_{3u} & \{2,8\} \\
60 & B_{3g} & \{\} \\
60 & B_{2u} & \{2,8\} \\
60 & B_{2g} & \{\} \\
60 & B_{1u} & \{2,8\} \\
60 & B_{1g} & \{\} \\
60 & A_u & \{2,8\} \\
60 & A_g & \{\} \\
61 & B_{3u} & \{8\} \\
61 & B_{3g} & \{\} \\
61 & B_{2u} & \{8\} \\
61 & B_{2g} & \{\} \\
61 & B_{1u} & \{8\} \\
61 & B_{1g} & \{\} \\
61 & A_u & \{8\} \\
61 & A_g & \{\} \\
62 & B_{3u} & \{2,8\} \\
62 & B_{3g} & \{\} \\
62 & B_{2u} & \{2,8\} \\
62 & B_{2g} & \{\} \\
62 & B_{1u} & \{2,8\} \\
62 & B_{1g} & \{\} \\
62 & A_u & \{2,8\} \\
62 & A_g & \{\} \\
63 & B_{3u} & \{2,4,8\} \\
\end{array}&&
\begin{array}{lll}
\mbox{SG}&\mbox{Irrep}&X^{\text{BdG}}\\
\hline 
63 & B_{3g} & \{\} \\
63 & B_{2u} & \{2,4,8\} \\
63 & B_{2g} & \{\} \\
63 & B_{1u} & \{2,4,8\} \\
63 & B_{1g} & \{\} \\
63 & A_u & \{2,4,8\} \\
63 & A_g & \{\} \\
64 & B_{3u} & \{4,8\} \\
64 & B_{3g} & \{\} \\
64 & B_{2u} & \{4,8\} \\
64 & B_{2g} & \{\} \\
64 & B_{1u} & \{4,8\} \\
64 & B_{1g} & \{\} \\
64 & A_u & \{4,8\} \\
64 & A_g & \{\} \\
65 & B_{3u} & \{2,2,4,4,8\} \\
65 & B_{3g} & \{\} \\
65 & B_{2u} & \{2,2,4,4,8\} \\
65 & B_{2g} & \{\} \\
65 & B_{1u} & \{2,2,4,4,8\} \\
65 & B_{1g} & \{\} \\
65 & A_u & \{2,2,4,4,8\} \\
65 & A_g & \{\} \\
66 & B_{3u} & \{2,2,4,8\} \\
66 & B_{3g} & \{\} \\
66 & B_{2u} & \{2,2,4,8\} \\
66 & B_{2g} & \{\} \\
66 & B_{1u} & \{2,2,4,8\} \\
66 & B_{1g} & \{\} \\
66 & A_u & \{2,2,4,8\} \\
66 & A_g & \{\} \\
67 & B_{3u} & \{2,4,4,8\} \\
67 & B_{3g} & \{\} \\
67 & B_{2u} & \{2,4,4,8\} \\
67 & B_{2g} & \{\} \\
67 & B_{1u} & \{2,4,4,8\} \\
67 & B_{1g} & \{\} \\
67 & A_u & \{2,4,4,8\} \\
67 & A_g & \{\} \\
68 & B_{3u} & \{4,8\} \\
68 & B_{3g} & \{\} \\
68 & B_{2u} & \{4,8\} \\
68 & B_{2g} & \{\} \\
68 & B_{1u} & \{4,8\} \\
68 & B_{1g} & \{\} \\
68 & A_u & \{4,8\} \\
68 & A_g & \{\} \\
69 & B_{3u} & \{2,4,4,8\} \\
69 & B_{3g} & \{\} \\
\end{array}&&
\begin{array}{lll}
\mbox{SG}&\mbox{Irrep}&X^{\text{BdG}}\\
\hline 
69 & B_{2u} & \{2,4,4,8\} \\
69 & B_{2g} & \{\} \\
69 & B_{1u} & \{2,4,4,8\} \\
69 & B_{1g} & \{\} \\
69 & A_u & \{2,4,4,8\} \\
69 & A_g & \{\} \\
70 & B_{3u} & \{2,8\} \\
70 & B_{3g} & \{\} \\
70 & B_{2u} & \{2,8\} \\
70 & B_{2g} & \{\} \\
70 & B_{1u} & \{2,8\} \\
70 & B_{1g} & \{\} \\
70 & A_u & \{2,8\} \\
70 & A_g & \{\} \\
71 & B_{3u} & \{2,2,4,8\} \\
71 & B_{3g} & \{\} \\
71 & B_{2u} & \{2,2,4,8\} \\
71 & B_{2g} & \{\} \\
71 & B_{1u} & \{2,2,4,8\} \\
71 & B_{1g} & \{\} \\
71 & A_u & \{2,2,4,8\} \\
71 & A_g & \{\} \\
72 & B_{3u} & \{2,4,8\} \\
72 & B_{3g} & \{\} \\
72 & B_{2u} & \{2,4,8\} \\
72 & B_{2g} & \{\} \\
72 & B_{1u} & \{2,4,8\} \\
72 & B_{1g} & \{\} \\
72 & A_u & \{2,4,8\} \\
72 & A_g & \{\} \\
73 & B_{3u} & \{4,8\} \\
73 & B_{3g} & \{\} \\
73 & B_{2u} & \{4,8\} \\
73 & B_{2g} & \{\} \\
73 & B_{1u} & \{4,8\} \\
73 & B_{1g} & \{\} \\
73 & A_u & \{4,8\} \\
73 & A_g & \{\} \\
74 & B_{3u} & \{2,2,4,8\} \\
74 & B_{3g} & \{\} \\
74 & B_{2u} & \{2,2,4,8\} \\
74 & B_{2g} & \{\} \\
74 & B_{1u} & \{2,2,4,8\} \\
74 & B_{1g} & \{\} \\
74 & A_u & \{2,2,4,8\} \\
74 & A_g & \{\} \\
75 & B & \{2\} \\
75 & A & \{\} \\
76 & B & \{\} \\
\end{array}
\end{align*}
\end{table}

\begin{table}
\begin{align*}
\begin{array}{lll}
\mbox{SG}&\mbox{Irrep}&X^{\text{BdG}}\\
\hline 
76 & A & \{\} \\
77 & B & \{\} \\
77 & A & \{\} \\
78 & B & \{\} \\
78 & A & \{\} \\
79 & B & \{\} \\
79 & A & \{\} \\
80 & B & \{\} \\
80 & A & \{\} \\
81 & B & \{2,2,4\} \\
81 & A & \{\} \\
82 & B & \{2,2,4\} \\
82 & A & \{\} \\
83 & B_u & \{2,4,4,8\} \\
83 & B_g & \{2,2,2\} \\
83 & A_u & \{2,2,2,2,4,8,16\} \\
83 & A_g & \{\} \\
84 & B_u & \{2,4,4\} \\
84 & B_g & \{2\} \\
84 & A_u & \{2,2,2,4,8\} \\
84 & A_g & \{\} \\
85 & B_u & \{4,4\} \\
85 & B_g & \{2,2\} \\
85 & A_u & \{2,4,8\} \\
85 & A_g & \{\} \\
86 & B_u & \{2,4\} \\
86 & B_g & \{2\} \\
86 & A_u & \{2,4,8\} \\
86 & A_g & \{\} \\
87 & B_u & \{2,4,4\} \\
87 & B_g & \{2,2\} \\
87 & A_u & \{2,2,2,4,16\} \\
87 & A_g & \{\} \\
88 & B_u & \{2,4\} \\
88 & B_g & \{2\} \\
88 & A_u & \{2,2,8\} \\
88 & A_g & \{\} \\
89 & B_2 & \{2\} \\
89 & B_1 & \{2\} \\
89 & A_2 & \{\} \\
89 & A_1 & \{\} \\
90 & B_2 & \{2\} \\
90 & B_1 & \{2\} \\
90 & A_2 & \{\} \\
90 & A_1 & \{\} \\
91 & B_2 & \{\} \\
91 & B_1 & \{\} \\
91 & A_2 & \{\} \\
91 & A_1 & \{\} \\
\end{array}&&
\begin{array}{lll}
\mbox{SG}&\mbox{Irrep}&X^{\text{BdG}}\\
\hline 
92 & B_2 & \{\} \\
92 & B_1 & \{\} \\
92 & A_2 & \{\} \\
92 & A_1 & \{\} \\
93 & B_2 & \{\} \\
93 & B_1 & \{\} \\
93 & A_2 & \{\} \\
93 & A_1 & \{\} \\
94 & B_2 & \{\} \\
94 & B_1 & \{\} \\
94 & A_2 & \{\} \\
94 & A_1 & \{\} \\
95 & B_2 & \{\} \\
95 & B_1 & \{\} \\
95 & A_2 & \{\} \\
95 & A_1 & \{\} \\
96 & B_2 & \{\} \\
96 & B_1 & \{\} \\
96 & A_2 & \{\} \\
96 & A_1 & \{\} \\
97 & B_2 & \{\} \\
97 & B_1 & \{\} \\
97 & A_2 & \{\} \\
97 & A_1 & \{\} \\
98 & B_2 & \{\} \\
98 & B_1 & \{\} \\
98 & A_2 & \{\} \\
98 & A_1 & \{\} \\
99 & B_2 & \{2\} \\
99 & B_1 & \{2\} \\
99 & A_2 & \{\} \\
99 & A_1 & \{\} \\
100 & B_2 & \{2\} \\
100 & B_1 & \{2\} \\
100 & A_2 & \{\} \\
100 & A_1 & \{\} \\
101 & B_2 & \{\} \\
101 & B_1 & \{\} \\
101 & A_2 & \{\} \\
101 & A_1 & \{\} \\
102 & B_2 & \{\} \\
102 & B_1 & \{\} \\
102 & A_2 & \{\} \\
102 & A_1 & \{\} \\
103 & B_2 & \{2\} \\
103 & B_1 & \{2\} \\
103 & A_2 & \{\} \\
103 & A_1 & \{\} \\
104 & B_2 & \{\} \\
\end{array}&&
\begin{array}{lll}
\mbox{SG}&\mbox{Irrep}&X^{\text{BdG}}\\
\hline 
104 & B_1 & \{\} \\
104 & A_2 & \{\} \\
104 & A_1 & \{\} \\
105 & B_2 & \{\} \\
105 & B_1 & \{\} \\
105 & A_2 & \{\} \\
105 & A_1 & \{\} \\
106 & B_2 & \{\} \\
106 & B_1 & \{\} \\
106 & A_2 & \{\} \\
106 & A_1 & \{\} \\
107 & B_2 & \{\} \\
107 & B_1 & \{\} \\
107 & A_2 & \{\} \\
107 & A_1 & \{\} \\
108 & B_2 & \{2\} \\
108 & B_1 & \{2\} \\
108 & A_2 & \{\} \\
108 & A_1 & \{\} \\
109 & B_2 & \{\} \\
109 & B_1 & \{\} \\
109 & A_2 & \{\} \\
109 & A_1 & \{\} \\
110 & B_2 & \{\} \\
110 & B_1 & \{\} \\
110 & A_2 & \{\} \\
110 & A_1 & \{\} \\
111 & B_2 & \{2,2,4\} \\
111 & A_2 & \{\} \\
111 & B_1 & \{2,2,4\} \\
111 & A_1 & \{\} \\
112 & B_2 & \{2,4\} \\
112 & A_2 & \{\} \\
112 & B_1 & \{2,4\} \\
112 & A_1 & \{\} \\
113 & B_2 & \{2,4\} \\
113 & A_2 & \{\} \\
113 & B_1 & \{2,4\} \\
113 & A_1 & \{\} \\
114 & B_2 & \{4\} \\
114 & A_2 & \{\} \\
114 & B_1 & \{4\} \\
114 & A_1 & \{\} \\
115 & B_2 & \{2,2,4\} \\
115 & A_2 & \{\} \\
115 & B_1 & \{2,2,4\} \\
115 & A_1 & \{\} \\
116 & B_2 & \{2,4\} \\
116 & A_2 & \{\} \\
\end{array}&&
\begin{array}{lll}
\mbox{SG}&\mbox{Irrep}&X^{\text{BdG}}\\
\hline
116 & B_1 & \{2,4\} \\
116 & A_1 & \{\} \\
117 & B_2 & \{2,4\} \\
117 & A_2 & \{\} \\
117 & B_1 & \{2,4\} \\
117 & A_1 & \{\} \\
118 & B_2 & \{2,4\} \\
118 & A_2 & \{\} \\
118 & B_1 & \{2,4\} \\
118 & A_1 & \{\} \\
119 & B_2 & \{2,2,4\} \\
119 & A_2 & \{\} \\
119 & B_1 & \{2,2,4\} \\
119 & A_1 & \{\} \\
120 & B_2 & \{2,4\} \\
120 & A_2 & \{\} \\
120 & B_1 & \{2,4\} \\
120 & A_1 & \{\} \\
121 & B_2 & \{2,4\} \\
121 & A_2 & \{\} \\
121 & B_1 & \{2,4\} \\
121 & A_1 & \{\} \\
122 & B_2 & \{4\} \\
122 & A_2 & \{\} \\
122 & B_1 & \{4\} \\
122 & A_1 & \{\} \\
123 & B_{2u} & \{2,4,4,8\} \\
123 & B_{2g} & \{2,2,2\} \\
123 & B_{1u} & \{2,4,4,8\} \\
123 & B_{1g} & \{2,2,2\} \\
123 & A_{2u} & \{2,2,2,2,4,8,16\} \\
123 & A_{2g} & \{\} \\
123 & A_{1u} & \{2,2,2,2,4,8,16\} \\
123 & A_{1g} & \{\} \\
124 & B_{2u} & \{2,4,4\} \\
124 & B_{2g} & \{2,2\} \\
124 & B_{1u} & \{2,4,4\} \\
124 & B_{1g} & \{2,2\} \\
124 & A_{2u} & \{2,2,2,4,16\} \\
124 & A_{2g} & \{\} \\
124 & A_{1u} & \{2,2,2,4,16\} \\
124 & A_{1g} & \{\} \\
125 & B_{2u} & \{4,4\} \\
125 & B_{2g} & \{2,2\} \\
125 & B_{1u} & \{4,4\} \\
125 & B_{1g} & \{2,2\} \\
125 & A_{2u} & \{2,4,8\} \\
125 & A_{2g} & \{\} \\
125 & A_{1u} & \{2,4,8\} \\
\end{array}
\end{align*}
\end{table}

\begin{table}
\begin{align*}
\begin{array}{lll}
\mbox{SG}&\mbox{Irrep}&X^{\text{BdG}}\\
\hline
125 & A_{1g} & \{\} \\
126 & B_{2u} & \{4\} \\
126 & B_{2g} & \{2\} \\
126 & B_{1u} & \{4\} \\
126 & B_{1g} & \{2\} \\
126 & A_{2u} & \{2,8\} \\
126 & A_{2g} & \{\} \\
126 & A_{1u} & \{2,8\} \\
126 & A_{1g} & \{\} \\
127 & B_{2u} & \{2,4,8\} \\
127 & B_{2g} & \{2,2,2\} \\
127 & B_{1u} & \{2,4,8\} \\
127 & B_{1g} & \{2,2,2\} \\
127 & A_{2u} & \{2,2,8,16\} \\
127 & A_{2g} & \{\} \\
127 & A_{1u} & \{2,2,8,16\} \\
127 & A_{1g} & \{\} \\
128 & B_{2u} & \{2,2,4\} \\
128 & B_{2g} & \{2,2\} \\
128 & B_{1u} & \{2,2,4\} \\
128 & B_{1g} & \{2,2\} \\
128 & A_{2u} & \{2,2,2,16\} \\
128 & A_{2g} & \{\} \\
128 & A_{1u} & \{2,2,2,16\} \\
128 & A_{1g} & \{\} \\
129 & B_{2u} & \{4,4\} \\
129 & B_{2g} & \{2,2\} \\
129 & B_{1u} & \{4,4\} \\
129 & B_{1g} & \{2,2\} \\
129 & A_{2u} & \{2,4,8\} \\
129 & A_{2g} & \{\} \\
129 & A_{1u} & \{2,4,8\} \\
129 & A_{1g} & \{\} \\
130 & B_{2u} & \{2,4\} \\
130 & B_{2g} & \{2,2\} \\
130 & B_{1u} & \{2,4\} \\
130 & B_{1g} & \{2,2\} \\
130 & A_{2u} & \{2,8\} \\
130 & A_{2g} & \{\} \\
130 & A_{1u} & \{2,8\} \\
130 & A_{1g} & \{\} \\
131 & B_{2u} & \{2,4,4\} \\
131 & B_{2g} & \{2\} \\
131 & B_{1u} & \{2,4,4\} \\
131 & B_{1g} & \{2\} \\
131 & A_{2u} & \{2,2,2,4,8\} \\
131 & A_{2g} & \{\} \\
131 & A_{1u} & \{2,2,2,4,8\} \\
131 & A_{1g} & \{\} \\
\end{array}&&
\begin{array}{lll}
\mbox{SG}&\mbox{Irrep}&X^{\text{BdG}}\\
\hline
132 & B_{2u} & \{4,4\} \\
132 & B_{2g} & \{2\} \\
132 & B_{1u} & \{4,4\} \\
132 & B_{1g} & \{2\} \\
132 & A_{2u} & \{2,2,4,8\} \\
132 & A_{2g} & \{\} \\
132 & A_{1u} & \{2,2,4,8\} \\
132 & A_{1g} & \{\} \\
133 & B_{2u} & \{4\} \\
133 & B_{2g} & \{2\} \\
133 & B_{1u} & \{4\} \\
133 & B_{1g} & \{2\} \\
133 & A_{2u} & \{2,8\} \\
133 & A_{2g} & \{\} \\
133 & A_{1u} & \{2,8\} \\
133 & A_{1g} & \{\} \\
134 & B_{2u} & \{2,4\} \\
134 & B_{2g} & \{2\} \\
134 & B_{1u} & \{2,4\} \\
134 & B_{1g} & \{2\} \\
134 & A_{2u} & \{2,4,8\} \\
134 & A_{2g} & \{\} \\
134 & A_{1u} & \{2,4,8\} \\
134 & A_{1g} & \{\} \\
135 & B_{2u} & \{2,4\} \\
135 & B_{2g} & \{2\} \\
135 & B_{1u} & \{2,4\} \\
135 & B_{1g} & \{2\} \\
135 & A_{2u} & \{2,2,8\} \\
135 & A_{2g} & \{\} \\
135 & A_{1u} & \{2,2,8\} \\
135 & A_{1g} & \{\} \\
136 & B_{2u} & \{2,4\} \\
136 & B_{2g} & \{2\} \\
136 & B_{1u} & \{2,4\} \\
136 & B_{1g} & \{2\} \\
136 & A_{2u} & \{2,2,8\} \\
136 & A_{2g} & \{\} \\
136 & A_{1u} & \{2,2,8\} \\
136 & A_{1g} & \{\} \\
137 & B_{2u} & \{4\} \\
137 & B_{2g} & \{2\} \\
137 & B_{1u} & \{4\} \\
137 & B_{1g} & \{2\} \\
137 & A_{2u} & \{2,8\} \\
137 & A_{2g} & \{\} \\
137 & A_{1u} & \{2,8\} \\
137 & A_{1g} & \{\} \\
138 & B_{2u} & \{2,4\} \\
\end{array}&&
\begin{array}{lll}
\mbox{SG}&\mbox{Irrep}&X^{\text{BdG}}\\
\hline
138 & B_{2g} & \{2\} \\
138 & B_{1u} & \{2,4\} \\
138 & B_{1g} & \{2\} \\
138 & A_{2u} & \{2,2,8\} \\
138 & A_{2g} & \{\} \\
138 & A_{1u} & \{2,2,8\} \\
138 & A_{1g} & \{\} \\
139 & B_{2u} & \{2,4,4\} \\
139 & B_{2g} & \{2,2\} \\
139 & B_{1u} & \{2,4,4\} \\
139 & B_{1g} & \{2,2\} \\
139 & A_{2u} & \{2,2,2,4,16\} \\
139 & A_{2g} & \{\} \\
139 & A_{1u} & \{2,2,2,4,16\} \\
139 & A_{1g} & \{\} \\
140 & B_{2u} & \{2,4,4\} \\
140 & B_{2g} & \{2,2\} \\
140 & B_{1u} & \{2,4,4\} \\
140 & B_{1g} & \{2,2\} \\
140 & A_{2u} & \{2,2,4,16\} \\
140 & A_{2g} & \{\} \\
140 & A_{1u} & \{2,2,4,16\} \\
140 & A_{1g} & \{\} \\
141 & B_{2u} & \{2,4\} \\
141 & B_{2g} & \{2\} \\
141 & B_{1u} & \{2,4\} \\
141 & B_{1g} & \{2\} \\
141 & A_{2u} & \{2,2,8\} \\
141 & A_{2g} & \{\} \\
141 & A_{1u} & \{2,2,8\} \\
141 & A_{1g} & \{\} \\
142 & B_{2u} & \{4\} \\
142 & B_{2g} & \{2\} \\
142 & B_{1u} & \{4\} \\
142 & B_{1g} & \{2\} \\
142 & A_{2u} & \{2,8\} \\
142 & A_{2g} & \{\} \\
142 & A_{1u} & \{2,8\} \\
142 & A_{1g} & \{\} \\
143 & A & \{\} \\
144 & A & \{\} \\
145 & A & \{\} \\
146 & A & \{\} \\
147 & A_u & \{2,2,4,8\} \\
147 & A_g & \{\} \\
148 & A_u & \{2,2,4,8\} \\
148 & A_g & \{\} \\
149 & A_2 & \{\} \\
149 & A_1 & \{\} \\
\end{array}&&
\begin{array}{lll}
\mbox{SG}&\mbox{Irrep}&X^{\text{BdG}}\\
\hline 
150 & A_2 & \{\} \\
150 & A_1 & \{\} \\
151 & A_2 & \{\} \\
151 & A_1 & \{\} \\
152 & A_2 & \{\} \\
152 & A_1 & \{\} \\
153 & A_2 & \{\} \\
153 & A_1 & \{\} \\
154 & A_2 & \{\} \\
154 & A_1 & \{\} \\
155 & A_2 & \{\} \\
155 & A_1 & \{\} \\
156 & A_2 & \{\} \\
156 & A_1 & \{\} \\
157 & A_2 & \{\} \\
157 & A_1 & \{\} \\
158 & A_2 & \{\} \\
158 & A_1 & \{\} \\
159 & A_2 & \{\} \\
159 & A_1 & \{\} \\
160 & A_2 & \{\} \\
160 & A_1 & \{\} \\
161 & A_2 & \{\} \\
161 & A_1 & \{\} \\
162 & A_{2u} & \{2,2,4,8\} \\
162 & A_{2g} & \{\} \\
162 & A_{1u} & \{2,2,4,8\} \\
162 & A_{1g} & \{\} \\
163 & A_{2u} & \{2,2,8\} \\
163 & A_{2g} & \{\} \\
163 & A_{1u} & \{2,2,8\} \\
163 & A_{1g} & \{\} \\
164 & A_{2u} & \{2,2,4,8\} \\
164 & A_{2g} & \{\} \\
164 & A_{1u} & \{2,2,4,8\} \\
164 & A_{1g} & \{\} \\
165 & A_{2u} & \{2,2,8\} \\
165 & A_{2g} & \{\} \\
165 & A_{1u} & \{2,2,8\} \\
165 & A_{1g} & \{\} \\
166 & A_{2u} & \{2,2,4,8\} \\
166 & A_{2g} & \{\} \\
166 & A_{1u} & \{2,2,4,8\} \\
166 & A_{1g} & \{\} \\
167 & A_{2u} & \{2,2,8\} \\
167 & A_{2g} & \{\} \\
167 & A_{1u} & \{2,2,8\} \\
167 & A_{1g} & \{\} \\
168 & B & \{\} \\
\end{array}
\end{align*}
\end{table}

\begin{table}
\begin{align*}
\begin{array}{lll}
\mbox{SG}&\mbox{Irrep}&X^{\text{BdG}}\\
\hline
168 & A & \{\} \\
169 & B & \{\} \\
169 & A & \{\} \\
170 & B & \{\} \\
170 & A & \{\} \\
171 & B & \{\} \\
171 & A & \{\} \\
172 & B & \{\} \\
172 & A & \{\} \\
173 & B & \{\} \\
173 & A & \{\} \\
174 & A'' & \{2,6,6\} \\
174 & A' & \{\} \\
175 & B_u & \{2,4,8\} \\
175 & B_g & \{3,6\} \\
175 & A_u & \{2,2,2,2,12,24\} \\
175 & A_g & \{\} \\
176 & B_u & \{2,8\} \\
176 & B_g & \{6\} \\
176 & A_u & \{2,2,2,24\} \\
176 & A_g & \{\} \\
177 & B_1 & \{\} \\
177 & B_2 & \{\} \\
177 & A_2 & \{\} \\
177 & A_1 & \{\} \\
178 & B_1 & \{\} \\
178 & B_2 & \{\} \\
178 & A_2 & \{\} \\
178 & A_1 & \{\} \\
179 & B_1 & \{\} \\
179 & B_2 & \{\} \\
179 & A_2 & \{\} \\
179 & A_1 & \{\} \\
180 & B_1 & \{\} \\
180 & B_2 & \{\} \\
180 & A_2 & \{\} \\
180 & A_1 & \{\} \\
181 & B_1 & \{\} \\
181 & B_2 & \{\} \\
181 & A_2 & \{\} \\
181 & A_1 & \{\} \\
182 & B_1 & \{\} \\
182 & B_2 & \{\} \\
182 & A_2 & \{\} \\
182 & A_1 & \{\} \\
183 & B_2 & \{\} \\
183 & B_1 & \{\} \\
183 & A_2 & \{\} \\
183 & A_1 & \{\} \\
\end{array}&&
\begin{array}{lll}
\mbox{SG}&\mbox{Irrep}&X^{\text{BdG}}\\
\hline
184 & B_2 & \{\} \\
184 & B_1 & \{\} \\
184 & A_2 & \{\} \\
184 & A_1 & \{\} \\
185 & B_2 & \{\} \\
185 & B_1 & \{\} \\
185 & A_2 & \{\} \\
185 & A_1 & \{\} \\
186 & B_2 & \{\} \\
186 & B_1 & \{\} \\
186 & A_2 & \{\} \\
186 & A_1 & \{\} \\
187 & A''_2 & \{2,6,6\} \\
187 & A'_2 & \{\} \\
187 & A''_1 & \{2,6,6\} \\
187 & A'_1 & \{\} \\
188 & A''_2 & \{2,2,6\} \\
188 & A'_2 & \{\} \\
188 & A''_1 & \{2,2,6\} \\
188 & A'_1 & \{\} \\
189 & A''_2 & \{6,6\} \\
189 & A'_2 & \{\} \\
189 & A''_1 & \{6,6\} \\
189 & A'_1 & \{\} \\
190 & A''_2 & \{2,6\} \\
190 & A'_2 & \{\} \\
190 & A''_1 & \{2,6\} \\
190 & A'_1 & \{\} \\
191 & B_{1u} & \{2,4,8\} \\
191 & B_{1g} & \{3,6\} \\
191 & B_{2u} & \{2,4,8\} \\
191 & B_{2g} & \{3,6\} \\
191 & A_{2u} & \{2,2,2,2,12,24\} \\
191 & A_{2g} & \{\} \\
191 & A_{1u} & \{2,2,2,2,12,24\} \\
191 & A_{1g} & \{\} \\
192 & B_{1u} & \{2,8\} \\
192 & B_{1g} & \{6\} \\
192 & B_{2u} & \{2,8\} \\
192 & B_{2g} & \{6\} \\
192 & A_{2u} & \{2,2,2,2,24\} \\
192 & A_{2g} & \{\} \\
192 & A_{1u} & \{2,2,2,2,24\} \\
192 & A_{1g} & \{\} \\
193 & B_{1u} & \{2,8\} \\
193 & B_{1g} & \{6\} \\
193 & B_{2u} & \{2,8\} \\
193 & B_{2g} & \{6\} \\
193 & A_{2u} & \{2,2,2,24\} \\
\end{array}&&
\begin{array}{lll}
\mbox{SG}&\mbox{Irrep}&X^{\text{BdG}}\\
\hline 
193 & A_{2g} & \{\} \\
193 & A_{1u} & \{2,2,2,24\} \\
193 & A_{1g} & \{\} \\
194 & B_{1u} & \{2,8\} \\
194 & B_{1g} & \{6\} \\
194 & B_{2u} & \{2,8\} \\
194 & B_{2g} & \{6\} \\
194 & A_{2u} & \{2,2,2,24\} \\
194 & A_{2g} & \{\} \\
194 & A_{1u} & \{2,2,2,24\} \\
194 & A_{1g} & \{\} \\
195 & A & \{\} \\
196 & A & \{\} \\
197 & A & \{\} \\
198 & A & \{\} \\
199 & A & \{\} \\
200 & A_u & \{2,2,4,8\} \\
200 & A_g & \{\} \\
201 & A_u & \{2,4,8\} \\
201 & A_g & \{\} \\
202 & A_u & \{2,2,8\} \\
202 & A_g & \{\} \\
203 & A_u & \{2,2,8\} \\
203 & A_g & \{\} \\
204 & A_u & \{2,4,8\} \\
204 & A_g & \{\} \\
205 & A_u & \{2,8\} \\
205 & A_g & \{\} \\
206 & A_u & \{2,4,8\} \\
206 & A_g & \{\} \\
207 & A_2 & \{2\} \\
207 & A_1 & \{\} \\
208 & A_2 & \{\} \\
208 & A_1 & \{\} \\
209 & A_2 & \{\} \\
209 & A_1 & \{\} \\
210 & A_2 & \{\} \\
210 & A_1 & \{\} \\
211 & A_2 & \{\} \\
211 & A_1 & \{\} \\
212 & A_2 & \{\} \\
212 & A_1 & \{\} \\
213 & A_2 & \{\} \\
213 & A_1 & \{\} \\
214 & A_2 & \{\} \\
214 & A_1 & \{\} \\
215 & A_2 & \{2,2,4\} \\
215 & A_1 & \{\} \\
216 & A_2 & \{2,2,4\} \\
\end{array}&&
\begin{array}{lll}
\mbox{SG}&\mbox{Irrep}&X^{\text{BdG}}\\
\hline
216 & A_1 & \{\} \\
217 & A_2 & \{2,4\} \\
217 & A_1 & \{\} \\
218 & A_2 & \{2,4\} \\
218 & A_1 & \{\} \\
219 & A_2 & \{2,4\} \\
219 & A_1 & \{\} \\
220 & A_2 & \{4\} \\
220 & A_1 & \{\} \\
221 & A_{2u} & \{2,4,8\} \\
221 & A_{2g} & \{2,2,2\} \\
221 & A_{1u} & \{2,2,2,2,8,16\} \\
221 & A_{1g} & \{\} \\
222 & A_{2u} & \{2,4\} \\
222 & A_{2g} & \{2\} \\
222 & A_{1u} & \{2,2,8\} \\
222 & A_{1g} & \{\} \\
223 & A_{2u} & \{2,4\} \\
223 & A_{2g} & \{2\} \\
223 & A_{1u} & \{2,2,2,8\} \\
223 & A_{1g} & \{\} \\
224 & A_{2u} & \{2,2,4\} \\
224 & A_{2g} & \{2\} \\
224 & A_{1u} & \{2,2,4,8\} \\
224 & A_{1g} & \{\} \\
225 & A_{2u} & \{2,2,4\} \\
225 & A_{2g} & \{2,2\} \\
225 & A_{1u} & \{2,2,2,2,16\} \\
225 & A_{1g} & \{\} \\
226 & A_{2u} & \{2,2,4\} \\
226 & A_{2g} & \{2,2\} \\
226 & A_{1u} & \{2,2,2,16\} \\
226 & A_{1g} & \{\} \\
227 & A_{2u} & \{2,2,4\} \\
227 & A_{2g} & \{2\} \\
227 & A_{1u} & \{2,2,2,8\} \\
227 & A_{1g} & \{\} \\
228 & A_{2u} & \{2,4\} \\
228 & A_{2g} & \{2\} \\
228 & A_{1u} & \{2,2,8\} \\
228 & A_{1g} & \{\} \\
229 & A_{2u} & \{2,4,4\} \\
229 & A_{2g} & \{2,2\} \\
229 & A_{1u} & \{2,2,2,4,16\} \\
229 & A_{1g} & \{\} \\
230 & A_{2u} & \{2,4\} \\
230 & A_{2g} & \{2\} \\
230 & A_{1u} & \{2,2,8\} \\
230 & A_{1g} & \{\} \\
\end{array}
\end{align*}
\end{table}

\begin{table}
\caption{The list of $X^{\text{BdG}}$ for spinless electrons with time-reversal symmetry.}
\label{tab:TRS_spinless}
\begin{align*}
\begin{array}{ccc}
\mbox{SG}&\mbox{Irrep}&X^{\text{BdG}}\\
\hline 
1 & A & \{\} \\
2 & A_u & \{2,2,2,4,4,4,8\} \\
2 & A_g & \{\} \\
3 & B & \{2,2,4\} \\
3 & A & \{\} \\
4 & B & \{\} \\
4 & A & \{\} \\
5 & B & \{2\} \\
5 & A & \{\} \\
6 & A'' & \{2\} \\
6 & A' & \{\} \\
7 & A'' & \{\} \\
7 & A' & \{\} \\
8 & A'' & \{\} \\
8 & A' & \{\} \\
9 & A'' & \{\} \\
9 & A' & \{\} \\
10 & B_u & \{2,2,2,2,2,2,4,4,4\} \\
10 & B_g & \{2\} \\
10 & A_u & \{2,2,2,2,2,4\} \\
10 & A_g & \{\} \\
11 & B_u & \{2,2,4\} \\
11 & B_g & \{\} \\
11 & A_u & \{2,2,2,4\} \\
11 & A_g & \{\} \\
12 & B_u & \{2,2,2,2,4,4\} \\
12 & B_g & \{\} \\
12 & A_u & \{2,2,4,4\} \\
12 & A_g & \{\} \\
13 & B_u & \{2,2,2,4,4\} \\
13 & B_g & \{2\} \\
13 & A_u & \{2,2,4\} \\
13 & A_g & \{\} \\
14 & B_u & \{2,2,4\} \\
14 & B_g & \{\} \\
14 & A_u & \{2,2,4\} \\
14 & A_g & \{\} \\
15 & B_u & \{2,2,4,4\} \\
15 & B_g & \{\} \\
15 & A_u & \{2,4,4\} \\
15 & A_g & \{\} \\
16 & B_3 & \{2,2,2,2\} \\
16 & B_2 & \{2,2,2,2\} \\
16 & B_1 & \{2,2,2,2\} \\
16 & A & \{\} \\
17 & B_3 & \{2,2,2\} \\
17 & B_2 & \{2\} \\
17 & B_1 & \{2\} \\
17 & A & \{\} \\
\end{array}&&
\begin{array}{ccc}
\mbox{SG}&\mbox{Irrep}&X^{\text{BdG}}\\
\hline 
18 & B_3 & \{\} \\
18 & B_2 & \{2\} \\
18 & B_1 & \{2\} \\
18 & A & \{\} \\
19 & B_3 & \{\} \\
19 & B_2 & \{\} \\
19 & B_1 & \{\} \\
19 & A & \{\} \\
20 & B_3 & \{2\} \\
20 & B_2 & \{\} \\
20 & B_1 & \{\} \\
20 & A & \{\} \\
21 & B_3 & \{2,2\} \\
21 & B_2 & \{2,2\} \\
21 & B_1 & \{2,2\} \\
21 & A & \{\} \\
22 & B_3 & \{2\} \\
22 & B_2 & \{2\} \\
22 & B_1 & \{2\} \\
22 & A & \{\} \\
23 & B_3 & \{2\} \\
23 & B_2 & \{2\} \\
23 & B_1 & \{2\} \\
23 & A & \{\} \\
24 & B_3 & \{2\} \\
24 & B_2 & \{2\} \\
24 & B_1 & \{2\} \\
24 & A & \{\} \\
25 & B_1 & \{2,2,2\} \\
25 & B_2 & \{2,2,2\} \\
25 & A_2 & \{\} \\
25 & A_1 & \{\} \\
26 & B_1 & \{2\} \\
26 & B_2 & \{\} \\
26 & A_2 & \{2\} \\
26 & A_1 & \{\} \\
27 & B_1 & \{2,2,4\} \\
27 & B_2 & \{2,2,4\} \\
27 & A_2 & \{\} \\
27 & A_1 & \{\} \\
28 & B_1 & \{2,2\} \\
28 & B_2 & \{2\} \\
28 & A_2 & \{\} \\
28 & A_1 & \{\} \\
29 & B_1 & \{\} \\
29 & B_2 & \{\} \\
29 & A_2 & \{\} \\
29 & A_1 & \{\} \\
30 & B_1 & \{2\} \\
\end{array}&&
\begin{array}{ccc}
\mbox{SG}&\mbox{Irrep}&X^{\text{BdG}}\\
\hline 
30 & B_2 & \{2\} \\
30 & A_2 & \{\} \\
30 & A_1 & \{\} \\
31 & B_1 & \{\} \\
31 & B_2 & \{\} \\
31 & A_2 & \{\} \\
31 & A_1 & \{\} \\
32 & B_1 & \{2\} \\
32 & B_2 & \{2\} \\
32 & A_2 & \{\} \\
32 & A_1 & \{\} \\
33 & B_1 & \{\} \\
33 & B_2 & \{\} \\
33 & A_2 & \{\} \\
33 & A_1 & \{\} \\
34 & B_1 & \{2\} \\
34 & B_2 & \{2\} \\
34 & A_2 & \{\} \\
34 & A_1 & \{\} \\
35 & B_1 & \{2,2\} \\
35 & B_2 & \{2,2\} \\
35 & A_2 & \{\} \\
35 & A_1 & \{\} \\
36 & B_1 & \{\} \\
36 & B_2 & \{\} \\
36 & A_2 & \{\} \\
36 & A_1 & \{\} \\
37 & B_1 & \{2,4\} \\
37 & B_2 & \{2,4\} \\
37 & A_2 & \{\} \\
37 & A_1 & \{\} \\
38 & B_1 & \{2,2\} \\
38 & B_2 & \{\} \\
38 & A_2 & \{\} \\
38 & A_1 & \{\} \\
39 & B_1 & \{2\} \\
39 & B_2 & \{2,2\} \\
39 & A_2 & \{\} \\
39 & A_1 & \{\} \\
40 & B_1 & \{2\} \\
40 & B_2 & \{\} \\
40 & A_2 & \{\} \\
40 & A_1 & \{\} \\
41 & B_1 & \{\} \\
41 & B_2 & \{\} \\
41 & A_2 & \{\} \\
41 & A_1 & \{\} \\
42 & B_1 & \{2\} \\
42 & B_2 & \{2\} \\
\end{array}&&
\begin{array}{ccc}
\mbox{SG}&\mbox{Irrep}&X^{\text{BdG}}\\
\hline 
42 & A_2 & \{\} \\
42 & A_1 & \{\} \\
43 & B_1 & \{\} \\
43 & B_2 & \{\} \\
43 & A_2 & \{\} \\
43 & A_1 & \{\} \\
44 & B_1 & \{2\} \\
44 & B_2 & \{2\} \\
44 & A_2 & \{\} \\
44 & A_1 & \{\} \\
45 & B_1 & \{2\} \\
45 & B_2 & \{2\} \\
45 & A_2 & \{\} \\
45 & A_1 & \{\} \\
46 & B_1 & \{2\} \\
46 & B_2 & \{\} \\
46 & A_2 & \{\} \\
46 & A_1 & \{\} \\
47 & B_{3u} & \{2,2,2,2,2,2,2,2,2\} \\
47 & B_{3g} & \{\} \\
47 & B_{2u} & \{2,2,2,2,2,2,2,2,2\} \\
47 & B_{2g} & \{\} \\
47 & B_{1u} & \{2,2,2,2,2,2,2,2,2\} \\
47 & B_{1g} & \{\} \\
47 & A_u & \{2\} \\
47 & A_g & \{\} \\
48 & B_{3u} & \{2,2,4\} \\
48 & B_{3g} & \{2\} \\
48 & B_{2u} & \{2,2,4\} \\
48 & B_{2g} & \{2\} \\
48 & B_{1u} & \{2,2,4\} \\
48 & B_{1g} & \{2\} \\
48 & A_u & \{2\} \\
48 & A_g & \{\} \\
49 & B_{3u} & \{2,2,2,2,2\} \\
49 & B_{3g} & \{\} \\
49 & B_{2u} & \{2,2,2,2,2,4\} \\
49 & B_{2g} & \{2,2\} \\
49 & B_{1u} & \{2,2,2,2,2,4\} \\
49 & B_{1g} & \{2,2\} \\
49 & A_u & \{2\} \\
49 & A_g & \{\} \\
50 & B_{3u} & \{2,2,2,4\} \\
50 & B_{3g} & \{2,2\} \\
50 & B_{2u} & \{2,2,4\} \\
50 & B_{2g} & \{2\} \\
50 & B_{1u} & \{2,2,4\} \\
50 & B_{1g} & \{2\} \\
50 & A_u & \{2\} \\
\end{array}
\end{align*}
\end{table}

\begin{table}
\begin{align*}
\begin{array}{ccc}
\mbox{SG}&\mbox{Irrep}&X^{\text{BdG}}\\
\hline 
50 & A_g & \{\} \\
51 & B_{3u} & \{2,2,2\} \\
51 & B_{3g} & \{\} \\
51 & B_{2u} & \{2,2,2,2\} \\
51 & B_{2g} & \{\} \\
51 & B_{1u} & \{2,2,2,2,2,2\} \\
51 & B_{1g} & \{2\} \\
51 & A_u & \{2,2\} \\
51 & A_g & \{\} \\
52 & B_{3u} & \{2,4\} \\
52 & B_{3g} & \{\} \\
52 & B_{2u} & \{2,2,4\} \\
52 & B_{2g} & \{2\} \\
52 & B_{1u} & \{2,4\} \\
52 & B_{1g} & \{\} \\
52 & A_u & \{4\} \\
52 & A_g & \{\} \\
53 & B_{3u} & \{2,2,2,2,4\} \\
53 & B_{3g} & \{2\} \\
53 & B_{2u} & \{2,2,2,4\} \\
53 & B_{2g} & \{\} \\
53 & B_{1u} & \{2,2,4\} \\
53 & B_{1g} & \{\} \\
53 & A_u & \{2,4\} \\
53 & A_g & \{\} \\
54 & B_{3u} & \{2,2\} \\
54 & B_{3g} & \{\} \\
54 & B_{2u} & \{2,2,4\} \\
54 & B_{2g} & \{2\} \\
54 & B_{1u} & \{2,2,2,4\} \\
54 & B_{1g} & \{2,2\} \\
54 & A_u & \{2\} \\
54 & A_g & \{\} \\
55 & B_{3u} & \{2,2,2\} \\
55 & B_{3g} & \{\} \\
55 & B_{2u} & \{2,2,2\} \\
55 & B_{2g} & \{\} \\
55 & B_{1u} & \{2,2,2\} \\
55 & B_{1g} & \{\} \\
55 & A_u & \{2,2,2\} \\
55 & A_g & \{\} \\
56 & B_{3u} & \{2\} \\
56 & B_{3g} & \{\} \\
56 & B_{2u} & \{2,2,4\} \\
56 & B_{2g} & \{2\} \\
56 & B_{1u} & \{2,2,4\} \\
56 & B_{1g} & \{2\} \\
56 & A_u & \{2\} \\
56 & A_g & \{\} \\
\end{array}&&
\begin{array}{ccc}
\mbox{SG}&\mbox{Irrep}&X^{\text{BdG}}\\
\hline 
57 & B_{3u} & \{2,2,2\} \\
57 & B_{3g} & \{\} \\
57 & B_{2u} & \{2,2\} \\
57 & B_{2g} & \{2\} \\
57 & B_{1u} & \{2\} \\
57 & B_{1g} & \{\} \\
57 & A_u & \{2,2\} \\
57 & A_g & \{\} \\
58 & B_{3u} & \{2,4\} \\
58 & B_{3g} & \{\} \\
58 & B_{2u} & \{2,2,2,4\} \\
58 & B_{2g} & \{\} \\
58 & B_{1u} & \{2,2,2,4\} \\
58 & B_{1g} & \{\} \\
58 & A_u & \{2,4\} \\
58 & A_g & \{\} \\
59 & B_{3u} & \{2\} \\
59 & B_{3g} & \{\} \\
59 & B_{2u} & \{2,2,2\} \\
59 & B_{2g} & \{2\} \\
59 & B_{1u} & \{2,2,2\} \\
59 & B_{1g} & \{2\} \\
59 & A_u & \{2,2\} \\
59 & A_g & \{\} \\
60 & B_{3u} & \{2,4\} \\
60 & B_{3g} & \{\} \\
60 & B_{2u} & \{4\} \\
60 & B_{2g} & \{\} \\
60 & B_{1u} & \{2,4\} \\
60 & B_{1g} & \{\} \\
60 & A_u & \{4\} \\
60 & A_g & \{\} \\
61 & B_{3u} & \{2\} \\
61 & B_{3g} & \{\} \\
61 & B_{2u} & \{2\} \\
61 & B_{2g} & \{\} \\
61 & B_{1u} & \{2\} \\
61 & B_{1g} & \{\} \\
61 & A_u & \{2\} \\
61 & A_g & \{\} \\
62 & B_{3u} & \{2\} \\
62 & B_{3g} & \{\} \\
62 & B_{2u} & \{2,2\} \\
62 & B_{2g} & \{\} \\
62 & B_{1u} & \{2\} \\
62 & B_{1g} & \{\} \\
62 & A_u & \{2,2\} \\
62 & A_g & \{\} \\
63 & B_{3u} & \{2,2,2,2\} \\
\end{array}&&
\begin{array}{ccc}
\mbox{SG}&\mbox{Irrep}&X^{\text{BdG}}\\
\hline 
63 & B_{3g} & \{\} \\
63 & B_{2u} & \{2,2\} \\
63 & B_{2g} & \{\} \\
63 & B_{1u} & \{2,2\} \\
63 & B_{1g} & \{\} \\
63 & A_u & \{2,2\} \\
63 & A_g & \{\} \\
64 & B_{3u} & \{2,2,2\} \\
64 & B_{3g} & \{2\} \\
64 & B_{2u} & \{2,2\} \\
64 & B_{2g} & \{\} \\
64 & B_{1u} & \{2,2,2\} \\
64 & B_{1g} & \{\} \\
64 & A_u & \{2,2\} \\
64 & A_g & \{\} \\
65 & B_{3u} & \{2,2,2,2,2,2\} \\
65 & B_{3g} & \{\} \\
65 & B_{2u} & \{2,2,2,2,2,2\} \\
65 & B_{2g} & \{\} \\
65 & B_{1u} & \{2,2,2,2,2,2\} \\
65 & B_{1g} & \{\} \\
65 & A_u & \{2,2\} \\
65 & A_g & \{\} \\
66 & B_{3u} & \{2,2,2,2\} \\
66 & B_{3g} & \{\} \\
66 & B_{2u} & \{2,2,2,2,4\} \\
66 & B_{2g} & \{2\} \\
66 & B_{1u} & \{2,2,2,2,4\} \\
66 & B_{1g} & \{2\} \\
66 & A_u & \{2,2\} \\
66 & A_g & \{\} \\
67 & B_{3u} & \{2,2,2,2\} \\
67 & B_{3g} & \{2\} \\
67 & B_{2u} & \{2,2,2,2,2\} \\
67 & B_{2g} & \{\} \\
67 & B_{1u} & \{2,2,2,2,2\} \\
67 & B_{1g} & \{\} \\
67 & A_u & \{2\} \\
67 & A_g & \{\} \\
68 & B_{3u} & \{2,2\} \\
68 & B_{3g} & \{\} \\
68 & B_{2u} & \{2,2,4\} \\
68 & B_{2g} & \{2\} \\
68 & B_{1u} & \{2,2,4\} \\
68 & B_{1g} & \{2\} \\
68 & A_u & \{2\} \\
68 & A_g & \{\} \\
69 & B_{3u} & \{2,2,2,2,2\} \\
69 & B_{3g} & \{\} \\
\end{array}&&
\begin{array}{ccc}
\mbox{SG}&\mbox{Irrep}&X^{\text{BdG}}\\
\hline 
69 & B_{2u} & \{2,2,2,2,2\} \\
69 & B_{2g} & \{\} \\
69 & B_{1u} & \{2,2,2,2,2\} \\
69 & B_{1g} & \{\} \\
69 & A_u & \{2,2\} \\
69 & A_g & \{\} \\
70 & B_{3u} & \{2,4\} \\
70 & B_{3g} & \{\} \\
70 & B_{2u} & \{2,4\} \\
70 & B_{2g} & \{\} \\
70 & B_{1u} & \{2,4\} \\
70 & B_{1g} & \{\} \\
70 & A_u & \{4\} \\
70 & A_g & \{\} \\
71 & B_{3u} & \{2,2,2,2,2\} \\
71 & B_{3g} & \{\} \\
71 & B_{2u} & \{2,2,2,2,2\} \\
71 & B_{2g} & \{\} \\
71 & B_{1u} & \{2,2,2,2,2\} \\
71 & B_{1g} & \{\} \\
71 & A_u & \{2,2\} \\
71 & A_g & \{\} \\
72 & B_{3u} & \{2,2,2,2\} \\
72 & B_{3g} & \{\} \\
72 & B_{2u} & \{2,2,2\} \\
72 & B_{2g} & \{2\} \\
72 & B_{1u} & \{2,2,2\} \\
72 & B_{1g} & \{2\} \\
72 & A_u & \{2,2\} \\
72 & A_g & \{\} \\
73 & B_{3u} & \{2,2,2\} \\
73 & B_{3g} & \{2\} \\
73 & B_{2u} & \{2,2,2\} \\
73 & B_{2g} & \{2\} \\
73 & B_{1u} & \{2,2,2\} \\
73 & B_{1g} & \{2\} \\
73 & A_u & \{2\} \\
73 & A_g & \{\} \\
74 & B_{3u} & \{2,2,2\} \\
74 & B_{3g} & \{\} \\
74 & B_{2u} & \{2,2,2,2\} \\
74 & B_{2g} & \{\} \\
74 & B_{1u} & \{2,2,2,2\} \\
74 & B_{1g} & \{\} \\
74 & A_u & \{2,2\} \\
74 & A_g & \{\} \\
75 & B & \{2\} \\
75 & A & \{\} \\
76 & B & \{\} \\
\end{array}
\end{align*}
\end{table}

\begin{table}
\begin{align*}
\begin{array}{ccc}
\mbox{SG}&\mbox{Irrep}&X^{\text{BdG}}\\
\hline 
76 & A & \{\} \\
77 & B & \{\} \\
77 & A & \{\} \\
78 & B & \{\} \\
78 & A & \{\} \\
79 & B & \{\} \\
79 & A & \{\} \\
80 & B & \{\} \\
80 & A & \{\} \\
81 & B & \{2,2,4\} \\
81 & A & \{\} \\
82 & B & \{2,4\} \\
82 & A & \{\} \\
83 & B_u & \{2,2,2,2\} \\
83 & B_g & \{2,2,2\} \\
83 & A_u & \{2,2,2,2,2,2,4\} \\
83 & A_g & \{\} \\
84 & B_u & \{2,2\} \\
84 & B_g & \{2\} \\
84 & A_u & \{2,2,2,2\} \\
84 & A_g & \{\} \\
85 & B_u & \{2\} \\
85 & B_g & \{2,2\} \\
85 & A_u & \{2,2\} \\
85 & A_g & \{\} \\
86 & B_u & \{2\} \\
86 & B_g & \{2\} \\
86 & A_u & \{2,2\} \\
86 & A_g & \{\} \\
87 & B_u & \{2,2,2\} \\
87 & B_g & \{2,2\} \\
87 & A_u & \{2,2,2,2,4\} \\
87 & A_g & \{\} \\
88 & B_u & \{4\} \\
88 & B_g & \{2\} \\
88 & A_u & \{2,4\} \\
88 & A_g & \{\} \\
89 & B_2 & \{2\} \\
89 & B_1 & \{\} \\
89 & A_2 & \{2,2,2\} \\
89 & A_1 & \{\} \\
90 & B_2 & \{\} \\
90 & B_1 & \{2\} \\
90 & A_2 & \{2\} \\
90 & A_1 & \{\} \\
91 & B_2 & \{2\} \\
91 & B_1 & \{\} \\
91 & A_2 & \{2,2\} \\
91 & A_1 & \{\} \\
\end{array}&&
\begin{array}{ccc}
\mbox{SG}&\mbox{Irrep}&X^{\text{BdG}}\\
\hline 
92 & B_2 & \{\} \\
92 & B_1 & \{\} \\
92 & A_2 & \{\} \\
92 & A_1 & \{\} \\
93 & B_2 & \{2\} \\
93 & B_1 & \{\} \\
93 & A_2 & \{2,2,2\} \\
93 & A_1 & \{\} \\
94 & B_2 & \{\} \\
94 & B_1 & \{\} \\
94 & A_2 & \{\} \\
94 & A_1 & \{\} \\
95 & B_2 & \{2\} \\
95 & B_1 & \{\} \\
95 & A_2 & \{2,2\} \\
95 & A_1 & \{\} \\
96 & B_2 & \{\} \\
96 & B_1 & \{\} \\
96 & A_2 & \{\} \\
96 & A_1 & \{\} \\
97 & B_2 & \{2\} \\
97 & B_1 & \{2\} \\
97 & A_2 & \{2,2\} \\
97 & A_1 & \{\} \\
98 & B_2 & \{\} \\
98 & B_1 & \{\} \\
98 & A_2 & \{2\} \\
98 & A_1 & \{\} \\
99 & B_2 & \{\} \\
99 & B_1 & \{\} \\
99 & A_2 & \{\} \\
99 & A_1 & \{\} \\
100 & B_2 & \{\} \\
100 & B_1 & \{\} \\
100 & A_2 & \{\} \\
100 & A_1 & \{\} \\
101 & B_2 & \{\} \\
101 & B_1 & \{\} \\
101 & A_2 & \{\} \\
101 & A_1 & \{\} \\
102 & B_2 & \{\} \\
102 & B_1 & \{\} \\
102 & A_2 & \{\} \\
102 & A_1 & \{\} \\
103 & B_2 & \{2\} \\
103 & B_1 & \{2\} \\
103 & A_2 & \{\} \\
103 & A_1 & \{\} \\
104 & B_2 & \{\} \\
\end{array}&&
\begin{array}{ccc}
\mbox{SG}&\mbox{Irrep}&X^{\text{BdG}}\\
\hline 
104 & B_1 & \{\} \\
104 & A_2 & \{\} \\
104 & A_1 & \{\} \\
105 & B_2 & \{\} \\
105 & B_1 & \{\} \\
105 & A_2 & \{\} \\
105 & A_1 & \{\} \\
106 & B_2 & \{\} \\
106 & B_1 & \{\} \\
106 & A_2 & \{\} \\
106 & A_1 & \{\} \\
107 & B_2 & \{\} \\
107 & B_1 & \{\} \\
107 & A_2 & \{\} \\
107 & A_1 & \{\} \\
108 & B_2 & \{\} \\
108 & B_1 & \{\} \\
108 & A_2 & \{\} \\
108 & A_1 & \{\} \\
109 & B_2 & \{\} \\
109 & B_1 & \{\} \\
109 & A_2 & \{\} \\
109 & A_1 & \{\} \\
110 & B_2 & \{\} \\
110 & B_1 & \{\} \\
110 & A_2 & \{\} \\
110 & A_1 & \{\} \\
111 & B_2 & \{2,2,2,2\} \\
111 & A_2 & \{2\} \\
111 & B_1 & \{2\} \\
111 & A_1 & \{\} \\
112 & B_2 & \{2,2,2\} \\
112 & A_2 & \{2\} \\
112 & B_1 & \{2\} \\
112 & A_1 & \{\} \\
113 & B_2 & \{2\} \\
113 & A_2 & \{\} \\
113 & B_1 & \{2\} \\
113 & A_1 & \{\} \\
114 & B_2 & \{2\} \\
114 & A_2 & \{\} \\
114 & B_1 & \{2\} \\
114 & A_1 & \{\} \\
115 & B_2 & \{2,2,2\} \\
115 & A_2 & \{\} \\
115 & B_1 & \{2\} \\
115 & A_1 & \{\} \\
116 & B_2 & \{2,2\} \\
116 & A_2 & \{\} \\
\end{array}&&
\begin{array}{ccc}
\mbox{SG}&\mbox{Irrep}&X^{\text{BdG}}\\
\hline 
116 & B_1 & \{2\} \\
116 & A_1 & \{\} \\
117 & B_2 & \{2,2\} \\
117 & A_2 & \{2\} \\
117 & B_1 & \{2\} \\
117 & A_1 & \{\} \\
118 & B_2 & \{2\} \\
118 & A_2 & \{\} \\
118 & B_1 & \{2\} \\
118 & A_1 & \{\} \\
119 & B_2 & \{2,2,2\} \\
119 & A_2 & \{\} \\
119 & B_1 & \{2,2\} \\
119 & A_1 & \{\} \\
120 & B_2 & \{2,2,2\} \\
120 & A_2 & \{2\} \\
120 & B_1 & \{2\} \\
120 & A_1 & \{\} \\
121 & B_2 & \{2,2\} \\
121 & A_2 & \{\} \\
121 & B_1 & \{2\} \\
121 & A_1 & \{\} \\
122 & B_2 & \{2\} \\
122 & A_2 & \{\} \\
122 & B_1 & \{2\} \\
122 & A_1 & \{\} \\
123 & B_{2u} & \{2,2,2\} \\
123 & B_{2g} & \{\} \\
123 & B_{1u} & \{2\} \\
123 & B_{1g} & \{\} \\
123 & A_{2u} & \{2,2,2,2,2,2,2,2,2\} \\
123 & A_{2g} & \{\} \\
123 & A_{1u} & \{2,2,2\} \\
123 & A_{1g} & \{\} \\
124 & B_{2u} & \{2,2,2\} \\
124 & B_{2g} & \{2\} \\
124 & B_{1u} & \{2,2\} \\
124 & B_{1g} & \{2\} \\
124 & A_{2u} & \{2,2,2,2,2,2\} \\
124 & A_{2g} & \{\} \\
124 & A_{1u} & \{2,2,2\} \\
124 & A_{1g} & \{\} \\
125 & B_{2u} & \{2\} \\
125 & B_{2g} & \{2\} \\
125 & B_{1u} & \{2\} \\
125 & B_{1g} & \{2\} \\
125 & A_{2u} & \{2,2,2,2\} \\
125 & A_{2g} & \{2\} \\
125 & A_{1u} & \{2\} \\
\end{array}
\end{align*}
\end{table}

\begin{table}
\begin{align*}
\begin{array}{ccc}
\mbox{SG}&\mbox{Irrep}&X^{\text{BdG}}\\
\hline 
125 & A_{1g} & \{\} \\
126 & B_{2u} & \{2,2\} \\
126 & B_{2g} & \{2,2\} \\
126 & B_{1u} & \{2\} \\
126 & B_{1g} & \{2\} \\
126 & A_{2u} & \{2,2,2\} \\
126 & A_{2g} & \{2\} \\
126 & A_{1u} & \{2\} \\
126 & A_{1g} & \{\} \\
127 & B_{2u} & \{2\} \\
127 & B_{2g} & \{\} \\
127 & B_{1u} & \{2,2,2\} \\
127 & B_{1g} & \{\} \\
127 & A_{2u} & \{2,2,2,2,2\} \\
127 & A_{2g} & \{\} \\
127 & A_{1u} & \{2,2,2\} \\
127 & A_{1g} & \{\} \\
128 & B_{2u} & \{2,2\} \\
128 & B_{2g} & \{2\} \\
128 & B_{1u} & \{2,2,2\} \\
128 & B_{1g} & \{2\} \\
128 & A_{2u} & \{2,2,2,4\} \\
128 & A_{2g} & \{\} \\
128 & A_{1u} & \{2,2,4\} \\
128 & A_{1g} & \{\} \\
129 & B_{2u} & \{\} \\
129 & B_{2g} & \{\} \\
129 & B_{1u} & \{2\} \\
129 & B_{1g} & \{2\} \\
129 & A_{2u} & \{2,2,2\} \\
129 & A_{2g} & \{\} \\
129 & A_{1u} & \{2\} \\
129 & A_{1g} & \{\} \\
130 & B_{2u} & \{2\} \\
130 & B_{2g} & \{2\} \\
130 & B_{1u} & \{2,2\} \\
130 & B_{1g} & \{2,2\} \\
130 & A_{2u} & \{2,2\} \\
130 & A_{2g} & \{\} \\
130 & A_{1u} & \{2\} \\
130 & A_{1g} & \{\} \\
131 & B_{2u} & \{2,2,2\} \\
131 & B_{2g} & \{\} \\
131 & B_{1u} & \{\} \\
131 & B_{1g} & \{\} \\
131 & A_{2u} & \{2,2,2,2,2,2\} \\
131 & A_{2g} & \{\} \\
131 & A_{1u} & \{2\} \\
131 & A_{1g} & \{\} \\
\end{array}&&
\begin{array}{ccc}
\mbox{SG}&\mbox{Irrep}&X^{\text{BdG}}\\
\hline 
132 & B_{2u} & \{2\} \\
132 & B_{2g} & \{\} \\
132 & B_{1u} & \{2\} \\
132 & B_{1g} & \{\} \\
132 & A_{2u} & \{2,2,2,2,2\} \\
132 & A_{2g} & \{\} \\
132 & A_{1u} & \{2\} \\
132 & A_{1g} & \{\} \\
133 & B_{2u} & \{2,2\} \\
133 & B_{2g} & \{2,2\} \\
133 & B_{1u} & \{2\} \\
133 & B_{1g} & \{2\} \\
133 & A_{2u} & \{2,2,2,2\} \\
133 & A_{2g} & \{2,2\} \\
133 & A_{1u} & \{2\} \\
133 & A_{1g} & \{\} \\
134 & B_{2u} & \{2\} \\
134 & B_{2g} & \{2\} \\
134 & B_{1u} & \{2\} \\
134 & B_{1g} & \{2\} \\
134 & A_{2u} & \{2,2,2,2\} \\
134 & A_{2g} & \{2\} \\
134 & A_{1u} & \{2\} \\
134 & A_{1g} & \{\} \\
135 & B_{2u} & \{2\} \\
135 & B_{2g} & \{\} \\
135 & B_{1u} & \{2,2\} \\
135 & B_{1g} & \{\} \\
135 & A_{2u} & \{2,2,2\} \\
135 & A_{2g} & \{\} \\
135 & A_{1u} & \{2,2\} \\
135 & A_{1g} & \{\} \\
136 & B_{2u} & \{2\} \\
136 & B_{2g} & \{\} \\
136 & B_{1u} & \{2,2\} \\
136 & B_{1g} & \{\} \\
136 & A_{2u} & \{2,2,2\} \\
136 & A_{2g} & \{\} \\
136 & A_{1u} & \{2,2\} \\
136 & A_{1g} & \{\} \\
137 & B_{2u} & \{\} \\
137 & B_{2g} & \{2\} \\
137 & B_{1u} & \{2\} \\
137 & B_{1g} & \{2\} \\
137 & A_{2u} & \{2,2\} \\
137 & A_{2g} & \{\} \\
137 & A_{1u} & \{2,2\} \\
137 & A_{1g} & \{\} \\
138 & B_{2u} & \{\} \\
\end{array}&&
\begin{array}{ccc}
\mbox{SG}&\mbox{Irrep}&X^{\text{BdG}}\\
\hline 
138 & B_{2g} & \{\} \\
138 & B_{1u} & \{2\} \\
138 & B_{1g} & \{2\} \\
138 & A_{2u} & \{2,2\} \\
138 & A_{2g} & \{\} \\
138 & A_{1u} & \{2\} \\
138 & A_{1g} & \{\} \\
139 & B_{2u} & \{2\} \\
139 & B_{2g} & \{\} \\
139 & B_{1u} & \{2\} \\
139 & B_{1g} & \{\} \\
139 & A_{2u} & \{2,2,2,2,2,2\} \\
139 & A_{2g} & \{\} \\
139 & A_{1u} & \{2,2\} \\
139 & A_{1g} & \{\} \\
140 & B_{2u} & \{\} \\
140 & B_{2g} & \{\} \\
140 & B_{1u} & \{2,2\} \\
140 & B_{1g} & \{\} \\
140 & A_{2u} & \{2,2,2,2,2\} \\
140 & A_{2g} & \{\} \\
140 & A_{1u} & \{2,2\} \\
140 & A_{1g} & \{\} \\
141 & B_{2u} & \{2\} \\
141 & B_{2g} & \{2\} \\
141 & B_{1u} & \{2\} \\
141 & B_{1g} & \{2\} \\
141 & A_{2u} & \{2,2,2\} \\
141 & A_{2g} & \{\} \\
141 & A_{1u} & \{2,2\} \\
141 & A_{1g} & \{\} \\
142 & B_{2u} & \{2\} \\
142 & B_{2g} & \{2\} \\
142 & B_{1u} & \{2\} \\
142 & B_{1g} & \{2\} \\
142 & A_{2u} & \{2,2,2\} \\
142 & A_{2g} & \{2\} \\
142 & A_{1u} & \{2\} \\
142 & A_{1g} & \{\} \\
143 & A & \{\} \\
144 & A & \{\} \\
145 & A & \{\} \\
146 & A & \{\} \\
147 & A_u & \{2,2,4,8\} \\
147 & A_g & \{\} \\
148 & A_u & \{2,2,4,8\} \\
148 & A_g & \{\} \\
149 & A_2 & \{2\} \\
149 & A_1 & \{\} \\
\end{array}&&
\begin{array}{ccc}
\mbox{SG}&\mbox{Irrep}&X^{\text{BdG}}\\
\hline 
150 & A_2 & \{2\} \\
150 & A_1 & \{\} \\
151 & A_2 & \{2\} \\
151 & A_1 & \{\} \\
152 & A_2 & \{2\} \\
152 & A_1 & \{\} \\
153 & A_2 & \{2\} \\
153 & A_1 & \{\} \\
154 & A_2 & \{2\} \\
154 & A_1 & \{\} \\
155 & A_2 & \{2\} \\
155 & A_1 & \{\} \\
156 & A_2 & \{\} \\
156 & A_1 & \{\} \\
157 & A_2 & \{\} \\
157 & A_1 & \{\} \\
158 & A_2 & \{\} \\
158 & A_1 & \{\} \\
159 & A_2 & \{\} \\
159 & A_1 & \{\} \\
160 & A_2 & \{\} \\
160 & A_1 & \{\} \\
161 & A_2 & \{\} \\
161 & A_1 & \{\} \\
162 & A_{2u} & \{2,2,2,2,4\} \\
162 & A_{2g} & \{\} \\
162 & A_{1u} & \{2,2,4\} \\
162 & A_{1g} & \{\} \\
163 & A_{2u} & \{2,2,4\} \\
163 & A_{2g} & \{\} \\
163 & A_{1u} & \{2,4\} \\
163 & A_{1g} & \{\} \\
164 & A_{2u} & \{2,2,2,2,4\} \\
164 & A_{2g} & \{\} \\
164 & A_{1u} & \{2,2,4\} \\
164 & A_{1g} & \{\} \\
165 & A_{2u} & \{2,2,4\} \\
165 & A_{2g} & \{\} \\
165 & A_{1u} & \{2,4\} \\
165 & A_{1g} & \{\} \\
166 & A_{2u} & \{2,2,2,2,4\} \\
166 & A_{2g} & \{\} \\
166 & A_{1u} & \{2,2,4\} \\
166 & A_{1g} & \{\} \\
167 & A_{2u} & \{2,2,4\} \\
167 & A_{2g} & \{\} \\
167 & A_{1u} & \{2,4\} \\
167 & A_{1g} & \{\} \\
168 & B & \{4\} \\
\end{array}
\end{align*}
\end{table}

\begin{table}
\begin{align*}
\begin{array}{ccc}
\mbox{SG}&\mbox{Irrep}&X^{\text{BdG}}\\
\hline 
168 & A & \{\} \\
169 & B & \{\} \\
169 & A & \{\} \\
170 & B & \{\} \\
170 & A & \{\} \\
171 & B & \{4\} \\
171 & A & \{\} \\
172 & B & \{4\} \\
172 & A & \{\} \\
173 & B & \{\} \\
173 & A & \{\} \\
174 & A'' & \{2,2,2,2\} \\
174 & A' & \{\} \\
175 & B_u & \{4,4,4\} \\
175 & B_g & \{2,2\} \\
175 & A_u & \{2,2,2,2,2,2,4\} \\
175 & A_g & \{\} \\
176 & B_u & \{4\} \\
176 & B_g & \{2\} \\
176 & A_u & \{2,2,2,4\} \\
176 & A_g & \{\} \\
177 & B_1 & \{2\} \\
177 & B_2 & \{2\} \\
177 & A_2 & \{2,2\} \\
177 & A_1 & \{\} \\
178 & B_1 & \{\} \\
178 & B_2 & \{\} \\
178 & A_2 & \{2\} \\
178 & A_1 & \{\} \\
179 & B_1 & \{\} \\
179 & B_2 & \{\} \\
179 & A_2 & \{2\} \\
179 & A_1 & \{\} \\
180 & B_1 & \{2\} \\
180 & B_2 & \{2\} \\
180 & A_2 & \{2,2\} \\
180 & A_1 & \{\} \\
181 & B_1 & \{2\} \\
181 & B_2 & \{2\} \\
181 & A_2 & \{2,2\} \\
181 & A_1 & \{\} \\
182 & B_1 & \{\} \\
182 & B_2 & \{\} \\
182 & A_2 & \{2\} \\
182 & A_1 & \{\} \\
183 & B_2 & \{2\} \\
183 & B_1 & \{2\} \\
183 & A_2 & \{\} \\
183 & A_1 & \{\} \\
\end{array}&&
\begin{array}{ccc}
\mbox{SG}&\mbox{Irrep}&X^{\text{BdG}}\\
\hline 
184 & B_2 & \{4\} \\
184 & B_1 & \{4\} \\
184 & A_2 & \{\} \\
184 & A_1 & \{\} \\
185 & B_2 & \{\} \\
185 & B_1 & \{\} \\
185 & A_2 & \{\} \\
185 & A_1 & \{\} \\
186 & B_2 & \{\} \\
186 & B_1 & \{\} \\
186 & A_2 & \{\} \\
186 & A_1 & \{\} \\
187 & A''_2 & \{2,2,2,2,2\} \\
187 & A'_2 & \{\} \\
187 & A''_1 & \{2,2,2\} \\
187 & A'_1 & \{\} \\
188 & A''_2 & \{2,2,2,2\} \\
188 & A'_2 & \{\} \\
188 & A''_1 & \{2,2,2\} \\
188 & A'_1 & \{\} \\
189 & A''_2 & \{2,2,2,2\} \\
189 & A'_2 & \{\} \\
189 & A''_1 & \{2,2\} \\
189 & A'_1 & \{\} \\
190 & A''_2 & \{2,2,2\} \\
190 & A'_2 & \{\} \\
190 & A''_1 & \{2,2\} \\
190 & A'_1 & \{\} \\
191 & B_{1u} & \{2,2,2\} \\
191 & B_{1g} & \{2\} \\
191 & B_{2u} & \{2,2,2\} \\
191 & B_{2g} & \{2\} \\
191 & A_{2u} & \{2,2,2,2,2,2,2,2\} \\
191 & A_{2g} & \{\} \\
191 & A_{1u} & \{2,2,2,2\} \\
191 & A_{1g} & \{\} \\
192 & B_{1u} & \{2,4\} \\
192 & B_{1g} & \{2,2\} \\
192 & B_{2u} & \{2,4\} \\
192 & B_{2g} & \{2,2\} \\
192 & A_{2u} & \{2,2,2,2,2,2\} \\
192 & A_{2g} & \{\} \\
192 & A_{1u} & \{2,2,2,2\} \\
192 & A_{1g} & \{\} \\
193 & B_{1u} & \{2\} \\
193 & B_{1g} & \{2\} \\
193 & B_{2u} & \{2\} \\
193 & B_{2g} & \{2\} \\
193 & A_{2u} & \{2,2,2,2,2\} \\
\end{array}&&
\begin{array}{ccc}
\mbox{SG}&\mbox{Irrep}&X^{\text{BdG}}\\
\hline 
193 & A_{2g} & \{\} \\
193 & A_{1u} & \{2,2,2\} \\
193 & A_{1g} & \{\} \\
194 & B_{1u} & \{2\} \\
194 & B_{1g} & \{2\} \\
194 & B_{2u} & \{2\} \\
194 & B_{2g} & \{2\} \\
194 & A_{2u} & \{2,2,2,2,2\} \\
194 & A_{2g} & \{\} \\
194 & A_{1u} & \{2,2,2\} \\
194 & A_{1g} & \{\} \\
195 & A & \{\} \\
196 & A & \{\} \\
197 & A & \{\} \\
198 & A & \{\} \\
199 & A & \{\} \\
200 & A_u & \{2,2\} \\
200 & A_g & \{\} \\
201 & A_u & \{2,2\} \\
201 & A_g & \{\} \\
202 & A_u & \{2,2,2\} \\
202 & A_g & \{\} \\
203 & A_u & \{2,4\} \\
203 & A_g & \{\} \\
204 & A_u & \{2,2,2\} \\
204 & A_g & \{\} \\
205 & A_u & \{2,2\} \\
205 & A_g & \{\} \\
206 & A_u & \{2,2\} \\
206 & A_g & \{\} \\
207 & A_2 & \{\} \\
207 & A_1 & \{\} \\
208 & A_2 & \{\} \\
208 & A_1 & \{\} \\
209 & A_2 & \{2\} \\
209 & A_1 & \{\} \\
210 & A_2 & \{\} \\
210 & A_1 & \{\} \\
211 & A_2 & \{2\} \\
211 & A_1 & \{\} \\
212 & A_2 & \{\} \\
212 & A_1 & \{\} \\
213 & A_2 & \{\} \\
213 & A_1 & \{\} \\
214 & A_2 & \{\} \\
214 & A_1 & \{\} \\
215 & A_2 & \{2\} \\
215 & A_1 & \{\} \\
216 & A_2 & \{2,2\} \\
\end{array}&&
\begin{array}{ccc}
\mbox{SG}&\mbox{Irrep}&X^{\text{BdG}}\\
\hline 
216 & A_1 & \{\} \\
217 & A_2 & \{2\} \\
217 & A_1 & \{\} \\
218 & A_2 & \{2\} \\
218 & A_1 & \{\} \\
219 & A_2 & \{2\} \\
219 & A_1 & \{\} \\
220 & A_2 & \{2\} \\
220 & A_1 & \{\} \\
221 & A_{2u} & \{2,2\} \\
221 & A_{2g} & \{\} \\
221 & A_{1u} & \{2,2,2,2\} \\
221 & A_{1g} & \{\} \\
222 & A_{2u} & \{2,2\} \\
222 & A_{2g} & \{2\} \\
222 & A_{1u} & \{2,2\} \\
222 & A_{1g} & \{\} \\
223 & A_{2u} & \{2\} \\
223 & A_{2g} & \{\} \\
223 & A_{1u} & \{2,2\} \\
223 & A_{1g} & \{\} \\
224 & A_{2u} & \{2,2\} \\
224 & A_{2g} & \{2\} \\
224 & A_{1u} & \{2,2\} \\
224 & A_{1g} & \{\} \\
225 & A_{2u} & \{2,2\} \\
225 & A_{2g} & \{\} \\
225 & A_{1u} & \{2,2,2\} \\
225 & A_{1g} & \{\} \\
226 & A_{2u} & \{2\} \\
226 & A_{2g} & \{\} \\
226 & A_{1u} & \{2,2,2\} \\
226 & A_{1g} & \{\} \\
227 & A_{2u} & \{2,2\} \\
227 & A_{2g} & \{2\} \\
227 & A_{1u} & \{2,2,2\} \\
227 & A_{1g} & \{\} \\
228 & A_{2u} & \{2,2\} \\
228 & A_{2g} & \{2\} \\
228 & A_{1u} & \{2,2\} \\
228 & A_{1g} & \{\} \\
229 & A_{2u} & \{2,2\} \\
229 & A_{2g} & \{\} \\
229 & A_{1u} & \{2,2,2\} \\
229 & A_{1g} & \{\} \\
230 & A_{2u} & \{2,2\} \\
230 & A_{2g} & \{2\} \\
230 & A_{1u} & \{2,2\} \\
230 & A_{1g} & \{\} \\
\end{array}
\end{align*}
\end{table}

\begin{table}
\caption{The list of $X^{\text{BdG}}$ for spinful electrons without time-reversal symmetry.}
\label{tab:noTRS_spinful}
\begin{align*}
\begin{array}{ccc}
\mbox{SG}&\mbox{Irrep}&X^{\text{BdG}}\\
\hline 
1 & A & \{\} \\
2 & A_u & \{2,2,2,4,4,4,8\} \\
2 & A_g & \{\} \\
3 & B & \{\} \\
3 & A & \{2,2,4\} \\
4 & B & \{\} \\
4 & A & \{\} \\
5 & B & \{\} \\
5 & A & \{2\} \\
6 & A'' & \{\} \\
6 & A' & \{2\} \\
7 & A'' & \{\} \\
7 & A' & \{\} \\
8 & A'' & \{\} \\
8 & A' & \{\} \\
9 & A'' & \{\} \\
9 & A' & \{\} \\
10 & B_u & \{2,2,2,2,2,4\} \\
10 & B_g & \{\} \\
10 & A_u & \{2,2,2,2,2,2,4,4,4\} \\
10 & A_g & \{2\} \\
11 & B_u & \{2,2,2,4\} \\
11 & B_g & \{\} \\
11 & A_u & \{2,2,4\} \\
11 & A_g & \{\} \\
12 & B_u & \{2,2,4,4\} \\
12 & B_g & \{\} \\
12 & A_u & \{2,2,2,2,4,4\} \\
12 & A_g & \{\} \\
13 & B_u & \{2,2,4\} \\
13 & B_g & \{\} \\
13 & A_u & \{2,2,2,4,4\} \\
13 & A_g & \{2\} \\
14 & B_u & \{2,2,4\} \\
14 & B_g & \{\} \\
14 & A_u & \{2,2,4\} \\
14 & A_g & \{\} \\
15 & B_u & \{2,4,4\} \\
15 & B_g & \{\} \\
15 & A_u & \{2,2,4,4\} \\
15 & A_g & \{\} \\
16 & B_3 & \{\} \\
16 & B_2 & \{\} \\
16 & B_1 & \{\} \\
16 & A & \{\} \\
17 & B_3 & \{\} \\
17 & B_2 & \{2\} \\
17 & B_1 & \{2\} \\
17 & A & \{2,2,2\} \\
\end{array}&&
\begin{array}{ccc}
\mbox{SG}&\mbox{Irrep}&X^{\text{BdG}}\\
\hline 
18 & B_3 & \{2\} \\
18 & B_2 & \{\} \\
18 & B_1 & \{\} \\
18 & A & \{2\} \\
19 & B_3 & \{\} \\
19 & B_2 & \{\} \\
19 & B_1 & \{\} \\
19 & A & \{\} \\
20 & B_3 & \{\} \\
20 & B_2 & \{\} \\
20 & B_1 & \{\} \\
20 & A & \{2\} \\
21 & B_3 & \{2\} \\
21 & B_2 & \{\} \\
21 & B_1 & \{\} \\
21 & A & \{2\} \\
22 & B_3 & \{\} \\
22 & B_2 & \{\} \\
22 & B_1 & \{\} \\
22 & A & \{\} \\
23 & B_3 & \{\} \\
23 & B_2 & \{\} \\
23 & B_1 & \{\} \\
23 & A & \{\} \\
24 & B_3 & \{\} \\
24 & B_2 & \{\} \\
24 & B_1 & \{\} \\
24 & A & \{2,2\} \\
25 & B_1 & \{\} \\
25 & B_2 & \{\} \\
25 & A_2 & \{\} \\
25 & A_1 & \{\} \\
26 & B_1 & \{\} \\
26 & B_2 & \{\} \\
26 & A_2 & \{\} \\
26 & A_1 & \{\} \\
27 & B_1 & \{\} \\
27 & B_2 & \{\} \\
27 & A_2 & \{\} \\
27 & A_1 & \{\} \\
28 & B_1 & \{\} \\
28 & B_2 & \{\} \\
28 & A_2 & \{2\} \\
28 & A_1 & \{2,2\} \\
29 & B_1 & \{\} \\
29 & B_2 & \{\} \\
29 & A_2 & \{\} \\
29 & A_1 & \{\} \\
30 & B_1 & \{\} \\
\end{array}%
\end{align*}
\end{table}

\begin{table}
\caption{The list of $X^{\text{BdG}}$ for spinless electrons without time-reversal symmetry.}
\label{tab:noTRS_spinless}
\begin{align*}
%
\end{align*}
\end{table}

\clearpage

\section{Key space groups}
\label{KeySGs}
As explained in the main text, most of the nonzero entries in the Tables in Appendix~\ref{XBSTable} can be understood by focusing on their key space groups. In this section we provide details on this point for class DIII systems.

In Tables~\ref{tab:XBS} and \ref{tab:XBS2}, the key space group for each space group is listed.  
The quantities that characterize elements of $\XBSb$ are also listed in these Tables.  These Tables are unfortunately not complete yet in the sense that they do not cover all the nonzero entries of Table~\ref{tab:XBS_DIII} (46 out of in total 399 nontrivial entries are still lacking interpretation), but nonetheless they are still useful in the practical application.
To understand the meaning of indices in These tables, let us go over them one by one. 

\begin{table}[b]
	\begin{center}
		\caption{\label{tab:XBS} The list of key space groups and their super groups. In the right-most column, key space groups are underlined.}
		\begin{tabular}{cc|cc}\hline \hline
			Key Space Group&Key Indices&$\XBSb$&Space Groups\\\hline
			\multirow{40}{*}{$P\bar{1}$ ($A_u$)} & \multirow{40}{*}{$\nu_{i}^{1\text{D}}, \nu_{i}^{2\text{D}}, \kappa_1$} & $(\mathbb{Z}_2)^3\times(\mathbb{Z}_4)^3\times\mathbb{Z}_8$ &  \underline{$2(A_u)$}, $10(A_u, B_u)$, $47(A_{u}, B_{1u}, B_{2u}, B_{3u})$, \\
			&&$(\mathbb{Z}_2)^2\times(\mathbb{Z}_4)^2\times\mathbb{Z}_8$ & $12(A_u, B_u)$, $65(A_{u}, B_{1u}, B_{2u}, B_{3u})$.\\&\\
			&&	\multirow{3}{*}{$\mathbb{Z}_2\times(\mathbb{Z}_4)^2\times\mathbb{Z}_8$} & $11(A_u, B_u),13(A_u, B_u), 49 (A_{u}, B_{1u}, B_{2u}, B_{3u})$,\\
			&&&$51 (A_{u}, B_{1u}, B_{2u}, B_{3u}),67 (A_{u}, B_{1u}, B_{2u}, B_{3u}), $\\
			&&&$69 (A_{u}, B_{1u}, B_{2u}, B_{3u}) $.\\&\\
			&&\multirow{2}{*}{$(\mathbb{Z}_2)^2\times\mathbb{Z}_4\times\mathbb{Z}_8$} & $15(A_u, B_u), 66 (A_{u}, B_{1u}, B_{2u}, B_{3u}),$ \\
			&&&$71 (A_{u}, B_{1u}, B_{2u}, B_{3u}) , 74 (A_{u}, B_{1u}, B_{2u}, B_{3u}).$\\
			&&\multirow{3}{*}{$\mathbb{Z}_2\times\mathbb{Z}_4\times\mathbb{Z}_8$} 
			&  $14(A_u,B_u), 53(A_{u}, B_{1u}, B_{2u}, B_{3u}),$\\
			&&& $55(A_{u}, B_{1u}, B_{2u}, B_{3u}), 63(A_{u}, B_{1u}, B_{2u}, B_{3u}),$\\
			&&&$72(A_{u}, B_{1u}, B_{2u}, B_{3u}), 175(B_u), 191(B_{1u}, B_{2u}).$ \\
			&&\multirow{4}{*}{$\mathbb{Z}_4\times\mathbb{Z}_8$} 
			&  $48(A_u, B_{1u}, B_{2u}, B_{3u}), 50(A_u, B_{1u}, B_{2u}, B_{3u}),$\\ 
			&&&$54(A_u, B_{1u}, B_{2u}, B_{3u}), 57(A_u, B_{1u}, B_{2u}, B_{3u}),$\\
			&&&$59(A_u, B_{1u}, B_{2u}, B_{3u}), 64(A_u, B_{1u}, B_{2u}, B_{3u}), $\\
			&&&$68(A_u, B_{1u}, B_{2u}, B_{3u}), 73(A_u, B_{1u}, B_{2u}, B_{3u}).$\\&\\
			&&$(\mathbb{Z}_2)^2\times\mathbb{Z}_8$ & $58(A_u, B_{1u}, B_{2u}, B_{3u}).$\\&\\
			&&\multirow{4}{*}{$\mathbb{Z}_2\times\mathbb{Z}_8$} & $52(A_u, B_{1u}, B_{2u}, B_{3u}), 56(A_u, B_{1u}, B_{2u}, B_{3u}),$ \\
			&&&$60(A_u, B_{1u}, B_{2u}, B_{3u}),  62(A_u, B_{1u}, B_{2u}, B_{3u}),$\\
			&&&$70(A_u, B_{1u}, B_{2u}, B_{3u}), 176(B_u), 192(B_{1u}, B_{2u}), $\\
			&&&$193(B_{1u}, B_{2u}), 194(B_{1u}, B_{2u}).$\\&\\
			&& $\mathbb{Z}_8$
			& $61(A_u, B_{1u}, B_{2u}, B_{3u})$.\\
			&& \multirow{1}{*}{$\mathbb{Z}_2\times(\mathbb{Z}_{4})^2$ }& $84(B_u), 87(B_u), 131(B_{1u}, B_{2u}),139(B_{1u}, B_{2u}), 140 (B_{1u}, B_{2u}).$\\
			&& $(\mathbb{Z}_{4})^2$ & $85(B_u), 125(B_{1u}, B_{2u}), 129(B_{1u}, B_{2u}), 132(B_{1u}, B_{2u}).$\\
			&& \multirow{2}{*}{$\mathbb{Z}_2\times\mathbb{Z}_{4}$} & $86(B_u), 88(B_u), 134(B_{1u}, B_{2u}), 135(B_{1u}, B_{2u}),$\\
			&&&$136(B_{1u}, B_{2u}), 138(B_{1u}, B_{2u}), 141(B_{1u}, B_{2u}).$\\
			&& \multirow{2}{*}{$\mathbb{Z}_4$} & $126(B_{1u}, B_{2u}), 133(B_{1u}, B_{2u}),$ \\
			&&&$137(B_{1u}, B_{2u}),142(B_{1u}, B_{2u}).$\\
			\hline
			\multirow{2}{*}{$P4, I4$(B)} & \multirow{2}{*}{$\nu^{2\text{D}}_{C_{4}}$} & \multirow{2}{*}{$\mathbb{Z}_2$} &\underline{$75(B)$}, $89(B_1, B_2), 90(B_1, B_2), 99(B_1, B_2),$ \\
			&&&$100(B_1, B_2), 103(B_1,B_2), 108(B_1, B_2), 207(A_2).$\\ 
			\hline
			\multirow{4}{*}{$P\bar{4}$ ($B$)} & \multirow{4}{*}{$\nu_{S_4}^{1\text{D}}, \nu_{S_4}^{2\text{D}}, \kappa_{4}$} & $(\mathbb{Z}_2)^2\times\mathbb{Z}_4$ & \underline{$81(B)$}, $111(B_1, B_2),115(B_1, B_2), 215(A_2)$.\\
			&&\multirow{2}{*}{$\mathbb{Z}_2\times\mathbb{Z}_4$} & $112(B_1, B_2), 113(B_1, B_2), 116(B_1, B_2), $\\
			&&&$117(B_1, B_2),  118(B_1, B_2), 218(A_2). $\\
			&& $\mathbb{Z}_4$ & $114(B_1, B_2), 122(B_1, B_2), 220(A_2).$\\
			\hline
			\multirow{4}{*}{$P4/m$ ($A_{u}$)}  & \multirow{2}{*}{$\nu^{1\text{D}, \Gamma \rightarrow Z}_{1/2, 3/2}, \nu^{1\text{D}, M \rightarrow A}_{1/2}$, $\nu^{1\text{D}}$}& $(\mathbb{Z}_2)^4\times\mathbb{Z}_4\times\mathbb{Z}_{8}\times\mathbb{Z}_{16}$ & \underline{$83(A_u)$}, $123(A_{1u}, A_{2u}).$\\
			&\multirow{2}{*}{$\nu^{2\text{D}}$, $z_{8,\pi}$ $z_{16}$}&$(\mathbb{Z}_2)^3\times\mathbb{Z}_4\times\mathbb{Z}_{16}$ & $124(A_{1u}, A_{2u}).$\\
			&&$(\mathbb{Z}_2)^2\times\mathbb{Z}_8\times\mathbb{Z}_{16}$ &$127(A_{1u}, A_{2u}).$\\
			&&$(\mathbb{Z}_2)^3\times\mathbb{Z}_{16}$ & $128(A_{1u}, A_{2u}).$\\
			\multirow{2}{*}{$I4/m$ (A$_{u}$)} &\multirow{2}{*}{$\nu^{1\text{D}, \Gamma \rightarrow Z}_{1/2, 3/2}$, $\nu^{1\text{D}}$, $\nu^{2\text{D}}$, $z_{16}$}&$(\mathbb{Z}_2)^3\times\mathbb{Z}_4\times\mathbb{Z}_{16}$  &$87(A_u), 139(A_{1u}, A_{2u}).$\\
			&&$(\mathbb{Z}_2)^2\times\mathbb{Z}_4\times\mathbb{Z}_{16}$  &$140(A_{1u}, A_{2u}).$\\
			\multirow{2}{*}{$P4/m$ ($B_{g}$)} &\multirow{2}{*}{$\nu_{C_4}^{+,2\text{D}},\nu_{C_4}^{-,2\text{D}},\kappa_4$} & $(\mathbb{Z}_2)^3$& \underline{$83(B_g)$}, $123(B_{1g}, B_{2g}),  127(B_{1g}, B_{2g}), 221(A_{2g}).$\\
			&&$(\mathbb{Z}_2)^2$& $124(B_{1g}, B_{2g}), 128(B_{1g}, B_{2g}).$\\
			\multirow{1}{*}{$I4/m$ ($B_{g}$)} &$\nu_{S_4}^{+},\nu_{S_4}^{-},\kappa_4$&$(\mathbb{Z}_2)^2$&$87(B_g), 139(B_{1g}, B_{2g}), 140(B_{1g}, B_{2g}), 225(A_{2g}), 229(A_{2g}).$\\
			\multirow{2}{*}{$P4/m$ ($B_{u}$)} &\multirow{2}{*}{$\nu^{1\text{D},\Gamma\rightarrow X}, \nu^{2\text{D}}_{1}, z_{8}, \kappa_1$} & $\mathbb{Z}_2\times(\mathbb{Z}_4)^2\times\mathbb{Z}_8$& \underline{$83(B_u)$}, $123(B_{1u}, B_{2u}).$\\
			&&$\mathbb{Z}_2\times\mathbb{Z}_4 \times \mathbb{Z}_8$& $127(B_{1u}, B_{2u}).$\\
			&&$\mathbb{Z}_2\times(\mathbb{Z}_4)^2$& $124(B_{1u}, B_{2u}).$\\
			&&$(\mathbb{Z}_2)^2 \times \mathbb{Z}_4$& $128(B_{1u}, B_{2u}).$\\
			\hline\hline
		\end{tabular}
	\end{center}
\end{table}

\begin{table*}
	\begin{center}
		\caption{\label{tab:XBS2}(Continued from the previous page)}
		\begin{tabular}{cc|cc}
			\hline\hline
			\multirow{3}{*}{$P4_2/m$ ($A_{u}$)} &\multirow{3}{*}{$\nu_{1/2,3/2}^{1\text{D}, \Gamma\rightarrow Z}+\nu_{1/2,3/2}^{1\text{D}, M\rightarrow A},\nu_{i}^{1\text{D}}, \nu_{x}^{2\text{D}}, \kappa_1$}& $(\mathbb{Z}_2)^3\times\mathbb{Z}_4\times\mathbb{Z}_8$ & \underline{$84(A_u)$}, $131(A_{1u}, A_{2u}).$\\
			& & $(\mathbb{Z}_2)^2\times\mathbb{Z}_4\times\mathbb{Z}_8$ & $132(A_{1u}, A_{2u}).$\\
			&&$(\mathbb{Z}_2)^2\times\mathbb{Z}_8$ & $135(A_{1u}, A_{2u}), 136(A_{1u}, A_{2u}).$\\
			\multirow{2}{*}{$P4/n$ ($A_{u}$)} &\multirow{2}{*}{$\nu^{1\text{D}, \Gamma \rightarrow Z}_{1/2, 3/2}, \nu_{z}^{2\text{D}}, \kappa_1$}& $\mathbb{Z}_2\times\mathbb{Z}_4\times\mathbb{Z}_8$ & \underline{$85(A_u)$}, $125(A_{1u}, A_{2u}), 129(A_{1u}, A_{2u}).$\\
			& & $\mathbb{Z}_2\times\mathbb{Z}_8$ & $126(A_{1u}, A_{2u}), 130(A_{1u}, A_{2u}).$\\
			\multirow{3}{*}{$P4_2/n$ ($A_{u}$)} &\multirow{3}{*}{$\nu_{1/2,3/2}^{\Gamma \rightarrow Z}+\nu_{1/2,3/2}^{M \rightarrow A},\nu_{z}^{2\text{D}},\kappa_1$}& $\mathbb{Z}_ 2 \times \mathbb{Z}_4 \times \mathbb{Z}_8$  & \underline{$86(A_u)$}, $134(A_{1u}, A_{2u}).$\\
			&&$\mathbb{Z}_ 2  \times \mathbb{Z}_8$  & $133(A_{1u}, A_{2u}), 137(A_{1u}, A_{2u}).$\\
			&&$(\mathbb{Z}_ 2)^2 \times \mathbb{Z}_8$  & $138(A_{1u}, A_{2u}).$\\
			\multirow{2}{*}{$I4_1/a$ ($A_{u}$)} &\multirow{2}{*}{$\nu_{1/2}^{\Gamma \rightarrow Z}+\tfrac{1}{2}\nu_{S_4}^{P,P'}, \nu_{1}^{1\text{D}}, \kappa_1$}& $(\mathbb{Z}_ 2)^2 \times \mathbb{Z}_8$ & \underline{$88(A_u)$}, $141(A_{1u}, A_{2u}).$\\
			&&$\mathbb{Z}_ 2 \times \mathbb{Z}_8$ & $142(A_{1u}, A_{2u}).$\\
			\multirow{2}{*}{$P4_2/m$ ($B_{g}$)} &\multirow{2}{*}{$\kappa_4$}& \multirow{2}{*}{$\mathbb{Z}_ 2$} & \underline{$84(B_{g})$}, $131(B_{1g},B_{2g}), 132(B_{1g},B_{2g}),$\\
			&&& $135(B_{1g},B_{2g}), 136(B_{1g},B_{2g}).$\\
			\multirow{2}{*}{$P4/n$ ($B_{g}$)} &\multirow{2}{*}{$\nu_{S_4}^{2\text{D}},\kappa_4$}& $(\mathbb{Z}_ 2)^2$ & \underline{$85(B_g)$}, $125(B_{1g},B_{2g}), 129(B_{1g},B_{2g})$. 
			\\
			&&$\mathbb{Z}_ 2$ & $126(B_{1g},B_{2g}).$\\
			\multirow{2}{*}{$P4_2/n$ ($B_{g}$)} &\multirow{2}{*}{$\kappa_4$}& \multirow{2}{*}{$\mathbb{Z}_ 2$} & \underline{$86(B_g)$}, $133(B_{1g},B_{2g}), 134(B_{1g},B_{2g})$, \\
			&&&$137(B_{1g},B_{2g}), 138(B_{1g},B_{2g}).$\\
			\multirow{1}{*}{$I4_1/a$ ($B_{g}$)} &\multirow{1}{*}{$\kappa_4$}& $\mathbb{Z}_ 2$ & \underline{$88(B_g)$}, $141(B_{1g},B_{2g}), 142(B_{1g},B_{2g}).$\\
			\hline
			\multirow{2}{*}{$P\bar{3}$ ($A_{u}$)} &\multirow{2}{*}{$\nu^{1\text{D}}_{1/2, 3/2}, \nu^{2\text{D}}, \kappa_1$}& $(\mathbb{Z}_2)^2\times\mathbb{Z}_4\times\mathbb{Z}_ 8$ & \underline{$147(A_u)$}, $162(A_{1u},A_{2u}), 164(A_{1u},A_{2u}).$\\
			&& $(\mathbb{Z}_ 2)^2 \times \mathbb{Z}_8$ &$163(A_{1u},A_{2u}), 165(A_{1u},A_{2u}).$\\
			\multirow{2}{*}{$R\bar{3}$ ($A_{u}$)} &\multirow{2}{*}{$\nu^{1\text{D}}_{1/2, 3/2}, \nu^{2\text{D}}, \kappa_1$}& $(\mathbb{Z}_2)^2\times\mathbb{Z}_4\times\mathbb{Z}_ 8$ & $148(A_u), 166(A_{1u},A_{2u}).$\\
			&& $(\mathbb{Z}_2)^2\times\mathbb{Z}_ 8$ & $167(A_{1u},A_{2u}).$\\
			\hline
			\multirow{4}{*}{$P\bar{6}$} &\multirow{4}{*}{$\nu^{1\text{D}}_{S_{6}}, \nu^{'}_{S_{6}}, \nu^{''}_{S_{6}}, z_{3}^{k_z=0, \pi}$}& $\mathbb{Z}_2\times(\mathbb{Z}_6)^2$ & \underline{$174(A'')$}, $187(A''_1, A''_2)).$\\
			&& $(\mathbb{Z}_6)^2$ & $189(A''_1, A''_2).$\\
			&& $(\mathbb{Z}_2)^2\times\mathbb{Z}_6$ & $188(A''_1, A''_2).$\\
			&& $\mathbb{Z}_2\times\mathbb{Z}_ 6$ & $190(A''_1, A''_2).$\\
			\hline
			$P6/m$ ($A_{u}$)& $\nu_{1/2, 3/2, 5/2}^{\Gamma\rightarrow A}, \nu_{S_6}^{1\text{D}, K\rightarrow H} z_{12}^{k_z=0, \pi}, z_{24} $& $(\mathbb{Z}_2)^{4}\times\mathbb{Z}_{12}\times\mathbb{Z}_{24}$ & \underline{$175(A_u)$}, $191(A_{1u}, A_{2u}).$ \\
			&& $(\mathbb{Z}_2)^{4}\times\mathbb{Z}_{24}$ & $192(A_{1u}, A_{2u}).$ \\
			$P6/m$ ($B_{g}$)&$\nu^{1\text{D}}_{S_{6}}, z_{6}^{k_z=0, \pi}$& $ \mathbb{Z}_3\times\mathbb{Z}_{6}$ &  \underline{$175(B_g)$}, $191(B_{1g}, B_{2g}).$\\
			&& $\mathbb{Z}_6$ & $192(B_{1g}, B_{2g}).$\\
			$P6_3/m$ ($A_{u}$)&$\nu_{1/2}^{\Gamma\rightarrow A}+\nu_{5/2}^{\Gamma\rightarrow A}, \nu_{3/2}^{\Gamma\rightarrow A}, \nu_{S_{6}}^{'}, z_{12}^{k_z=0}, z_{24}$& $(\mathbb{Z}_2)^{3}\times\mathbb{Z}_{24}$ & \underline{$176(A_u)$}, $193(A_{1u}, A_{2u}),  194(A_{1u}, A_{2u}).$ \\
			$P6_3/m$ ($B_{g}$)&$\nu^{'}_{S_{6}}, \nu^{''}_{S_{6}}, z_{6}^{k_z=0}$ & $\mathbb{Z}_{6}$ & \underline{$176(B_g)$}, $193(B_{1g}, B_{2g}), 194(B_{1g}, B_{2g}). $\\
			\hline\hline
		\end{tabular}
	\end{center}
\end{table*}

\subsection{$P\bar{1}$}
Let us start with $P\bar{1}$, which is the simplest key SG. Although this space group has been studied in details in Ref.~\onlinecite{PhysRevX.8.031070,1811.08712, Skurativska2019}, we review the results briefly as a preparation for the discussion of other key space groups. There are two one-dimensional representations of inversion symmetry, $A_g$ and $A_u$, corresponding to even parity ($\chi_I=+1$) and odd parity ($\chi_I=-1$) SCs. Since $\XBSb$ is trivial for the even parity case, here we focus on the odd parity case. 
\begin{table}[H]
	\begin{center}
		\caption{\label{tab:p1bar} $\XBSb$ for $P\bar{1}$.}
		\begin{tabular}{c|c|c|c}
			\hline\hline
			SG  (rep of $\Delta_{\bm{k}}$) & 1D & 2D & 3D \\
			\hline
			$P\bar{1}$  ($A_u$)  &$\mathbb{Z}_2$ &$(\mathbb{Z}_2)^2 \times \mZ_4$ &$(\mathbb{Z}_2)^3 \times (\mathbb{Z}_4)^3 \times \mZ_8$ \\
			\hline\hline
		\end{tabular}
	\end{center}
\end{table}

In 1D, the $\mathbb{Z}_2$ topological invariant can be determined by the sum of the inversion parities at $k=0$ and $\pi$:
\begin{align}
\label{inv_1D}
\nu^{1\text{D}} &\equiv \frac{1}{4}\sum_{\bm{k}\in 1\text{D TRIMs}}\sum_{\alpha=\pm 1}\alpha N_{\bk}^{\alpha} \mod 2,
\end{align}
where $\nu^{1\text{D}}=+1$ mod $2$ corresponds to the nontrivial phase.  

In 2D, the $(\mZ_2)^2$ part in $X^{\text{BdG}}$ (see Table~\ref{tab:p1bar}) indicates the weak topology, i.e., stacking of Kitaev chains. As pointed out by Ref.~\onlinecite{Skurativska2019}, the remaining $\mZ_4$ factor, corresponding to
\begin{align}
\label{inv_2D}
\nu^{2\text{D}} &\equiv \frac{1}{4}\sum_{\bm{k}\in 2\text{D TRIMs}}\sum_{\alpha=\pm 1}\alpha N_{\bk}^{\alpha}\mod 4,
\end{align}
which diagnoses the strong TSC ($\nu^{2\text{D}} =1$ or $3$ mod $4$) and the higher-order TSC  ($\nu^{2\text{D}} =2$ mod $4$). Although the classification of 2D TSC in class DIII is $\mZ_2$, the higher-order TSC is stabilized by the inversion symmetry and can be realized by stacking two 2D helical TSCs.

Similarly, in 3D the $(\mZ_2)^3\times (\mZ_4)^3$ part represents the weak indices. The remaining $\mathbb{Z}_8$ index can be computed by
\begin{align}
\label{kappa1_app}
\kappa_1 &= \frac{1}{4}\sum_{\bm{k}\in 3\text{D TRIMs}}\sum_{\alpha=\pm 1}\alpha N_{\bk}^{\alpha}\mod 8, 
\end{align}
which agrees with the 3D winding number for class AIII
\begin{align}
\label{3D-winding}
\nu_{\text{w}} = \frac{1}{48\pi^2}\int d^3\bk \epsilon^{ijk}\mathrm{tr}\left[U_\Gamma \left(H^{-1}_{\bk}\partial_i H_{\bk}\right) \left(H^{-1}_{\bk}\partial_j H_{\bk}\right) \left(H^{-1}_{\bk}\partial_k H_{\bk}\right) \right]\in\mathbb{Z}
\end{align}
modulo 2. Here, $U_\Gamma$ is the matrix representation of the chiral symmetry. When $\nu_{\text{w}} = 0$, $\kappa_1 = 2$ mod $4$ corresponds to the second-order TSC with 1D hinge states~\cite{1811.08712}, and $\kappa_1 = 4$ mod $8$ corresponds to the third-order TSC~\cite{Skurativska2019}.

\subsection{$P\bar{4}$ with $B$ representation}
\label{rotoinversion}
The space group $P\bar{4}$ has four-fold rotoinversion symmetry $S_4\equiv IC_4$ that maps $(x,y,z)$ to $(y, -x, -z)$. Here we consider the nontrivial 1D representation $B$ with $\chi_{S_4}=-1$. 
\begin{table}[H]
	\begin{center}
		\caption{\label{tab:p4bar} $\XBSb$ for $P\bar{4}$.}
		\begin{tabular}{c|c|c|c}
		\hline\hline
		SG (rep of $\Delta_{\bm{k}}$)  & 1D & 2D & 3D \\
		\hline
		$P\bar{4}$ ($B$) &$\mathbb{Z}_2$ &$\mathbb{Z}_2$ &$(\mathbb{Z}_2)^2 \times \mathbb{Z}_4$ \\
		\hline\hline
		\end{tabular}
	\end{center}
\end{table}

For 1D chains along the rotoinversion axis, $X^{\text{BdG}}=\mathbb{Z}_2$ is given by
\begin{align}
\nu_{S_4}^{1\text{D}} \equiv \frac{1}{2\sqrt{2}}\sum_{k_z=0,\pi}\sum_{\alpha=\pm1,\pm3}e^{i\frac{\pi\alpha}{4}} N_{\bk}^{\alpha}\mod 2,
\end{align}
where $N_{\bm{k}}^{\alpha}$ represents the number of irreducible representations with the four-fold rotoinversion eigenvalue $e^{i \frac{\pi\alpha}{4}}$. 

2D systems orthogonal to the rotoinversion axis can be understood in the same way as the $P4$ symmetry case discussed in the Sec.~IV of the main text. We can simply use a similar formula to diagnose the $\mathbb{Z}_2$ QSH index:
\begin{align}
\nu_{S_4}^{2\text{D}} &\equiv \frac{1}{2\sqrt{2}}\sum_{\Gamma, M}\sum_{\alpha=\pm1,\pm3}e^{i\frac{\pi\alpha}{4}} N_{\bk}^{\alpha}\mod 2.
\end{align}

Finally for 3D systems, in addition to the weak $(\mZ_2)^2$ factor that stems from 1D and 2D, there exists a strong $\mathbb{Z}_4$ factor (see Table~\ref{tab:p4bar}), which can be computed by~\cite{PhysRevX.8.031070}
\begin{align}
\kappa_4 = \frac{1}{2\sqrt{2}}\sum_{\bm{k}\in K_4}\sum_{\alpha=\pm1,\pm3}e^{i\frac{\pi\alpha}{4}} N_{\bk}^{\alpha}\mod 4,
\end{align}
where $K_4$ represents the set of four high-symmetry momenta invariant under $S_4$. This quantity is defined mod 4 because all vectors in $\AIb$ takes $\kappa_4=0$ mod 4. 
This $\kappa_4$ agrees with the 3D winding number mod 2~\cite{PhysRevX.8.031070, 1811.08712}. Furthermore, $\kappa_4 = 2$ mod 4 indicates second-order TSCs when $\nu_{\text{w}}=0$. To see this from the surface theory~\cite{PhysRevX.8.031070,PhysRevB.97.205136}, let us introduce the surface Hamiltonian of a 3D TSC and the corresponding representation matrices for the time-reversal, particle-hole, and the rotoinversion symmetry.
\begin{align}
\label{surface_Ham}
h_{\bm{r}, \bm{k}}^{\xi} &= -\xi (\bm{k} \times \bm{n}_{\bm{r}})\cdot \bm{\sigma},\\
\label{surface_TRS}
u_{\calT} &= -i \sigma_y,\\
\label{surface_PHS}
u_{\Xi}^{\xi} &= \xi (\bm{n}_{\bm{r}}\cdot \bm{\sigma})\sigma_y,\\
u_{S_4}^{\xi} &= -\xi e^{-i \tfrac{\pi}{4}\sigma_z},
\end{align}
where $\xi=\pm1$ represents the 3D winding number $\nu_{\text{w}}$. Let us consider the stacked Hamiltonian $h_{\bm{r}, \bm{k}}^{\mathcal{S}} =h_{\bm{r}, \bm{k}}^{+}\oplus h_{\bm{r}, \bm{k}}^{-} = -(\bm{k} \times \bm{n}_{\bm{r}})\cdot \bm{\sigma}\tau_z$ with $\nu_{\text{w}} = 1-1=0$. Since each of the two Hamiltonians has $\kappa_4 = 3$ mod 4, the stacked Hamiltonian has $\kappa_4 = 2$ mod 4. For this system, the mass term $M_{\br} = m_{\br}\sigma_0 \tau_x$ respects all symmetries when $m_{S_4\br} = -m_{\br}$. The one dimensional domain wall of the mass gap hosts a gapless helical Majorana mode.

\subsection{$P\bar{3}$}
The space group $P\bar{3}$ contains both the inversion and the three-fold rotation about the $z$ axis.  Here we consider $A_u$ representation defined by $\chi_{C_3} = +1$ and $\chi_I=-1$. $X^{\text{BdG}}$ can be mostly understood from $P\bar{1}$.  The classification of 1D Kitaev chains is promoted to $(\mZ_2)^2$ due to the three-fold rotation symmetry. They can be characterized by
\begin{align}
\nu_{1/2}^{1\text{D}} &= \frac{1}{4}\sum_{\bm{k}\in \Gamma, Z}\sum_{\alpha=\pm 1}\sum_{\beta=\pm1} \beta N_{\bm{k}}^{\alpha, \beta}\ \ \text{mod}\ \ 2,\\
\nu_{3/2}^{1\text{D}} &= \frac{1}{4}\sum_{\bm{k}\in \Gamma, Z}\sum_{\alpha=\pm 3}\sum_{\beta=\pm1} \beta N_{\bm{k}}^{\alpha, \beta}\ \ \text{mod}\ \ 2,
\end{align}
where $N_{\bm{k}}^{\alpha,\beta}$ is representation count with the three-fold rotation eigenvalue $e^{i \frac{\pi\alpha}{3}}$ and the inversion eigenvalue $\beta$ at $\Gamma$ and $Z$. 
\begin{table}[H]
\begin{center}
\caption{\label{tab:p3bar} $\XBSb$ for $P\bar{3}$.}
\begin{tabular}{c|c|c|c}
\hline\hline
			SG  (rep of $\Delta_{\bm{k}}$) & 1D & 2D & 3D \\
\hline
 $P\bar{3}$ ($A_u$) &$(\mathbb{Z}_2)^2$ &$\mathbb{Z}_4$ &$(\mathbb{Z}_2)^2 \times \mathbb{Z}_4 \times \mathbb{Z}_8 $ \\
\hline\hline
\end{tabular}
\end{center}
\end{table}

The remaining factors of $\mZ_4$ and $\mZ_8$ in $\XBSb$ are explained purely by the inversion symmetry. They can be computed by by $\nu^{2\text{D}}$ in Eq.~\eqref{inv_2D} for the $k_z=0$ plane and $\kappa_1$ in Eq.~\eqref{kappa1_app}.

\subsection{$P\bar{6}$}
The space group $P\bar{6}$ is generated from the six-fold rotoinversion symmetry $S_{6}^{z}=IC_{6}^{z}$. There are two one-dimensional representations, $A'$ and $A''$, corresponding to $\chi_{S_6} =+1$ and $\chi_{S_6}=-1$. Since $\XBSb$ is trivial for $A'$ representation, here we focus on $A''$ representation.

\begin{table}[H]
\begin{center}
\caption{\label{tab:p6bar} $\XBSb$ for $P\bar{6}$.}
\begin{tabular}{c|c|c|c}
\hline\hline
SG  (rep of $\Delta_{\bm{k}}$) & 1D & 2D & 3D \\
\hline
$P\bar{6}$ ($A''$)  &$\mathbb{Z}_2$ &$\mathbb{Z}_3$ &$\mathbb{Z}_2 \times(\mathbb{Z}_6)^2 $ \\
\hline\hline
\end{tabular}
\end{center}
\end{table}

Analogously to the four-fold rotoinversion case, we define a $\mathbb{Z}_2$ index for 1D chains by
\begin{align}
\label{S6-1D}
\nu_{S_6} &= \frac{1}{2\sqrt{3}}\sum_{k_z = 0, \pi}\sum_{\alpha=\pm1,\pm3,\pm5}e^{i\tfrac{\pi\alpha}{6}}N_{k_z}^{\alpha}=\frac{1}{2}\left(N_{\Gamma}^{1} - N_{\Gamma}^{5} + N_{A}^{1} - N_{A}^{5} \right)\quad \text{mod}\ 2,
\end{align}
where $N_{\bk}^{\alpha}$ is representation count with the six-fold rotoinversion eigenvalue $e^{i \frac{\pi\alpha}{6}}$, and $\Gamma$ and A represent $k_z=0$ and $\pi$. 

For 2D systems, we define a $\mathbb{Z}_3$ index
\begin{align}
\label{S6-2D}
z_{3} &= -\frac{1}{2}N_{K}^{5} +\frac{1}{2}N_{K}^{1} + \frac{3}{2}N_{K}^{9} -\frac{1}{2}N_{K'}^{5} +\frac{1}{2}N_{K'}^{1} + \frac{3}{2}N_{K'}^{9} + N_{\Gamma}^{5}- N_{\Gamma}^{1} \quad \text{mod}\ 3,
\end{align}
which agrees with the mirror Chern number mod $3$. 

For 3D systems, the $\mZ_2$ of $\XBSb$ in Table~\ref{tab:p6bar} is explained by $\nu_{S_6}$ in Eq.~\eqref{S6-1D}. The remaining $(\mZ_6)^2$ part can be characterized by
\begin{align}
z'_{6} &=\frac{1}{2}\left(4 N_{\Gamma}^{1} - 4 N_{\Gamma}^{5} -5 N_{K}^{3} +N_{K}^{5} -7 N_{K}^{9} - N_{K}^{11} + 2N_{H}^{1} + N_{H}^{3} - N_{H}^{5} -2 N_{H}^{7}-N_{H}^{9}+N_{H}^{11}\right)\quad \text{mod}\ 6,\\
z''_{6} &=\frac{1}{2}\left(4 N_{A}^{1} - 4 N_{A}^{5}+N_{K}^{1} - N_{K}^{3} -2N_{K}^{5} -N_{K}^{7}+ N_{K}^{9} +2 N_{K}^{11} -N_{H}^{1} -7 N_{H}^{3} + N_{H}^{7}-5N_{H}^{9}\right)\quad \text{mod}\ 6,
\end{align}

\subsection{$P6/m$}
The space group $P6/m$ is similar to $P4/m$. It has the inversion and the six-fold rotation symmetry.  There are four real one-dimensional representations 
$A_g$ ($\chi_{C_6} = +1$, $\chi_I=+1$), 
$A_u$ ($\chi_{C_6} = +1$, $\chi_I=-1$),
$B_g$ ($\chi_{C_6} = -1$, $\chi_I=+1$), and 
$B_u$ ($\chi_{C_6} = -1$, $\chi_I=-1$). 
$\XBSb$ for $A_g$ representation is trivial, and $\XBSb$ for $B_u$ representation can be understood by focusing on its inversion supergroup ($P\bar{1}$).  Thus here we discuss the remaining two representations.
\begin{table}[H]
\begin{center}
\caption{\label{tab:p6/m}$\XBSb$ for $P6/m$.}
\begin{tabular}{c|c|c|c}
\hline\hline
			SG  (rep of $\Delta_{\bm{k}}$) & 1D & 2D & 3D \\
\hline
$P6/m$ ($A_u$) &$(\mathbb{Z}_2)^3$ &$\mathbb{Z}_{12} $&$(\mathbb{Z}_2)^4 \times \mathbb{Z}_{12} \times \mathbb{Z}_{24} $ \\
\hline
$P6/m$ ($B_u$) &$\mZ_2$ &$ \mathbb{Z}_4$ &$ \mZ_2 \times \mZ_4 \times \mZ_8$ \\
\hline
$P6/m$ ($B_g$) &$\mZ_1$ &$ \mathbb{Z}_3$ &$\mathbb{Z}_3\times \mathbb{Z}_6$ \\
\hline\hline
\end{tabular}
\end{center}
\end{table}

\subsubsection{$A_u$ representation}
For 1D systems, we define indices $\nu_{1/2}^{1\text{D}}$, $\nu_{3/2}^{1\text{D}}$, and $\nu_{5/2}^{1\text{D}}$ in the same way as for $P4/m$ with the $A_u$ representation. They describe Majorana edge modes characterized by six-fold rotation eigenvalues.
\begin{align}
\label{1D-P6m-1}
\nu_{1/2}^{1\text{D}} &= \frac{1}{4}\sum_{\bm{k}\in \Gamma, Z}\sum_{\alpha=\pm 1}\sum_{\beta=\pm1} \beta N_{\bk}^{\alpha,\beta},\\
\label{1D-P6m-2}
\nu_{3/2}^{1\text{D}} &= \frac{1}{4}\sum_{\bm{k}\in \Gamma, Z}\sum_{\alpha=\pm 3}\sum_{\beta=\pm1} \beta N_{\bk}^{\alpha,\beta},\\
\label{1D-P6m-3}
\nu_{5/2}^{1\text{D}} &= \frac{1}{4}\sum_{\bm{k}\in \Gamma, Z}\sum_{\alpha=\pm 5}\sum_{\beta=\pm1} \beta N_{\bk}^{\alpha,\beta}.
\end{align}
Here, $N_{\bm{k}}^{\alpha,\beta}$ is representation count with the three-fold rotation eigenvalue $e^{i \frac{\pi\alpha}{6}}$ and the inversion eigenvalue $\beta$ at $\Gamma$ and $Z$. 

For 2D systems, the mirror Chern number $C_M$ mod $6$ is given by~\cite{PhysRevB.86.115112}
\begin{align}
z_{6} &= \frac{3}{2} (-N^{-3, +}_{\Gamma} +N^{ 3, -}_{\Gamma} )+ \frac{5}{2} (N^{ 5, +}_{\Gamma} - N^{ -5, -}_{\Gamma}) +\frac{1}{2}(N^{ 1, +}_{\Gamma} - N^{ -1, -}_{\Gamma})\notag \\
&\quad\quad\quad+N_{K}^{1} - N_{K}^{-1}+3N_{K}^{-3} +\frac{3}{2}(-N_{M}^{3,+}+N_{M}^{-3,-})\quad \text{mod}\  6.\label{eq21}
\end{align}
Because of the inversion symmetry, $\nu^{2\text{D}}$ in Eq~\eqref{inv_2D} is also well-defined.  Since $2z_6 = 3\nu^{2\text{D}}$ (mod $12$) always holds for all elements in $\{\text{AI}\}^{\text{BdG}}$, the following combination defines a $\mZ_{12}$ index:
\begin{align}
z_{12} = 2z_6 - 3\nu^{2\text{D}}\quad \text{mod}\ 12.
\label{2dindex}
\end{align}
There are two cases for $z_{12} = 6$ mod $12$: (a) $z_6=3$ mod $6$ and $\nu^{2\text{D}}=0$ mod 4, implying a nontrivial mirror Chern number, and (b) $z_6=0$ mod $6$ and $\nu^{2\text{D}}=2$ mod 4, indicating the second order TSC or mirror Chern TSC with $C_M = 6m$ ($m\neq0$). Indeed, one can construct an example of the case (b) with $C_M=0$ by a wire construction.

Finally, let us discuss 3D systems, for which $\XBS=(\mathbb{Z}_2)^4 \times \mathbb{Z}_{12} \times \mathbb{Z}_{24}$ as summarized in Table~\ref{tab:p6/m}.
The following $\mathbb{Z}_2$ index 
\begin{align}
\label{3D-P6m-1}
	\nu_{S_6}^{1\text{D},K\rightarrow H} &= \frac{1}{2\sqrt{3}}\sum_{\bk\in K, H}\sum_{\alpha=\pm1,\pm3,\pm5}e^{i\tfrac{\pi\alpha}{6}}N_{\bk}^{\alpha}\mod 2,
\end{align}
together with those in Eqs.~\eqref{1D-P6m-1}--\eqref{1D-P6m-3}, explain the $(\mZ_2)^4$ part. In this formula, $K$ and $H$ are the high-symmetry points whose little groups are $C_{3h}$.  The $\mathbb{Z}_{12}$ index can be explained by $z_{12}$ in Eq.~\eqref{2dindex} for the $k_z=\pi$ plane.
Furthermore, we define a $\mathbb{Z}_{24}$ index
\begin{align}
z_{24} &= 4\kappa_6 - 3\kappa_1\quad \text{mod}\  24,
\end{align}
where $\kappa_1$ is defined in Eq.~\eqref{kappa1_app} and $\kappa_6 = \frac{1}{2\sqrt{3}}\sum_{\alpha=\pm1,\pm3,\pm5}\sum_{\bk\in K_6}e^{i\frac{\pi\alpha}{6}}N_{\bk}^{\alpha}$ is the sum of $S_6$ eigenvalues.  $z_{12}$ for $k_z=0$ plane is automatically fixed by $z_{12}$ for $k_z=\pi$ and $z_{24}$. The implication of $z_{24}$ mod 12 is the same as the case of class AII~\cite{QuantitativeMappings, PhysRevX.8.031070}.
To investigate the meaning of $z_{24} = 12$ (mod 24), we again perform a wire construction as illustrated in Fig~\ref{WCP4m} of the main text. It produces a phase with $z_{24} = 12$ (mod $24$) that exhibits Majorana corner states upon breaking the translation symmetry.  From this exercise, we conclude that $z_{24} = 12$ (mod $24$) implies either TSCs with nontrivial mirror Chern numbers or third-order TSCs when $\nu_{\text{w}} = 0$.

\subsubsection{$B_g$ representation}
In 2D systems, $z_6$ in Eq.~\eqref{eq21} explains $\XBS=\mZ_3$ because $z_6$ is constrained to be even for this representation. 

In 3D systems, $\mZ_3$ in $\XBS$ in Table~\ref{tab:p6/m} is explained by $z_{6}$ for the $k_z=0$ plane. The remaining 
$\mZ_6$ factor can be explained by
\begin{align}
	z'''_{6} 
	&= \frac{1}{2}\left( 2N_{A}^{1,+}+6N_{A}^{3,-}+10N_{A}^{5,+}-10N_{A}^{-5,-}-6N_{A}^{-3,+}-2N_{A}^{-1,-} \right.\notag\\
	&\left. \quad \quad  -3N_{K}^{1}+3N_{K}^{5}+N_{H}^{1} +12 N_{H}^{3}-N_{H}^{5}-6N_{L}^{3,+}+6N_{L}^{-3,-}\right),
\end{align}
where 
$N_{\bk}^{\alpha,\beta}$ is the representation count with the rotation eigenvalue $e^{i \frac{\pi\alpha}{6}}$ and the inversion parity $\beta$ at $\bk= A$ and $L$, while
$N_{K}^{\alpha}$ is the same with the rotoinversion eigenvalue $e^{i \frac{\pi\alpha}{6}}$. In this formula, $A, K$ and $L$ are the high-symmetry points whose little groups are $C_{6h}, C_{3h}$, and $C_{2h}$.

\clearpage

\section{Wannierizable topological superconductors in $P4/m$}
\label{WTSC-P4/m}
In Sec.~V of the main text, we introduced the notion of Wannierizable topological superconductors and developed a scheme of classifying them. In this section, we present the details of calculations of $\mathcal{X}^{\text{WTSC}}$ and discuss its correspondence to $X^{\text{BdG}}$ for the space group $P4/m$.

\subsection{Lattice homotopy equivalence}
Let us first summarize the lattice homotopy equivalence of this space group. This part is independent of the choice of the pairing symmetries. 

For 1D systems, the lattice homotopy equivalence reads
\begin{align}
\label{eq:1DLH}
\sum_{\beta=\pm 1}  \psi_{a}^{\alpha, \beta} &\sim \sum_{\beta=\pm 1}  \psi_{b}^{\alpha, \beta},  
\end{align}
where the Wyckoff positions are denoted by $\mathcal{W}_a = \{(0,0,0) \}, \mathcal{W}_b= \{(0,0,1/2) \}$ and $\psi_{x}^{\alpha, \beta}$ ($\alpha=\pm1, \pm3$, $\beta=+1, -1$, and $x=a,b$) represents an atomic insulator at the Wyckoff position $\mathcal{W}_x$ with the $C_4$ eigenvalue $e^{i \frac{\alpha\pi}{4}}$ and the inversion parity $\beta$. 
Under these equivalence relations, atomic wave functions are labeled by $6$ integers as
\begin{align}
\left(n_{a}^{\pm1,+1},n_{a}^{\pm1,-1},n_{a}^{\pm3,+1}, n_{a}^{\pm3,-1}, n_{b}^{\pm1,+1},n_{b}^{\pm3,+1}\right) \in \mZ^{6}.
\end{align}
Here we denote the count for the states $\psi_{x}^{\alpha, \beta}$ by $n_{x}^{\alpha, \beta}$.  We have eliminated $n_{b}^{\pm1,-1}$ and $n_{b}^{\pm3,-1}$ from the list using the equivalence relations.

For 2D systems, the lattice homotopy relations are
\begin{align}
\label{eq:2DLH}
\sum_{\beta=\pm1}(\psi_{x}^{\pm 1, \beta} +  \psi_{x}^{\pm 3, \beta})\sim \sum_{\beta=\pm1}  \psi_{c}^\beta\quad\text{for $x=a,b$},
\end{align}
where 
\begin{align}
\mathcal{W}_a &= \{(0,0,0) \}, \\
\mathcal{W}_b&= \{(1/2,1/2,0) \},  \\
\mathcal{W}_c &= \{ (1/2,0,0), (0,1/2,0)\}.
\end{align}
Under these equivalence relations, atomic wave functions are labeled by $8$ integers as
\begin{equation}\begin{split}\label{eq:}
(n_a^{\pm1,+1},n_a^{\pm1,-1},n_{a}^{\pm3,+1}, n_b^{\pm1,+},n_b^{\pm1,-1},n_{b}^{\pm3,+1}, n_c^{+1},n_c^{-1})
\in \mathbb Z^8.
\end{split}\end{equation}

For 3D systems, the lattice homotopy relations are
\begin{align}
\label{eq:3DLH}
\sum_{\beta=\pm1}(\psi_{x}^{\pm 1, \beta} +  \psi_{x}^{\pm 3, \beta})\sim \sum_{\beta=\pm1}  \psi_{y}^\beta\quad\text{for $x=a,b,c,d$ and $y=e,f$},
\end{align}
and
\begin{align}
\sum_{\beta=\pm1}  \psi_{a}^{\alpha, \beta} &\sim \sum_{\beta=\pm1}  \psi_{b}^{\alpha, \beta},  \\
\sum_{\beta=\pm1}  \psi_{c}^{\alpha, \beta} &\sim \sum_{\beta=\pm1}  \psi_{d}^{\alpha, \beta}, 
\end{align}
where 
\begin{align}
\mathcal{W}_a &= \{(0,0,0) \}, \\
\mathcal{W}_b&= \{(0,0,1/2) \},  \\
\mathcal{W}_c&= \{(1/2,1/2,0) \},  \\
\mathcal{W}_d&= \{(1/2,1/2,1/2) \},  \\
\mathcal{W}_e &= \{ (1/2,0,0), (0,1/2,0)\},  \\
\mathcal{W}_f &= \{ (1/2,0,1/2), (0,1/2,1/2)\}.
\end{align}
Under these equivalence relations, atomic wave functions are labeled by $13$ integers. 
\begin{align}
\left(n_{a}^{\pm1,+1},n_{a}^{\pm1,-1},n_{a}^{\pm3,+1}, n_{b}^{\pm1,+1},n_{b}^{\pm3,+1}, n_{c}^{\pm1,+1},n_{c}^{\pm1,-1}, n_{c}^{\pm3,+1}, n_{d}^{\pm1,+1},n_{d}^{\pm3,+1}, n_{e}^{+1}, n_{e}^{-1},n_{f}^{+1} \right) \in \mZ^{13}.
\end{align}

\subsection{$A_u$ representation}
With this preparation, we follow the steps described in Sec.~V of the main text and compute $\mathcal{X}^{\text{WTSC}}$ for each setting.

\subsubsection{1D}
Let us start with 1D. We define the particle-hole operator on the atomic states by
\begin{align}
\Xi_\chi \left[\psi_{x}^{\alpha, \beta}\right] &= \psi_{x}^{\alpha, -\beta}\quad\text{for $x=a,b$}.
\end{align}
We find that the matrix representation of the particle-hole symmetry on the atomic states is
\begin{align}
\Xi_{\chi} = \begin{pmatrix}
0 & 1 & 0 & 0 & 1 & 0  \\
1 & 0 & 0 & 0 & 1 & 0  \\
0 & 0 & 0 & 1 & 0 & 1 \\
0 & 0 & 1 & 0 & 0 &  1\\
0 & 0 & 0 & 0 & -1 & 0 \\
0 & 0 & 0 & 0 & 0 & -1\\
\end{pmatrix}.
\end{align}
The kernel of $\openone + \Xi_{\chi}$ and the image of $\openone - \Xi_{\chi}$ are
\begin{align}
&\mathrm{Ker}(\openone + \Xi_{\chi}) ={\rm span}
\left\{
\begin{array}{cccccccc}
(0 & 0 & -1 & 0 & 0 & 1), \\
(-1 & 0 & 0 & 0 & 1 & 0), \\
(0 & 0 & -1 & 1 & 0 & 0), \\
(-1 & 1 & 0 & 0 & 0 & 0) \\
\end{array}
\right \}.
\end{align}
\begin{align}
&\mathrm{Im}(\openone - \Xi_{\chi}) ={\rm span}
\left\{
\begin{array}{cccccccc}
( 1 & -1 & 0 & 0 & 0 & 0), \\
(0 & 0 & 1 & -1 & 0 & 0), \\
(-1 & -1 & 0 & 0 & 2 & 0), \\
(0 & 0 & -1 & -1 & 0 & 2)
\end{array}
\right \}.
\end{align}
We find that $\mathcal{X}^{\text{WTSC}} = (\mathbb{Z}_2)^2$ and that generators of $\mathcal{X}^{\text{WTSC}}$ correspond to the generators of $X^{\text{BdG}}$. 

In the following, we will simply repeat the same analysis for different setting. We will only show equations when the correspondence to the above analysis is obvious.

\subsubsection{2D}
\begin{align}
\Xi_{\chi} \left[\psi_{x}^{\alpha, \beta}\right] &= \psi_{x}^{\alpha, -\beta}\quad\text{for $x=a,b$},\\
\Xi_{\chi}\left[\psi_{c}^\beta \right] &=\psi_{c}^{-\beta}.
\end{align}
\begin{align}
\label{PH-2DAu}
\Xi_{\chi} = \begin{pmatrix}
0 & 1 & -1 & 0 & 0 & 0 & 0 & 0 \\
1 & 0 & -1 & 0 & 0 & 0 & 0 & 0 \\
0 & 0 & -1 & 0 & 0 & 0 & 0 & 0 \\
0 & 0 & 0 & 0 & 1 & -1 & 0 & 0 \\
0 & 0 & 0 & 1 & 0 & -1 & 0 & 0 \\
0 & 0 & 0 & 0 & 0 & -1 & 0 & 0 \\
0 & 0 & 1 & 0 & 0 & 1 & 0 & 1 \\
0 & 0 & 1 & 1 & 0 & 1 & 1 & 0 \\
\end{pmatrix}.
\end{align}
\begin{align}
\mathrm{Ker}(\openone + \Xi_{\chi}) =&{\rm span}
\left\{
\begin{array}{cccccccc}
( -1 & 0 & -1 & 0 & 0 & 0 & 0 & 1), \\
( -1 & 0 & -1 & 0 & 0 & 0 & 1 & 0), \\
( -1 & 0 & -1 & 1 & 0 & 1 & 0 & 0), \\
( 0 & 0 & 0 & -1 & 1 & 0 & 0 & 0), \\
( -1 & 1 & 0 & 0 & 0 & 0 & 0 & 0)
\end{array}
\right \}.
\end{align}
\begin{align}
\mathrm{Im}(\openone - \Xi_{\chi}) =&{\rm span}
\left\{
\begin{array}{cccccccc}
( 0 & 0 & 0 & 0 & 0 & 0 & 1 & -1), \\
( 0 & 0 & 0 & 1 & 1 & 2 & -1 & -1), \\
( 0 & 0 & 0 & 1 & -1 & 0 & 0 & 0),\\
( 1 & 1 & 2 & 0 & 0 & 0 & -1 & -1), \\
( 1 & -1 & 0 & 0 & 0 & 0 & 0 & 0)
\end{array}
\right \}.
\end{align}
We get $\mathcal{X}^{\text{WTSC}} = (\mathbb{Z}_2)^2$.  Now we evaluate the representation count of generators of $\mathcal{X}^{\text{WTSC}}$.
\begin{equation}
\bm{B}=(N^{\pm 3,-1}_{\Gamma},N^{\pm3,+1}_{\Gamma},N^{\pm1,-1}_{\Gamma},N^{\pm1,+1}_{\Gamma}, N^{\pm  i,-1}_{X},N^{\pm i,+1}_{X}, N^{\pm 3,-1}_{M},N^{\pm3,+1}_{M},N^{\pm1,-1}_{M},N^{\pm1,+1}_{M}).
\end{equation}
The generator $(1,0,1,0,0,0,-1,0)\in\mathcal{X}^{\text{WTSC}}$ corresponds to the atomic insulator $\psi_{a}^{\pm 1, +1} + \psi_{a}^{\pm 3, +1} - \psi_{c}^{+1}$, which has the representation count $\bm{B}_1=(0, 0, 0, 0, -1, 1, -1, 1, -1, 1)$. Similarly. the generator $(0,0,0,1,0,1,-1,0)\in\mathcal{X}^{\text{WTSC}}$ corresponds to $ \psi_{b}^{\pm 1, +1} + \psi_{b}^{\pm 3, +1}- \psi_{c}^{+1}$ with the representation $\bm{B}_2=(0, 0, 0, 0, 1, -1, -1, 1, -1, 1)$. We find that $\bm{B}_1$ belongs to $(1,4)$ and $\bm{B}_2$ belongs to $(1,0)$ of $X^{\text{BdG}}= \mathbb{Z}_2 \times \mathbb{Z}_8$.

\subsubsection{3D}
\begin{align}
\Xi_{\chi} \left[\psi_{x}^{\alpha, \beta}\right] &= \psi_{x}^{\alpha, -\beta}\quad\text{for $x=a,b,c,d$},\\
\Xi_{\chi}\left[\psi_{y}^\beta\right] &= \psi_{y}^{-\beta}\quad\text{for $y=e,f$}.
\end{align}
\begin{align}
\Xi_{\chi} =\left(
\begin{array}{ccccccccccccccccc}
0 & 1 & -1 & 0 & 0 & 0 & 0 & 0 & 1 & -1 & 0 & 0 & 0 \\
1 & 0 & -1 & 0 & 0 & 0 & 0 & 0 & 1 & -1 & 0 & 0 & 0 \\
0 & 0 & -1 & 0 & 0 & 0 & 0 & 0 & 0 & 0 & 0 & 0 & 0 \\
0 & 0 & 0 & 0 & 1 & -1 & 0 & 0 & 0 & 0 & 1 & -1 & 0 \\
0 & 0 & 0 & 1 & 0 & -1 & 0 & 0 & 0 & 0 & 1 & -1 & 0 \\
0 & 0 & 0 & 0 & 0 & -1 & 0 & 0 & 0 & 0 & 0 & 0 & 0 \\
0 & 0 & 1 & 0 & 0 & 1 & 0 & 1 & 0 & 1 & 0 & 1 & 1 \\
0 & 0 & 1 & 0 & 0 & 1 & 1 & 0 & 0 & 1 & 0 & 1 & 1 \\
0 & 0 & 0 & 0 & 0 & 0 & 0 & 0 & -1 & 0 & 0 & 0 & 0 \\
0 & 0 & 0 & 0 & 0 & 0 & 0 & 0 & 0 & -1 & 0 & 0 & 0 \\
0 & 0 & 0 & 0 & 0 & 0 & 0 & 0 & 0 & 0 & -1 & 0 & 0 \\
0 & 0 & 0 & 0 & 0 & 0 & 0 & 0 & 0 & 0 & 0 & -1 & 0 \\
0 & 0 & 0 & 0 & 0 & 0 & 0 & 0 & 0 & 0 & 0 & 0 & -1
\end{array}
\right).
\end{align}
\begin{align}
\mathrm{Ker}(\openone + \Xi_{\chi}) =&{\rm span}
\left\{
\begin{array}{ccccccccccccccccc}
(-1 & 0 & -1 & 0 & 0 & 0 & 0 & 0 & 0 & 0 & 0 & 0 & 1), \\
(-1 & 0 & -1 & 1 & 0 & 0 & 0 & 0 & 0 & 0 & 0 & 1 & 0), \\
(0 & 0 & 0 & -1 & 0 & 0 & 0 & 0 & 0 & 0 & 1 & 0 & 0), \\
(0 & 0 & -1 & 0 & 0 & 0 & 0 & 0 & 0 & 1 & 0 & 0 & 0), \\
(-1 & 0 & 0 & 0 & 0 & 0 & 0 & 0 & 1 & 0 & 0 & 0 & 0), \\
(-1 & 0 & -1 & 0 & 0 & 0 & 0 & 1 & 0 & 0 & 0 & 0 & 0), \\
(-1 & 0 & -1 & 0 & 0 & 0 & 1 & 0 & 0 & 0 & 0 & 0 & 0), \\
(-1 & 0 & -1 & 1 & 0 & 1 & 0 & 0 & 0 & 0 & 0 & 0 & 0), \\
(0 & 0 & 0 & -1 & 1 & 0 & 0 & 0 & 0 & 0 & 0 & 0 & 0),\\
(-1 & 1 & 0 & 0 & 0 & 0 & 0 & 0 & 0 & 0 & 0 & 0 & 0)
\end{array}
\right \}.
\end{align}
\begin{align}
\mathrm{Im}(\openone - \Xi_{\chi}) =&{\rm span}
\left\{
\begin{array}{ccccccccccccccccc}
(1 & -1 & 0 & 0 & 0 & 0 & 0 & 0 & 0 & 0 & 0 & 0 & 0), \\
(1 & 1 & 2 & 0 & 0 & 0 & -1 & -1 & 0 & 0 & 0 & 0 & 0), \\
(0 & 0 & 0 & 1 & -1 & 0 & 0 & 0 & 0 & 0 & 0 & 0 & 0), \\
(0 & 0 & 0 & 1 & 1 & 2 & -1 & -1 & 0 & 0 & 0 & 0 & 0), \\
(0 & 0 & 0 & 0 & 0 & 0 & 1 & -1 & 0 & 0 & 0 & 0 & 0), \\
(-1 & -1 & 0 & 0 & 0 & 0 & 0 & 0 & 2 & 0 & 0 & 0 & 0), \\
(1 & 1 & 0 & 0 & 0 & 0 & -1 & -1 & 0 & 2 & 0 & 0 & 0), \\
(0 & 0 & 0 & -1 & -1 & 0 & 0 & 0 & 0 & 0 & 2 & 0 & 0), \\
(0 & 0 & 0 & 1 & 1 & 0 & -1 & -1 & 0 & 0 & 0 & 2 & 0), \\
(0 & 0 & 0 & 0 & 0 & 0 & -1 & -1 & 0 & 0 & 0 & 0 & 2) \\
\end{array}
\right \}.
\end{align}
We find $\mathcal{X}^{\text{WTSC}} = (\mathbb{Z}_{2})^{7}$. We evaluate the representation count of  generators of $\mathcal{X}^{\text{WTSC}}$
\begin{align}
&(1,0,1,0,0,-1,0,-1,0,0,0,0,0), \\
&(1,0,1,0,0,0,0,0,0,0,-1,0,0), \\
&(1,0,0,-1,0,0,0,0,0,0,0,0,0,0), \\
&(0,0,1,0,0,-1,0,0,0,0,0,0,0), \\
&(0,0,0,0,0,1,0,0,-1,0,0,0,0), \\
&(0,0,0,0,0,0,0,1,0,-1,0,0,0), \\
&(0,0,0,0,0,0,0,0,0,0,1,0,-1)
\end{align}
in the same way as in the 2D case and find their indices. The results can be summarized as
\begin{align}
\begin{array}{c|ccccccc}
\bm{B}_i &\nu^{\Gamma \rightarrow Z}_{1/2} & \nu^{\Gamma \rightarrow Z}_{3/2}& \nu^{M \rightarrow A}_{1/2} & \nu^{\Gamma \rightarrow X} &\nu^{2\text{D}}_{k_x=0} & z_8 & z_{16}  \\
\hline
\bm{B}_1& 1 & 1 & 1 & 0 & 0 & 0 & 8 \\
\bm{B}_2& 0 & 0 & 0 & 1 & 2 & 4 & 8 \\
\bm{B}_3& 1 & 0 & 1 & 0 & 2 & 0 & 0 \\
\bm{B}_4& 0 & 1 & 0 & 0 & 2 & 0 & 8 \\
\bm{B}_5& 1 & 0 & 0 & 0 & 0 & 0 & 0 \\
\bm{B}_6& 0 & 1 & 1 & 0 & 0 & 0 & 0 \\
\bm{B}_7& 1 & 1 & 1 & 0 & 2 & 0 & 0 
\end{array}
\end{align}

\subsection{$B_u$ representation}
\subsubsection{1D}
\begin{align}
\Xi_{\chi} \left[\psi_{x}^{\pm 1, \beta}\right] &= \psi_{x}^{\pm 3, -\beta}\quad\text{for $x=a,b$},\\
\Xi_{\chi} \left[\psi_{x}^{\pm 3, \beta} \right] &= \psi_{x}^{\pm 1, -\beta}\quad\text{for $x=a,b$}.
\end{align}
\begin{align}
\Xi_{\chi} = \begin{pmatrix}
0 & 0 & 0 & 1 & 0 & 1 \\
0 & 0 & 1 & 0 & 0 & 1 \\
0 & 1 & 0 & 0 & 1 & 0 \\
1 & 0 & 0 & 0 & 1 & 0 \\
0 & 0 & 0 & 0 & 0 & -1 \\
0 & 0 & 0 & 0 & -1 & 0 \\
\end{pmatrix}.
\end{align}
\begin{align}
\mathrm{Ker}(\openone + \Xi_{\chi}) =&{\rm span}
\left\{
\begin{array}{cccccccc}
(-1 & -1 & 0 & 0 & 1 & 1), \\
(-1 & 0 & 0 & 1 & 0 & 0), \\
(0 & -1 & 1 & 0 & 0 & 0) \\
\end{array}
\right \}.
\end{align}
\begin{align}
\mathrm{Im}(\openone - \Xi_{\chi}) =&{\rm span}
\left\{
\begin{array}{cccccccc}
( 1 & 0 & 0 & -1 & 0 & 0), \\
(0 & 1 & -1 & 0 & 0 & 0), \\
(0 & 0 & -1 & -1 & 1 & 1)
\end{array}
\right \}.
\end{align}
We find $\mathcal{X}^{\text{WTSC}}$ is trivial.

\subsubsection{2D}
\begin{align}
\Xi_{\chi} \left[\psi_{x}^{\pm 1, \beta}\right] &= \psi_{x}^{\pm 3, -\beta}\quad\text{for $x=a,b$},\nonumber\\
\Xi_{\chi} \left[\psi_{x}^{\pm 3, \beta} \right] &= \psi_{x}^{\pm 1, -\beta}\quad\text{for $x=a,b$},\nonumber\\
\Xi_{\chi} \left[\psi_{c}^\beta \right] &= \psi_{c}^{-\beta}.
\end{align}
\begin{align}
\Xi_{\chi} = 
\left(
\begin{array}{cccccccc}
-1 & 0 & 0 & 0 & 0 & 0 & 0 & 0 \\
-1 & 0 & 1 & 0 & 0 & 0 & 0 & 0 \\
-1 & 1 & 0 & 0 & 0 & 0 & 0 & 0 \\
0 & 0 & 0 & -1 & 0 & 0 & 0 & 0 \\
0 & 0 & 0 & -1 & 0 & 1 & 0 & 0 \\
0 & 0 & 0 & -1 & 1 & 0 & 0 & 0 \\
1 & 0 & 0 & 1 & 0 & 0 & 0 & 1 \\
1 & 0 & 0 & 1 & 0 & 0 & 1 & 0 \\
\end{array}
\right).
\end{align}
\begin{align}
\mathrm{Ker}(\openone + \Xi_{\chi} ) =&{\rm span}
\left\{
\begin{array}{cccccccc}
( -1 & -1 & 0 & 0 & 0 & 0 & 0 & 1), \\
( -1 & -1 & 0 & 0 & 0 & 0 & 1 & 0), \\
( -1 & -1 & 0 & 1 & 0 & 1 & 0 & 0), \\
( -1 & -1 & 0 & 1 & 1 & 0 & 0 & 0), \\
( 0 & -1 & 1 & 0 & 0 & 0 & 0 & 0)
\end{array}
\right \}.
\end{align}
\begin{align}
\mathrm{Im}(\openone - \Xi_{\chi} ) = & {\rm span}
\left\{
\begin{array}{cccccccc}
( 0 & 1 & -1 & 0 & 0 & 0 & 0 & 0), \\
( 0 & 0 & 0 & 0 & 1 & -1 & 0 & 0), \\
( 0 & 0 & 0 & 0 & 0 & 0 & 1 & -1), \\
( 2 & 1 & 1 & 0 & 0 & 0 & -1 & -1), \\
( 0 & 0 & 0 & 2 & 1 & 1 & -1 & -1)
\end{array}
\right \}.
\end{align}
$\mathcal{X}^{\text{WTSC}} = (\mathbb{Z}_2)^2$ and the generators $(1,0,1,0,0,0,-1,0)$ and $(0,0,0,1,0,1,-1,0)$ belong to $(1,4) $ and $(1,0)$ of $X^{\text{BdG}}=\mathbb{Z}_2 \times \mathbb{Z}_8$.

\subsubsection{3D}
\begin{align}
\Xi_{\chi} \left[\psi_{x}^{\pm 1, \beta}\right] &= \psi_{x}^{\pm 3, -\beta}\quad\text{for $x=a,b,c,d$},\\
\Xi_{\chi} \left[\psi_{x}^{\pm 3, \beta}\right] &= \psi_{x}^{\pm 1, -\beta}\quad\text{for $x=a,b,c,d$},\\
\Xi_{\chi}\left[\psi_{y}^\beta \right] &= \psi_{y}^{-\beta}\quad\text{for $y=e,f$}.
\end{align}
\begin{align}
\Xi_{\chi} =\left(
\begin{array}{ccccccccccccccccc}
-1 & 0 & 0 & -1 & 1 & 0 & 0 & 0 & 0 & 0 & 0 & 0 & 0 \\
-1 & 0 & 1 & -1 & 1 & 0 & 0 & 0 & 0 & 0 & 0 & 0 & 0 \\
-1 & 1 & 0 & -1 & 0 & 0 & 0 & 0 & 0 & 0 & 0 & 0 & 0 \\
0 & 0 & 0 & 0 & -1 & 0 & 0 & 0 & 0 & 0 & 0 & 0 & 0 \\
0 & 0 & 0 & 0 & 0 & 0 & 0 & 0 & 0 & 0 & 0 & 0 & 0 \\
0 & 0 & 0 & 0 & 0 & -1 & 0 & 0 & -1 & 1 & 0 & 0 & 0 \\
0 & 0 & 0 & 0 & 0 & -1 & 0 & 1 & -1 & 1 & 0 & 0 & 0 \\
0 & 0 & 0 & 0 & 0 & -1 & 1 & 0 & 0 & 0 & 0 & 0 & 0 \\
0 & 0 & 0 & 0 & 0 & 0 & 0 & 0 & 0 & -1 & 0 & 0 & 0 \\
0 & 0 & 0 & 0 & 0 & 0 & 0 & 0 & 0 & 0 & 0 & 0 & 0 \\
1 & 0 & 0 & 1 & 0 & 1 & 0 & 0 & 1 & 0 & 0 & 1 & 1 \\
1 & 0 & 0 & 1 & 0 & 1 & 0 & 0 & 1 & 0 & 1 & 0 & 1 \\
0 & 0 & 0 & 0 & 0 & 0 & 0 & 0 & 0 & 0 & 0 & 0 & -1
\end{array}
\right),
\end{align}
\begin{align}
\mathrm{Ker}(\openone + \Xi_{\chi}) =&{\rm span}
\left\{
\begin{array}{ccccccccccccccccc}
(-1 & -1 & 0 & 0 & 0 & 0 & 0 & 0 & 0 & 0 & 0 & 0 & 1), \\
(-1 & -1 & 0 & 0 & 0 & 0 & 0 & 0 & 0 & 0 & 0 & 1 & 0), \\
(-1 & -1 & 0 & 0 & 0 & 0 & 0 & 0 & 0 & 0 & 1 & 0 & 0), \\
(-1 & -1 & 0 & 0 & 0 & 0 & 0 & 0 & 1 & 1 & 0 & 0 & 0), \\
(-1 & -1 & 0 & 0 & 0 & 1 & 0 & 1 & 0 & 0 & 0 & 0 & 0), \\
(-1 & -1 & 0 & 0 & 0 & 1 & 1 & 0 & 0 & 0 & 0 & 0 & 0), \\
(-1 & -1 & 0 & 1 & 1 & 0 & 0 & 0 & 0 & 0 & 0 & 0 & 0), \\
(0 & -1 & 1 & 0 & 0 & 0 & 0 & 0 & 0 & 0 & 0 & 0 & 0)
\end{array}
\right \}.
\end{align}
\begin{align}
\mathrm{Im}(\openone - \Xi_{\chi}) =&{\rm span}
\left\{
\begin{array}{ccccccccccccccccc}
(2 & 1 & 1 & 0 & 0 & 0 & 0 & 0 & 0 & 0 & -1 & -1 & 0), \\
(0 & 1 & -1 & 0 & 0 & 0 & 0 & 0 & 0 & 0 & 0 & 0 & 0), \\
(1 & 1 & 0 & 1 & 1 & 0 & 0 & 0 & 0 & 0 & -1 & -1 & 0), \\
(0 & 0 & 0 & 0 & 0 & 2 & 1 & 1 & 0 & 0 & -1 & -1 & 0), \\
(0 & 0 & 0 & 0 & 0 & 0 & 1 & -1 & 0 & 0 & 0 & 0 & 0), \\
(0 & 0 & 0 & 0 & 0 & 1 & 1 & 0 & 1 & 1 & -1 & -1 & 0), \\
(0 & 0 & 0 & 0 & 0 & 0 & 0 & 0 & 0 & 0 & 1 & -1 & 0), \\
(0 & 0 & 0 & 0 & 0 & 0 & 0 & 0 & 0 & 0 & -1 & -1 & 2)
\end{array}
\right \}.
\end{align}
We find $\mathcal{X}^{\text{WTSC}} = (\mZ_{2})^{3}$.  We evaluate the representation count of generators $(0, 0, 0, 0, 0, 1, 0, 1, 0, 0, 0, 0, -1)$, $(0, 0, 0, 0, 0, 0, 0, 0, 0, 0, 1, 0, -1)$, and $(1, 1, 0, 0, 0, 1, 0, 1, 0, 0, -1, -1, 0) $ and find their indices. The results can be summarized as
\begin{align}
\begin{array}{c|cccc}
\bm{B}_i & \nu^{\Gamma \rightarrow X} &\nu^{2\text{D}}_{k_x=0} & z_8 & \kappa_1 \\
\hline
\bm{B}_1& 1 & 0 & 0 & 0 \\
\bm{B}_2& 0 & 2 & 0 & 0 \\
\bm{B}_3& 0 & 0 & 4 & 0 
\end{array}
\end{align}

\subsection{$B_g$ representation}
\subsubsection{1D}
\begin{align}
\Xi_\chi \left[\psi_{x}^{\pm 1, \beta}\right] &= \psi_{x}^{\pm 3, \beta}\quad\text{for $x=a,b$},\\
\Xi_\chi \left[\psi_{x}^{\pm 3, \beta} \right] &= \psi_{x}^{\pm 1, \beta}\quad\text{for $x=a,b$}.
\end{align}
\begin{align}
\Xi_\chi= \begin{pmatrix}
0 & 0 & 1 & 0 & 0 & 0 \\
0 & 0 & 0 & 1 & 0 & 0 \\
1 & 0 & 0 & 0 & 0 & 0 \\
0 & 1 & 0 & 0 & 0 & 0 \\
0 & 0 & 0 & 0 & 0 & 1 \\
0 & 0 & 0 & 0 & 1 & 0 \\
\end{pmatrix}.
\end{align}
\begin{align}
\mathrm{Ker}(\openone + \Xi_\chi)=&{\rm span}
\left\{
\begin{array}{cccccccc}
(0 & 0 & 0 & 0 & -1 & 1), \\
(0 & -1 & 0 & 1 & 0 & 0), \\
(-1 & 0 & 1 & 0 & 0 & 0) \\
\end{array}
\right \},
\end{align}
\begin{align}
\mathrm{Im}(\openone - \Xi_\chi)=&{\rm span}
\left\{
\begin{array}{cccccccc}
(1 & 0 & -1 & 0 & 0 & 0), \\
(0 & 1 & 0 & -1 & 0 & 0), \\
(0 & 0 & 0 & 0 & 1 & -1)
\end{array}
\right \}.
\end{align}
$\mathcal{X}^{\text{WTSC}}$ is trivial.

\subsubsection{2D}
\begin{align}
\Xi_\chi \left[\psi_{x}^{\pm 1, \beta}\right] &= \psi_{x}^{\pm 3, \beta}\quad\text{for $x=a,b$},\\
\Xi_\chi \left[\psi_{x}^{\pm 3, \beta} \right] &= \psi_{x}^{\pm 1, \beta}\quad\text{for $x=a,b$}, \\
\Xi_\chi \left[\psi_{c}^{\pm 1, \beta} \right] &= \psi_{c}^{\pm 1, \beta}.
\end{align}
\begin{align}
\Xi_\chi = 
\left(
\begin{array}{cccccccc}
0 & -1 & 1 & 0 & 0 & 0 & 0 & 0 \\
0 & -1 & 0 & 0 & 0 & 0 & 0 & 0 \\
1 & -1 & 0 & 0 & 0 & 0 & 0 & 0 \\
0 & 0 & 0 & 0 & -1 & 1 & 0 & 0 \\
0 & 0 & 0 & 0 & -1 & 0 & 0 & 0 \\
0 & 0 & 0 & 1 & -1 & 0 & 0 & 0 \\
0 & 1 & 0 & 0 & 1 & 0 & 1 & 0 \\
0 & 1 & 0 & 0 & 1 & 0 & 0 & 1 \\
\end{array}
\right).
\end{align}
\begin{align}
\mathrm{Ker}(\openone + \Xi_\chi ) =&{\rm span}
\left\{
\begin{array}{cccccccc}
(1, 0, -1, 0, 0, 0, 0, 0),\\
(0, 0, 0, 1, 0, -1, 0, 0),\\
(0, 1, 1,   0, -1, -1, 0, 0),\\
(0, -1, -1, 0, -1, -1, 1, 1)
\end{array}
\right\}.
\end{align}
\begin{align}
\mathrm{Im}(\openone -\Xi_\chi ) =&{\rm span}
\left\{
\begin{array}{cccccccc}
(1, 0, -1, 0, 0, 0, 0, 0),\\
(0, 0, 0, 1, 0, -1, 0, 0),\\
(1, 2, 1, 0, 0, 0, -1, -1),\\
(0, 0, 0, 1, 2, 1, -1, -1)
\end{array}
\right\}.
\end{align}
$\mathcal{X}^{\text{WTSC}} = \mathbb Z_2$ and the generator $(0,1,1,0,1,1,-1,-1)$ belongs to $(1,1)$ of $X^{\text{BdG}}=(\mathbb{Z}_2)^2$.

\subsubsection{3D}
\begin{align}
\Xi_\chi \left[\psi_{x}^{\pm 1, \beta}\right] &= \psi_{x}^{\pm 3, \beta}\quad\text{for $x=a,b,c,d$},\\
\Xi_\chi \left[\psi_{x}^{\pm 3, \beta}\right] &= \psi_{x}^{\pm 1, \beta}\quad\text{for $x=a,b,c,d$},\\
\Xi_\chi \left[\psi_{y}^\beta \right] &= \psi_{y}^\beta\quad\text{for $y=e,f$}.
\end{align}
\begin{align}
\Xi_\chi =\left(
\begin{array}{ccccccccccccccccc}
0 & -1 & 1 & 0 & 0 & 0 & 0 & 0 & 0 & 0 & 0 & 0 & 0 \\
0 & -1 & 0 & 0 & 0 & 0 & 0 & 0 & 0 & 0 & 0 & 0 & 0 \\
1 & -1 & 0 & 0 & 0 & 0 & 0 & 0 & 0 & 0 & 0 & 0 & 0 \\
0 & 0 & 0 & 0 & 1 & 0 & 0 & 0 & 0 & 0 & 0 & 0 & 0 \\
0 & 0 & 0 & 1 & 0 & 0 & 0 & 0 & 0 & 0 & 0 & 0 & 0 \\
0 & 0 & 0 & 0 & 0 & 0 & -1 & 1 & 0 & 0 & 0 & 0 & 0 \\
0 & 0 & 0 & 0 & 0 & 0 & -1 & 0 & 0 & 0 & 0 & 0 & 0 \\
0 & 0 & 0 & 0 & 0 & 1 & -1 & 0 & 0 & 0 & 0 & 0 & 0 \\
0 & 0 & 0 & 0 & 0 & 0 & 0 & 0 & 0 & 1 & 0 & 0 & 0 \\
0 & 0 & 0 & 0 & 0 & 0 & 0 & 0 & 1 & 0 & 0 & 0 & 0 \\
0 & 1 & 0 & 0 & 0 & 0 & 1 & 0 & 0 & 0 & 1 & 0 & 0 \\
0 & 1 & 0 & 0 & 0 & 0 & 1 & 0 & 0 & 0 & 0 & 1 & 0 \\
0 & 0 & 0 & 0 & 0 & 0 & 0 & 0 & 0 & 0 & 0 & 0 & 1
\end{array}
\right).
\end{align}
\begin{align}
\mathrm{Ker}(\openone + \Xi_\chi)=&{\rm span}
\left\{
\begin{array}{ccccccccccccccccc}
(-2 & -2 & 0 & 0 & 0 & 0 & 0 & 0 & 0 & 0 & 1 & 1 & 0), \\
(0 & 0 & 0 & 0 & 0 & 0 & 0 & 0 & -1 & 1 & 0 & 0 & 0), \\
(0 & 0 & 0 & 0 & 0 & -1 & 0 & 1 & 0 & 0 & 0 & 0 & 0), \\
(-1 & -1 & 0 & 0 & 0 & 1 & 1 & 0 & 0 & 0 & 0 & 0 & 0), \\
(0 & 0 & 0 & -1 & 1 & 0 & 0 & 0 & 0 & 0 & 0 & 0 & 0), \\
(-1 & 0 & 1 & 0 & 0 & 0 & 0 & 0 & 0 & 0 & 0 & 0 & 0) 
\end{array}
\right \}.
\end{align}
\begin{align}
\mathrm{Im}(\openone - \Xi_\chi)=&{\rm span}
\left\{
\begin{array}{ccccccccccccccccc}
(1 & 0 & -1 & 0 & 0 & 0 & 0 & 0 & 0 & 0 & 0 & 0 & 0), \\
(1 & 2 & 1 & 0 & 0 & 0 & 0 & 0 & 0 & 0 & -1 & -1 & 0), \\
(0 & 0 & 0 & 1 & -1 & 0 & 0 & 0 & 0 & 0 & 0 & 0 & 0), \\
(0 & 0 & 0 & 0 & 0 & 1 & 0 & -1 & 0 & 0 & 0 & 0 & 0), \\
(0 & 0 & 0 & 0 & 0 & 1 & 2 & 1 & 0 & 0 & -1 & -1 & 0), \\
(0 & 0 & 0 & 0 & 0 & 0 & 0 & 0 & 1 & -1 & 0 & 0 & 0) \\
\end{array}
\right \}.
\end{align}
$\mathcal{X}^{\text{WTSC}} = \mathbb Z_2$ and the generator $(1, 1, 0, 0, 0, 0, -1, -1, 0, 0, 0, 0, 0)$ belongs to $(1,1,0)$ of $X^{\text{BdG}}=(\mathbb{Z}_2)^3$.

\clearpage

\section{Symmetry indicator of $P4$ in the weak-pairing limit}
\label{P4_SI}
In the main text, we defined the index $\nu_{C_4}$ in the system  with $\chi_{C_4}=-1$ by
\begin{align}
\nu_{C_4} &= \frac{1}{2\sqrt{2}}\sum_{\alpha=\pm 1,\pm 3}\sum_{\bk \in \Gamma, M}e^{i\frac{\pi\alpha}{4}}N_{\bk}^{\alpha} = \frac{1}{2}\sum_{\bk \in \Gamma, M}\left(N_{\bk}^{1} - N_{\bk}^{3} \right).
\end{align}
In this section, we relate $\nu_{C_4}$ to the Fermi surfaces of the normal phase. In the weak-pairing limit, $N_{\bk}^{\alpha}$ is reduced 
\begin{align}
N_{\bk}^{1} &= n_{\bk}^{1} - n_{\bk}^{3},\\ 
N_{\bk}^{3} &= n_{\bk}^{3} - n_{\bk}^{1}.
\end{align}
Here $n_{\bk}^{\alpha}$ represents the number of occupied states in the normal phase which have the rotation eigenvalue $e^{i\tfrac{\pi\alpha}{4}}$. Hence, $\nu_{C_4}$ can be expressed as
\begin{align}
\nu_{C_4}&=\sum_{\bk \in \Gamma, M}\left(n_{\bk}^{1} - n_{\bk}^{3} \right).
\end{align}
Furthermore, the total number of occupied bands at the high-symmetry point is given by
\begin{align}
n_{\bk, \text{occ}} &= n_{\bk}^{1} + n_{\bk}^{-1} + n_{\bk}^{3} + n_{\bk}^{-3}= 2n_{\bk}^{1} + 2n_{\bk}^{3},
\end{align}
where we used $n_{\bk}^{\alpha} = n_{\bk}^{-\alpha}$ in the last line. From these relations, we get
\begin{align}
\nu_{C_4}&= \frac{n_{\Gamma, \text{occ}} + n_{M, \text{occ}}}{2} - 2n_{\Gamma}^{3} - 2n_{M}^{3} = \frac{n_{\Gamma, \text{occ}} + n_{M, \text{occ}}}{2} \mod 2.
\end{align}
We conclude that $\nu_{C_4}$ in the weak-pairing limit counts the number of filled Kramers pairs in the normal phase.

\end{document}